\documentclass[a4paper,11pt]{article}
\usepackage{jcappub} 

\usepackage[T1]{fontenc}

\usepackage{graphicx}
\usepackage{rotating}
\graphicspath{{figures/}}
\usepackage{amssymb,amsmath}
\usepackage{subfiles}
\usepackage{hyperref}
\usepackage{footmisc}
\usepackage{float}
\usepackage{multirow}
\usepackage[acronym,nonumberlist,nomain]{glossaries}

\newcommand{\ov}{\langle\sigma v\rangle_\mathrm{ann}}
\newcommand{\SEC}{Sec.~}
\newcommand{\FIG}{Fig.~}
\newcommand{\TAB}{Tab.~}
\newcommand{\EQ}{Eq.~}
\newcommand{\orcid}[1]{\unskip\protect\href{https://orcid.org/#1}{\protect\includegraphics[width=8pt,clip]{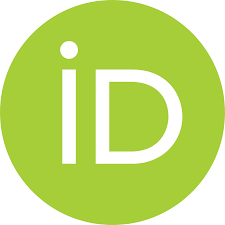}}}

\newcommand{\ctools}{{\sf ctools v$1.6.3$}}
\newcommand{\clumpy}{{\sf CLUMPY v$3.0.1$}}
\newcommand{\darksusy}{{\sf DarkSUSY v$6$}}

\newcommand{\Aqnu}{\tilde{q}_\nu}

\newcommand{\Anq}{\nu}
\newcommand{\Asigma}{\sigma_A}

\usepackage{lineno}
\usepackage{color}
\usepackage{bm}

\newcommand{\be}{\begin{equation}}
\newcommand{\ee}{\end{equation}}
\newcommand{\bea}{\begin{eqnarray}}
\newcommand{\eea}{\end{eqnarray}}

\title{Dark Matter Line Searches with the Cherenkov Telescope Array} 

\date{} 
\author{
\mbox{S.~Abe$^{\ref{AFFIL::JapanUTokyoICRR}}$\orcid{0000-0001-7250-3596}}, 
  \mbox{J.~Abhir$^{\ref{AFFIL::SwitzerlandETHZurich}}$\orcid{0000-0001-8215-4377}}, 
  \mbox{A.~Abhishek$^{\ref{AFFIL::ItalyUSienaandINFN}}$\orcid{0009-0005-5239-7905}}, 
  \mbox{F.~Acero$^{\ref{AFFIL::FranceCEAIRFUDAp},\ref{AFFIL::SpainFSLACIRLCNRSIAC}}$\orcid{0000-0002-6606-2816}}, 
  \mbox{A.~Acharyya$^{\ref{AFFIL::USAUAlabamaTuscaloosa}}$\orcid{0000-0002-2028-9230}}, 
  \mbox{R.~Adam$^{\ref{AFFIL::FranceOCotedAzur},\ref{AFFIL::FranceLLREcolePolytechnique}}$\orcid{0009-0000-5380-1109}}, 
  \mbox{A.~Aguasca-Cabot$^{\ref{AFFIL::SpainICCUB}}$\orcid{0000-0001-8816-4920}}, 
  \mbox{I.~Agudo$^{\ref{AFFIL::SpainIAACSIC}}$\orcid{0000-0002-3777-6182}}, 
  \mbox{A.~Aguirre-Santaella$^{\ref{AFFIL::UnitedKingdomICCUDurham}}$\orcid{0000-0002-9581-7288}}, 
  \mbox{J.~Alfaro$^{\ref{AFFIL::ChileUPontificiaCatolicadeChile}}$}, 
  \mbox{R.~Alfaro$^{\ref{AFFIL::MexicoUNAMMexico}}$}, 
  \mbox{N.~Alvarez-Crespo$^{\ref{AFFIL::SpainUCMAltasEnergias}}$}, 
  \mbox{R.~Alves~Batista$^{\ref{AFFIL::SpainIFTUAMCSIC}}$\orcid{0000-0003-2656-064X}}, 
  \mbox{J.-P.~Amans$^{\ref{AFFIL::FranceObservatoiredeParis}}$}, 
  \mbox{E.~Amato$^{\ref{AFFIL::ItalyOArcetri}}$\orcid{0000-0002-9881-8112}}, 
  \mbox{G.~Ambrosi$^{\ref{AFFIL::ItalyUPerugiaandINFN}}$}, 
  \mbox{L.~Angel$^{\ref{AFFIL::BrazilURioGrandedoNorteIIP},\ref{AFFIL::BrazilURioGrandedoNortePhys}}$\orcid{0000-0003-3684-6553}}, 
  \mbox{C.~Aramo$^{\ref{AFFIL::ItalyINFNNapoli}}$}, 
  \mbox{C.~Arcaro$^{\ref{AFFIL::ItalyINFNPadova}}$\orcid{0000-0002-1998-9707}}, 
  \mbox{T.~T.~H.~Arnesen$^{\ref{AFFIL::SpainIAC}}$\orcid{0009-0004-0816-0700}}, 
  \mbox{L.~Arrabito$^{\ref{AFFIL::FranceLUPMUMontpellier}}$\orcid{0000-0003-4727-7288}}, 
  \mbox{K.~Asano$^{\ref{AFFIL::JapanUTokyoICRR}}$\orcid{0000-0001-9064-160X}}, 
  \mbox{Y.~Ascasibar$^{\ref{AFFIL::SpainIFTUAMCSIC}}$\orcid{0000-0003-1577-2479}}, 
  \mbox{J.~Aschersleben$^{\ref{AFFIL::NetherlandsUGroningen}}$\orcid{0000-0002-6097-7898}}, 
  \mbox{H.~Ashkar$^{\ref{AFFIL::FranceLLREcolePolytechnique}}$\orcid{0000-0002-2153-1818}}, 
  \mbox{M.~Backes$^{\ref{AFFIL::NamibiaUNamibia},\ref{AFFIL::SouthAfricaNWU}}$\orcid{0000-0002-9326-6400}}, 
  \mbox{A.~Baktash$^{\ref{AFFIL::GermanyUHamburg}}$\orcid{0000-0002-5439-117X}}, 
  \mbox{C.~Balazs$^{\ref{AFFIL::AustraliaUMonash}}$\orcid{0000-0001-7154-1726}}, 
  \mbox{M.~Balbo$^{\ref{AFFIL::SwitzerlandUGenevaDPNC}}$\orcid{0000-0002-6556-3344}}, 
  \mbox{A.~Baquero~Larriva$^{\ref{AFFIL::SpainUCMAltasEnergias},\ref{AFFIL::EcuadorUAzuay}}$\orcid{0000-0002-1757-5826}}, 
  \mbox{V.~Barbosa~Martins$^{\ref{AFFIL::GermanyDESY}}$\orcid{0000-0002-5085-8828}}, 
  \mbox{U.~Barres~de~Almeida$^{\ref{AFFIL::BrazilCBPF},\ref{AFFIL::BrazilIAGUSaoPaulo}}$\orcid{0000-0001-7909-588X}}, 
  \mbox{J.~A.~Barrio$^{\ref{AFFIL::SpainUCMAltasEnergias}}$\orcid{0000-0002-0965-0259}}, 
  \mbox{I.~Batkovi\'c$^{\ref{AFFIL::ItalyUPadovaandINFN}}$\orcid{0000-0002-1209-2542}}, 
  \mbox{R.~Batzofin$^{\ref{AFFIL::GermanyUPotsdam}}$\orcid{0000-0002-5797-3386}}, 
  \mbox{J.~Baxter$^{\ref{AFFIL::JapanUTokyoICRR}}$\orcid{0009-0004-9545-794X}}, 
  \mbox{J.~Becerra~Gonz\'alez$^{\ref{AFFIL::SpainIAC}}$}, 
  \mbox{G.~Beck$^{\ref{AFFIL::SouthAfricaUWitwatersrand}}$\orcid{0000-0003-4916-4914}}, 
  \mbox{W.~Benbow$^{\ref{AFFIL::USACfAHarvardSmithsonian}}$\orcid{0000-0003-2098-170X}}, 
  \mbox{D.~Berge$^{\ref{AFFIL::GermanyUBerlin},\ref{AFFIL::GermanyDESY}}$\orcid{0000-0002-2918-1824}}, 
  \mbox{E.~Bernardini$^{\ref{AFFIL::ItalyUPadovaandINFN}}$\orcid{0000-0003-3108-1141}}, 
  \mbox{J.~Bernete$^{\ref{AFFIL::SpainCIEMAT}}$\orcid{0000-0002-8108-7552}}, 
  \mbox{K.~Bernl\"ohr$^{\ref{AFFIL::GermanyMPIK}}$\orcid{0000-0001-8065-3252}}, 
  \mbox{A.~Berti$^{\ref{AFFIL::GermanyMPP}}$\orcid{0000-0003-0396-4190}}, 
  \mbox{B.~Bertucci$^{\ref{AFFIL::ItalyUPerugiaandINFN}}$\orcid{0000-0001-7584-293X}}, 
  \mbox{P.~Bhattacharjee$^{\ref{AFFIL::FranceLAPPUSavoieMontBlanc}}$\orcid{0000-0002-0258-3831}}, 
  \mbox{S.~Bhattacharyya$^{\ref{AFFIL::SloveniaUNovaGoricaCAC}}$\orcid{0000-0002-6569-5953}}, 
  \mbox{C.~Bigongiari$^{\ref{AFFIL::ItalyORoma}}$}, 
  \mbox{A.~Biland$^{\ref{AFFIL::SwitzerlandETHZurich}}$}, 
  \mbox{E.~Bissaldi$^{\ref{AFFIL::ItalyPolitecnicoBari},\ref{AFFIL::ItalyINFNBari}}$\orcid{0000-0001-9935-8106}}, 
  \mbox{J.~Biteau$^{\ref{AFFIL::FranceIJCLab},\ref{AFFIL::FranceIUFInstitutuniversitairedeFrance}}$\orcid{0000-0002-4202-8939}}, 
  \mbox{O.~Blanch$^{\ref{AFFIL::SpainIFAEBIST}}$\orcid{0000-0002-8380-1633}}, 
  \mbox{J.~Blazek$^{\ref{AFFIL::CzechRepublicFZU}}$\orcid{0000-0002-5870-8947}}, 
  \mbox{F.~Bocchino$^{\ref{AFFIL::ItalyOPalermo}}$\orcid{0000-0002-2321-5616}}, 
  \mbox{C.~Boisson$^{\ref{AFFIL::FranceObservatoiredeParis}}$\orcid{0000-0001-5893-1797}}, 
  \mbox{J.~Bolmont$^{\ref{AFFIL::FranceLPNHEUSorbonne}}$\orcid{0000-0003-4739-8389}}, 
  \mbox{G.~Bonnoli$^{\ref{AFFIL::ItalyOBrera},\ref{AFFIL::ItalyINFNPisa}}$\orcid{0000-0003-2464-9077}}, 
  \mbox{A.~Bonollo$^{\ref{AFFIL::ItalyIUSSPaviaINAF},\ref{AFFIL::ItalyUTrento}}$\orcid{0009-0004-5418-6485}}, 
  \mbox{P.~Bordas$^{\ref{AFFIL::SpainICCUB}}$\orcid{0000-0002-0266-8536}}, 
  \mbox{Z.~Bosnjak$^{\ref{AFFIL::CroatiaUZagreb}}$\orcid{0000-0001-6536-0320}}, 
  \mbox{E.~Bottacini$^{\ref{AFFIL::ItalyUPadovaandINFN}}$}, 
  \mbox{M.~B\"ottcher$^{\ref{AFFIL::SouthAfricaNWU}}$\orcid{0000-0002-8434-5692}}, 
  \mbox{T.~Bringmann$^{\ref{AFFIL::NorwayUOslo}}$\orcid{0000-0002-0339-8144}}, 
  \mbox{E.~Bronzini$^{\ref{AFFIL::ItalyOASBologna}}$\orcid{0000-0001-8378-4303}}, 
  \mbox{R.~Brose$^{\ref{AFFIL::IrelandDCU},\ref{AFFIL::IrelandDIAS}}$\orcid{0000-0002-8312-6930}}, 
  \mbox{A.~M.~Brown$^{\ref{AFFIL::UnitedKingdomUDurham}}$\orcid{0000-0003-0259-3148}}, 
  \mbox{G.~Brunelli$^{\ref{AFFIL::ItalyOASBologna}}$}, 
  \mbox{A.~Bulgarelli$^{\ref{AFFIL::ItalyOASBologna}}$\orcid{0000-0001-6347-0649}}, 
  \mbox{T.~Bulik$^{\ref{AFFIL::PolandUWarsawPhysics}}$}, 
  \mbox{I.~Burelli$^{\ref{AFFIL::ItalyUUdineandINFNTrieste}}$}, 
  \mbox{L.~Burmistrov$^{\ref{AFFIL::SwitzerlandUGenevaDPNC}}$}, 
  \mbox{M.~Burton$^{\ref{AFFIL::UnitedKingdomArmaghObservatoryandPlanetarium},\ref{AFFIL::AustraliaUNewSouthWales}}$\orcid{0000-0001-7289-1998}}, 
  \mbox{M.~Buscemi$^{\ref{AFFIL::ItalyINFNCatania}}$}, 
  \mbox{T.~Bylund$^{\ref{AFFIL::FranceCEAIRFUDAp}}$\orcid{0000-0003-2946-1313}}, 
  \mbox{J.~Cailleux$^{\ref{AFFIL::FranceObservatoiredeParis}}$}, 
  \mbox{A.~Campoy-Ordaz$^{\ref{AFFIL::SpainUABandCERESIEEC}}$\orcid{0000-0001-9352-8936}}, 
  \mbox{B.~K.~Cantlay$^{\ref{AFFIL::ThailandUKasetsart},\ref{AFFIL::ThailandNARIT}}$\orcid{0009-0002-8750-6401}}, 
  \mbox{G.~Capasso$^{\ref{AFFIL::ItalyOCapodimonte}}$\orcid{0000-0002-4472-4858}}, 
  \mbox{A.~Caproni$^{\ref{AFFIL::BrazilUCidadeSPaulo}}$\orcid{0000-0001-9707-3895}}, 
  \mbox{R.~Capuzzo-Dolcetta$^{\ref{AFFIL::ItalyORoma},\ref{AFFIL::ItalyURomaSapienza}}$\orcid{0000-0002-6871-9519}}, 
  \mbox{P.~Caraveo$^{\ref{AFFIL::ItalyIASFMilano}}$\orcid{0000-0003-2478-8018}}, 
  \mbox{S.~Caroff$^{\ref{AFFIL::FranceLAPPUSavoieMontBlanc}}$\orcid{0000-0002-1103-130X}}, 
  \mbox{A.~Carosi$^{\ref{AFFIL::ItalyORoma}}$}, 
  \mbox{R.~Carosi$^{\ref{AFFIL::ItalyINFNPisa}}$\orcid{0000-0002-4137-4370}}, 
  \mbox{E.~Carquin$^{\ref{AFFIL::ChileUTecnicaFedericoSantaMaria}}$\orcid{0000-0002-7863-1166}}, 
  \mbox{M.-S.~Carrasco$^{\ref{AFFIL::FranceCPPMUAixMarseille}}$}, 
  \mbox{F.~Cassol$^{\ref{AFFIL::FranceCPPMUAixMarseille}}$\orcid{0000-0002-0372-1992}}, 
  \mbox{L.~Castaldini$^{\ref{AFFIL::ItalyOASBologna}}$\orcid{0009-0000-5501-4328}}, 
  \mbox{N.~Castrejon$^{\ref{AFFIL::SpainUAlcala}}$\orcid{0000-0001-6847-8594}}, 
  \mbox{A.~J.~Castro-Tirado$^{\ref{AFFIL::SpainIAACSIC}}$\orcid{0000-0003-2999-3563}}, 
  \mbox{D.~Cerasole$^{\ref{AFFIL::ItalyUandINFNBari}}$\orcid{0000-0003-2033-756X}}, 
  \mbox{M.~Cerruti$^{\ref{AFFIL::FranceAPCUParisCite}}$\orcid{0000-0001-7891-699X}}, 
  \mbox{P.~M.~Chadwick$^{\ref{AFFIL::UnitedKingdomUDurham}}$\orcid{0000-0002-1468-2685}}, 
  \mbox{S.~Chaty$^{\ref{AFFIL::FranceAPCUParisCite}}$\orcid{0000-0002-5769-8601}}, 
  \mbox{A.~W.~Chen$^{\ref{AFFIL::SouthAfricaUWitwatersrand}}$\orcid{0000-0001-6425-5692}}, 
  \mbox{M.~Chernyakova$^{\ref{AFFIL::IrelandDCU}}$\orcid{0000-0002-9735-3608}}, 
  \mbox{A.~Chiavassa$^{\ref{AFFIL::ItalyINFNTorino},\ref{AFFIL::ItalyUTorino}}$\orcid{0000-0001-6183-2589}}, 
  \mbox{J.~Chudoba$^{\ref{AFFIL::CzechRepublicFZU}}$\orcid{0000-0002-6425-2579}}, 
  \mbox{L.~Chytka$^{\ref{AFFIL::CzechRepublicFZU}}$\orcid{0000-0001-5741-259X}}, 
  \mbox{G.~M.~Cicciari$^{\ref{AFFIL::ItalyUPalermo}}$\orcid{0009-0007-3885-051X}}, 
  \mbox{A.~Cifuentes$^{\ref{AFFIL::SpainCIEMAT}}$\orcid{0000-0003-1033-5296}}, 
  \mbox{C.~H.~Coimbra~Araujo$^{\ref{AFFIL::BrazilUFPR}}$\orcid{0000-0003-3588-2587}}, 
  \mbox{M.~Colapietro$^{\ref{AFFIL::ItalyOCapodimonte}}$\orcid{0000-0002-8386-9726}}, 
  \mbox{V.~Conforti$^{\ref{AFFIL::ItalyOASBologna}}$\orcid{0000-0002-0007-3520}}, 
  \mbox{F.~Conte$^{\ref{AFFIL::GermanyMPIK}}$\orcid{0000-0002-3083-8539}}, 
  \mbox{J.~L.~Contreras$^{\ref{AFFIL::SpainUCMAltasEnergias}}$\orcid{0000-0001-7282-2394}}, 
  \mbox{A.~Costa$^{\ref{AFFIL::ItalyOCatania}}$\orcid{0000-0003-0344-8911}}, 
  \mbox{H.~Costantini$^{\ref{AFFIL::FranceCPPMUAixMarseille}}$\orcid{0000-0003-4027-3081}}, 
  \mbox{G.~Cotter$^{\ref{AFFIL::UnitedKingdomUOxford}}$\orcid{0000-0002-9975-1829}}, 
  \mbox{P.~Cristofari$^{\ref{AFFIL::FranceObservatoiredeParis}}$}, 
  \mbox{O.~Cuevas$^{\ref{AFFIL::ChileUdeValparaiso}}$}, 
  \mbox{Z.~Curtis-Ginsberg$^{\ref{AFFIL::USAUWisconsin}}$\orcid{0000-0002-0194-7576}}, 
  \mbox{G.~D'Amico$^{\ref{AFFIL::NorwayUBergen}}$\orcid{0000-0001-6472-8381}}, 
  \mbox{F.~D'Ammando$^{\ref{AFFIL::ItalyRadioastronomiaINAF}}$\orcid{0000-0001-7618-7527}}, 
  \mbox{S.~Dai$^{\ref{AFFIL::AustraliaUWesternSydney}}$\orcid{0000-0002-9618-2499}}, 
  \mbox{M.~Dalchenko$^{\ref{AFFIL::SwitzerlandUGenevaDPNC}}$\orcid{0000-0002-0137-136X}}, 
  \mbox{F.~Dazzi$^{\ref{AFFIL::ItalyINAF}}$\orcid{0000-0001-5409-6544}}, 
  \mbox{A.~De~Angelis$^{\ref{AFFIL::ItalyUPadovaandINFN}}$}, 
  \mbox{M.~de~Bony~de~Lavergne$^{\ref{AFFIL::FranceCEAIRFUDPhP}}$\orcid{0000-0002-4650-1666}}, 
  \mbox{V.~De~Caprio$^{\ref{AFFIL::ItalyOCapodimonte}}$\orcid{0000-0002-4587-8963}}, 
  \mbox{E.~M.~de~Gouveia~Dal~Pino$^{\ref{AFFIL::BrazilIAGUSaoPaulo}}$\orcid{0000-0001-8058-4752}}, 
  \mbox{B.~De~Lotto$^{\ref{AFFIL::ItalyUUdineandINFNTrieste}}$\orcid{0000-0003-3624-4480}}, 
  \mbox{M.~De~Lucia$^{\ref{AFFIL::ItalyINFNNapoli}}$\orcid{0000-0002-0519-9149}}, 
  \mbox{R.~de~Menezes$^{\ref{AFFIL::ItalyINFNTorino},\ref{AFFIL::ItalyUTorino}}$\orcid{0000-0001-5489-4925}}, 
  \mbox{M.~de~Naurois$^{\ref{AFFIL::FranceLLREcolePolytechnique}}$\orcid{0000-0002-7245-201X}}, 
  \mbox{V.~de~Souza$^{\ref{AFFIL::BrazilIFSCUSaoPaulo}}$\orcid{0000-0003-0865-233X}}, 
  \mbox{L.~del~Peral$^{\ref{AFFIL::SpainUAlcala}}$\orcid{0000-0003-2580-5668}}, 
  \mbox{M.~V.~del~Valle$^{\ref{AFFIL::BrazilIAGUSaoPaulo}}$\orcid{0000-0002-5444-0795}}, 
  \mbox{A.~G.~Delgado~Giler$^{\ref{AFFIL::BrazilIFSCUSaoPaulo},\ref{AFFIL::NetherlandsUGroningen}}$\orcid{0000-0003-2190-9857}}, 
  \mbox{J.~Delgado~Mengual$^{\ref{AFFIL::SpainPIC}}$\orcid{0000-0002-0166-5464}}, 
  \mbox{C.~Delgado$^{\ref{AFFIL::SpainCIEMAT}}$\orcid{0000-0002-7014-4101}}, 
  \mbox{M.~Dell'aiera$^{\ref{AFFIL::FranceLAPPUSavoieMontBlanc}}$\orcid{0000-0002-5221-0240}}, 
  \mbox{D.~della~Volpe$^{\ref{AFFIL::SwitzerlandUGenevaDPNC}}$\orcid{0000-0001-8530-7447}}, 
  \mbox{D.~Depaoli$^{\ref{AFFIL::GermanyMPIK}}$\orcid{0000-0002-2672-4141}}, 
  \mbox{T.~Di~Girolamo$^{\ref{AFFIL::ItalyUNapoli},\ref{AFFIL::ItalyINFNNapoli}}$\orcid{0000-0003-2339-4471}}, 
  \mbox{A.~Di~Piano$^{\ref{AFFIL::ItalyOASBologna},\ref{AFFIL::ItalyUModena}}$\orcid{0000-0002-9894-7491}}, 
  \mbox{F.~Di~Pierro$^{\ref{AFFIL::ItalyINFNTorino}}$\orcid{0000-0003-4861-432X}}, 
  \mbox{R.~Di~Tria$^{\ref{AFFIL::ItalyUandINFNBari}}$\orcid{0009-0007-1088-5307}}, 
  \mbox{L.~Di~Venere$^{\ref{AFFIL::ItalyINFNBari}}$}, 
  \mbox{C.~D{\'\i}az$^{\ref{AFFIL::SpainCIEMAT}}$}, 
  \mbox{S.~Diebold$^{\ref{AFFIL::GermanyIAAT}}$\orcid{0000-0002-8042-2443}}, 
  \mbox{A.~Dinesh$^{\ref{AFFIL::SpainUCMAltasEnergias}}$}, 
  \mbox{J.~Djuvsland$^{\ref{AFFIL::NorwayUBergen}}$\orcid{0000-0002-6488-8219}}, 
  \mbox{R.~M.~Dominik$^{\ref{AFFIL::GermanyUDortmundTU}}$\orcid{0000-0003-4168-7200}}, 
  \mbox{D.~Dominis~Prester$^{\ref{AFFIL::CroatiaURijeka}}$\orcid{0000-0002-9880-5039}}, 
  \mbox{A.~Donini$^{\ref{AFFIL::ItalyORoma}}$\orcid{0000-0002-3066-724X}}, 
  \mbox{D.~Dorner$^{\ref{AFFIL::GermanyUWurzburg},\ref{AFFIL::SwitzerlandETHZurich}}$\orcid{0000-0001-8823-479X}}, 
  \mbox{J.~D\"orner$^{\ref{AFFIL::GermanyUBochum}}$\orcid{0000-0001-6692-6293}}, 
  \mbox{M.~Doro$^{\ref{AFFIL::ItalyUPadovaandINFN}}$\orcid{0000-0001-9104-3214}}, 
  \mbox{J.-L.~Dournaux$^{\ref{AFFIL::FranceObservatoiredeParis}}$}, 
  \mbox{C.~Duangchan$^{\ref{AFFIL::GermanyUErlangenECAP},\ref{AFFIL::ThailandNARIT}}$\orcid{0009-0003-8227-6552}}, 
  \mbox{C.~Dubos$^{\ref{AFFIL::FranceIJCLab}}$}, 
  \mbox{L.~Ducci$^{\ref{AFFIL::GermanyIAAT}}$\orcid{0000-0002-9989-538X}}, 
  \mbox{V.~V.~Dwarkadas$^{\ref{AFFIL::USAUChicagoDAA}}$\orcid{0000-0002-4661-7001}}, 
  \mbox{J.~Ebr$^{\ref{AFFIL::CzechRepublicFZU}}$\orcid{0000-0001-8807-6162}}, 
  \mbox{C.~Eckner$^{\ref{AFFIL::SloveniaUNovaGoricaCAC},\ref{AFFIL::FranceLAPTh}}$\orcid{0000-0002-5135-2909}}, 
  \mbox{K.~Egberts$^{\ref{AFFIL::GermanyUPotsdam}}$}, 
  \mbox{S.~Einecke$^{\ref{AFFIL::AustraliaUAdelaide}}$\orcid{0000-0001-9687-8237}}, 
  \mbox{D.~Els\"asser$^{\ref{AFFIL::GermanyUDortmundTU}}$\orcid{0000-0001-6796-3205}}, 
  \mbox{G.~Emery$^{\ref{AFFIL::FranceCPPMUAixMarseille}}$\orcid{0000-0001-6155-4742}}, 
  \mbox{M.~Errando$^{\ref{AFFIL::USAWashingtonU}}$\orcid{0000-0002-1853-863X}}, 
  \mbox{C.~Escanuela$^{\ref{AFFIL::GermanyMPIK}}$\orcid{0000-0002-7297-8126}}, 
  \mbox{P.~Escarate$^{\ref{AFFIL::ChileEscIngElec},\ref{AFFIL::ChileUTecnicaFedericoSantaMaria}}$\orcid{0000-0002-6751-3842}}, 
  \mbox{M.~Escobar~Godoy$^{\ref{AFFIL::USASCIPP}}$}, 
  \mbox{J.~Escudero$^{\ref{AFFIL::SpainIAACSIC}}$\orcid{0000-0002-4131-655X}}, 
  \mbox{P.~Esposito$^{\ref{AFFIL::ItalyIUSSPaviaINAF},\ref{AFFIL::ItalyIASFMilano}}$\orcid{0000-0003-4849-5092}}, 
  \mbox{S.~Ettori$^{\ref{AFFIL::ItalyOASBologna}}$\orcid{0000-0003-4117-8617}}, 
  \mbox{D.~Falceta-Goncalves$^{\ref{AFFIL::BrazilEACHUSaoPaulo}}$\orcid{0000-0002-1914-6654}}, 
  \mbox{E.~Fedorova$^{\ref{AFFIL::ItalyORoma},\ref{AFFIL::UkraineAstObsofUKyiv}}$\orcid{0000-0002-8882-7496}}, 
  \mbox{S.~Fegan$^{\ref{AFFIL::FranceLLREcolePolytechnique}}$\orcid{0000-0002-9978-2510}}, 
  \mbox{Q.~Feng$^{\ref{AFFIL::USAUUtah}}$\orcid{0000-0001-6674-4238}}, 
  \mbox{G.~Ferrand$^{\ref{AFFIL::CanadaUManitoba},\ref{AFFIL::JapanRIKEN}}$\orcid{0000-0002-4231-8717}}, 
  \mbox{F.~Ferrarotto$^{\ref{AFFIL::ItalyINFNRomaLaSapienza}}$\orcid{0000-0001-5464-0378}}, 
  \mbox{E.~Fiandrini$^{\ref{AFFIL::ItalyUPerugiaandINFN}}$}, 
  \mbox{A.~Fiasson$^{\ref{AFFIL::FranceLAPPUSavoieMontBlanc}}$}, 
  \mbox{M.~Filipovic$^{\ref{AFFIL::AustraliaUWesternSydney}}$\orcid{0000-0002-4990-9288}}, 
  \mbox{V.~Fioretti$^{\ref{AFFIL::ItalyOASBologna}}$\orcid{0000-0002-6082-5384}}, 
  \mbox{M.~Fiori$^{\ref{AFFIL::ItalyOPadova}}$\orcid{0000-0002-7352-6818}}, 
  \mbox{L.~Foffano$^{\ref{AFFIL::ItalyIAPS}}$\orcid{0000-0002-0709-9707}}, 
  \mbox{L.~Font~Guiteras$^{\ref{AFFIL::SpainUABandCERESIEEC}}$\orcid{0000-0003-2109-5961}}, 
  \mbox{G.~Fontaine$^{\ref{AFFIL::FranceLLREcolePolytechnique}}$\orcid{0000-0002-6443-5025}}, 
  \mbox{S.~Fr\"ose$^{\ref{AFFIL::GermanyUDortmundTU}}$\orcid{0000-0003-1832-4129}}, 
  \mbox{Y.~Fukazawa$^{\ref{AFFIL::JapanUHiroshima}}$}, 
  \mbox{Y.~Fukui$^{\ref{AFFIL::JapanUNagoya}}$\orcid{0000-0002-8966-9856}}, 
  \mbox{A.~Furniss$^{\ref{AFFIL::USASCIPP}}$\orcid{0000-0003-1614-1273}}, 
  \mbox{G.~Galanti$^{\ref{AFFIL::ItalyIASFMilano}}$\orcid{0000-0001-7254-3029}}, 
  \mbox{G.~Galaz$^{\ref{AFFIL::ChileUPontificiaCatolicadeChile}}$\orcid{0000-0002-8835-0739}}, 
  \mbox{C.~Galelli$^{\ref{AFFIL::FranceObservatoiredeParis}}$\orcid{0000-0002-7372-9703}}, 
  \mbox{S.~Gallozzi$^{\ref{AFFIL::ItalyORoma}}$\orcid{0000-0003-4456-9875}}, 
  \mbox{V.~Gammaldi$^{\ref{AFFIL::SpainUniversidadSanPabloCEU},\ref{AFFIL::SpainIFTUAMCSIC}}$\orcid{0000-0003-1826-6117}}, 
  \mbox{M.~Garczarczyk$^{\ref{AFFIL::GermanyDESY}}$}, 
  \mbox{C.~Gasbarra$^{\ref{AFFIL::ItalyINFNRomaTorVergata}}$\orcid{0000-0001-8335-9614}}, 
  \mbox{D.~Gasparrini$^{\ref{AFFIL::ItalyINFNRomaTorVergata}}$\orcid{0000-0002-5064-9495}}, 
  \mbox{A.~Ghalumyan$^{\ref{AFFIL::ArmeniaNSLAlikhanyan}}$}, 
  \mbox{F.~Gianotti$^{\ref{AFFIL::ItalyOASBologna}}$\orcid{0000-0003-4666-119X}}, 
  \mbox{M.~Giarrusso$^{\ref{AFFIL::ItalyINFNCatania}}$\orcid{0000-0002-4453-1597}}, 
  \mbox{J.~G.~Giesbrecht~Formiga~Paiva$^{\ref{AFFIL::BrazilCBPF}}$\orcid{0000-0002-5817-2062}}, 
  \mbox{N.~Giglietto$^{\ref{AFFIL::ItalyPolitecnicoBari},\ref{AFFIL::ItalyINFNBari}}$\orcid{0000-0002-9021-2888}}, 
  \mbox{F.~Giordano$^{\ref{AFFIL::ItalyUandINFNBari}}$\orcid{0000-0002-8651-2394}}, 
  \mbox{R.~Giuffrida$^{\ref{AFFIL::ItalyOPalermo}}$}, 
  \mbox{J.-F.~Glicenstein$^{\ref{AFFIL::FranceCEAIRFUDPhP}}$}, 
  \mbox{J.~Glombitza$^{\ref{AFFIL::GermanyUErlangenECAP}}$\orcid{0000-0001-9683-4568}}, 
  \mbox{P.~Goldoni$^{\ref{AFFIL::FranceAPCUParisCiteCEAaffiliatedpersonnel}}$\orcid{0000-0001-5638-5817}}, 
  \mbox{J.~M.~Gonz\'alez$^{\ref{AFFIL::ChileUAndresBello}}$\orcid{0000-0002-2413-0681}}, 
  \mbox{M.~M.~Gonz\'alez$^{\ref{AFFIL::MexicoUNAMMexico}}$}, 
  \mbox{J.~Goulart~Coelho$^{\ref{AFFIL::BrazilUFES}}$\orcid{0000-0001-9386-1042}}, 
  \mbox{T.~Gradetzke$^{\ref{AFFIL::GermanyUDortmundTU}}$\orcid{0000-0003-0646-2495}}, 
  \mbox{J.~Granot$^{\ref{AFFIL::IsraelOpenUniversityofIsrael},\ref{AFFIL::USAGWUWashingtonDC}}$\orcid{0000-0001-8530-8941}}, 
  \mbox{D.~Grasso$^{\ref{AFFIL::ItalyINFNPisa}}$}, 
  \mbox{R.~Grau$^{\ref{AFFIL::SpainIFAEBIST}}$\orcid{0000-0002-1891-6290}}, 
  \mbox{L.~Gr\'eaux$^{\ref{AFFIL::FranceIJCLab}}$}, 
  \mbox{D.~Green$^{\ref{AFFIL::GermanyMPP}}$\orcid{0000-0003-0768-2203}}, 
  \mbox{J.~G.~Green$^{\ref{AFFIL::GermanyMPP}}$\orcid{0000-0002-1130-6692}}, 
  \mbox{G.~Grolleron$^{\ref{AFFIL::FranceLPNHEUSorbonne}}$}, 
  \mbox{L.~M.~V.~Guedes$^{\ref{AFFIL::BrazilURioGrandedoNortePhys}}$\orcid{0009-0001-4935-3355}}, 
  \mbox{O.~Gueta$^{\ref{AFFIL::ItalyCTAOBologna}}$\orcid{0000-0002-9440-2398}}, 
  \mbox{J.~Hackfeld$^{\ref{AFFIL::GermanyUBochum},\ref{AFFIL::GermanyUDortmundTU}}$\orcid{0000-0002-1003-6408}}, 
  \mbox{D.~Hadasch$^{\ref{AFFIL::JapanUTokyoICRR}}$\orcid{0000-0001-8663-6461}}, 
  \mbox{P.~Hamal$^{\ref{AFFIL::CzechRepublicFZU}}$\orcid{0000-0003-3139-7234}}, 
  \mbox{W.~Hanlon$^{\ref{AFFIL::USACfAHarvardSmithsonian}}$\orcid{0000-0002-0109-4737}}, 
  \mbox{S.~Hara$^{\ref{AFFIL::JapanUYamanashiGakuin}}$\orcid{0009-0001-1220-7717}}, 
  \mbox{V.~M.~Harvey$^{\ref{AFFIL::AustraliaUAdelaide}}$\orcid{0000-0001-9090-8415}}, 
  \mbox{T.~Hassan$^{\ref{AFFIL::SpainCIEMAT}}$\orcid{0000-0002-4758-9196}}, 
  \mbox{K.~Hayashi$^{\ref{AFFIL::JapanNITSendaiNatori},\ref{AFFIL::JapanUTokyoICRR}}$\orcid{0000-0002-8758-8139}}, 
  \mbox{B.~He{\ss}$^{\ref{AFFIL::GermanyIAAT}}$\orcid{0009-0004-9999-171X}}, 
  \mbox{L.~Heckmann$^{\ref{AFFIL::GermanyMPP},\ref{AFFIL::AustriaUInnsbruck}}$\orcid{0000-0002-6653-8407}}, 
  \mbox{M.~Heller$^{\ref{AFFIL::SwitzerlandUGenevaDPNC}}$}, 
  \mbox{S.~Hern\'andez~Cadena$^{\ref{AFFIL::MexicoUNAMMexico}}$\orcid{0000-0002-2565-8365}}, 
  \mbox{O.~Hervet$^{\ref{AFFIL::USASCIPP}}$\orcid{0000-0003-3878-1677}}, 
  \mbox{J.~Hinton$^{\ref{AFFIL::GermanyMPIK}}$}, 
  \mbox{N.~Hiroshima$^{\ref{AFFIL::JapanUTokyoICRR},\ref{AFFIL::JapanUToyama}}$}, 
  \mbox{B.~Hnatyk$^{\ref{AFFIL::UkraineAstObsofUKyiv}}$\orcid{0000-0001-7113-4709}}, 
  \mbox{R.~Hnatyk$^{\ref{AFFIL::UkraineAstObsofUKyiv}}$\orcid{0000-0002-6378-7678}}, 
  \mbox{W.~Hofmann$^{\ref{AFFIL::GermanyMPIK}}$}, 
  \mbox{J.~Holder$^{\ref{AFFIL::USAUDelaware}}$\orcid{0000-0002-6833-0474}}, 
  \mbox{D.~Horan$^{\ref{AFFIL::FranceLLREcolePolytechnique}}$\orcid{0000-0001-5574-2579}}, 
  \mbox{P.~Horvath$^{\ref{AFFIL::CzechRepublicUOlomouc}}$\orcid{0000-0002-6710-5339}}, 
  \mbox{T.~Hovatta$^{\ref{AFFIL::FinlandFINCA},\ref{AFFIL::FinlandUAalto}}$\orcid{0000-0002-2024-8199}}, 
  \mbox{M.~Hrabovsky$^{\ref{AFFIL::CzechRepublicUOlomouc}}$}, 
  \mbox{D.~Hrupec$^{\ref{AFFIL::CroatiaUOsijek}}$\orcid{0000-0002-7027-5021}}, 
  \mbox{M.~Iarlori$^{\ref{AFFIL::ItalyCETEMPSandUandINFNAquila}}$}, 
  \mbox{T.~Inada$^{\ref{AFFIL::JapanUTokyoICRR}}$\orcid{0000-0002-6923-9314}}, 
  \mbox{F.~Incardona$^{\ref{AFFIL::ItalyOCatania}}$}, 
  \mbox{S.~Inoue$^{\ref{AFFIL::JapanUChiba},\ref{AFFIL::JapanUTokyoICRR}}$\orcid{0000-0003-1096-9424}}, 
  \mbox{Y.~Inoue$^{\ref{AFFIL::JapanUOsaka}}$}, 
  \mbox{F.~Iocco$^{\ref{AFFIL::ItalyUNapoli},\ref{AFFIL::ItalyINFNNapoli}}$}, 
  \mbox{M.~Iori$^{\ref{AFFIL::ItalyINFNRomaLaSapienza}}$\orcid{0000-0002-6349-0380}}, 
  \mbox{K.~Ishio$^{\ref{AFFIL::PolandTorunInstituteofAstronomy}}$\orcid{0000-0003-3189-0766}}, 
  \mbox{M.~Jamrozy$^{\ref{AFFIL::PolandUJagiellonian}}$\orcid{0000-0002-0870-7778}}, 
  \mbox{P.~Janecek$^{\ref{AFFIL::CzechRepublicFZU}}$\orcid{0000-0003-3501-7163}}, 
  \mbox{F.~Jankowsky$^{\ref{AFFIL::GermanyLSW}}$}, 
  \mbox{P.~Jean$^{\ref{AFFIL::FranceIRAPUToulouse}}$\orcid{0000-0002-1757-9560}}, 
  \mbox{J.~Jimenez~Quiles$^{\ref{AFFIL::SpainIFAEBIST}}$\orcid{0009-0005-6729-5709}}, 
  \mbox{W.~Jin$^{\ref{AFFIL::USAUCLA}}$\orcid{0000-0002-1089-1754}}, 
  \mbox{C.~Juramy-Gilles$^{\ref{AFFIL::FranceLPNHEUSorbonne}}$\orcid{0000-0002-3145-9258}}, 
  \mbox{J.~Jurysek$^{\ref{AFFIL::CzechRepublicFZU}}$\orcid{0000-0002-3130-4168}}, 
  \mbox{M.~Kagaya$^{\ref{AFFIL::JapanNITSendaiHirose},\ref{AFFIL::JapanUTokyoICRR}}$}, 
  \mbox{O.~Kalekin$^{\ref{AFFIL::GermanyUErlangenECAP}}$}, 
  \mbox{V.~Karas$^{\ref{AFFIL::CzechRepublicASU}}$\orcid{0000-0002-5760-0459}}, 
  \mbox{H.~Katagiri$^{\ref{AFFIL::JapanUIbaraki}}$\orcid{0000-0003-2347-8819}}, 
  \mbox{J.~Kataoka$^{\ref{AFFIL::JapanUWaseda}}$}, 
  \mbox{S.~Kaufmann$^{\ref{AFFIL::UnitedKingdomUDurham}}$}, 
  \mbox{D.~Kazanas$^{\ref{AFFIL::GreeceUThessaloniki}}$}, 
  \mbox{D.~Kerszberg$^{\ref{AFFIL::SpainIFAEBIST}}$\orcid{0000-0002-5289-1509}}, 
  \mbox{D.~B.~Kieda$^{\ref{AFFIL::USAUUtah}}$\orcid{0000-0003-4785-0101}}, 
  \mbox{T.~Kleiner$^{\ref{AFFIL::GermanyDESY}}$\orcid{0000-0002-4260-9186}}, 
  \mbox{G.~Kluge$^{\ref{AFFIL::NorwayUOslo}}$\orcid{0009-0009-0384-0084}}, 
  \mbox{Y.~Kobayashi$^{\ref{AFFIL::JapanUChiba},\ref{AFFIL::JapanUTokyoICRR}}$\orcid{0009-0005-5680-6614}}, 
  \mbox{K.~Kohri$^{\ref{AFFIL::JapanNAOJ},\ref{AFFIL::JapanKEK}}$\orcid{0000-0003-3764-8612}}, 
  \mbox{N.~Komin$^{\ref{AFFIL::SouthAfricaUWitwatersrand}}$\orcid{0000-0003-3280-0582}}, 
  \mbox{P.~Kornecki$^{\ref{AFFIL::FranceObservatoiredeParis}}$\orcid{0000-0002-2706-7438}}, 
  \mbox{K.~Kosack$^{\ref{AFFIL::FranceCEAIRFUDAp}}$\orcid{0000-0001-8424-3621}}, 
  \mbox{G.~Kowal$^{\ref{AFFIL::BrazilEACHUSaoPaulo}}$\orcid{0000-0002-0176-9909}}, 
  \mbox{H.~Kubo$^{\ref{AFFIL::JapanUTokyoICRR}}$\orcid{0000-0001-9159-9853}}, 
  \mbox{J.~Kushida$^{\ref{AFFIL::JapanUTokai}}$\orcid{0000-0002-8002-8585}}, 
  \mbox{A.~La~Barbera$^{\ref{AFFIL::ItalyIASFPalermo}}$\orcid{0000-0002-5880-8913}}, 
  \mbox{N.~La~Palombara$^{\ref{AFFIL::ItalyIASFMilano}}$\orcid{0000-0001-7015-6359}}, 
  \mbox{M.~L\'ainez$^{\ref{AFFIL::SpainUCMAltasEnergias}}$\orcid{0000-0003-3848-922X}}, 
  \mbox{A.~Lamastra$^{\ref{AFFIL::ItalyORoma}}$\orcid{0000-0003-2403-913X}}, 
  \mbox{J.~Lapington$^{\ref{AFFIL::UnitedKingdomULeicester}}$}, 
  \mbox{P.~Laporte$^{\ref{AFFIL::FranceObservatoiredeParis}}$}, 
  \mbox{S.~Lazarevi\'c$^{\ref{AFFIL::AustraliaUWesternSydney}}$\orcid{0000-0001-6109-8548}}, 
  \mbox{J.~Lazendic-Galloway$^{\ref{AFFIL::AustraliaUMonash}}$}, 
  \mbox{M.~Lemoine-Goumard$^{\ref{AFFIL::FranceLP2IUBordeaux}}$\orcid{0000-0002-4462-3686}}, 
  \mbox{J.-P.~Lenain$^{\ref{AFFIL::FranceLPNHEUSorbonne}}$\orcid{0000-0001-7284-9220}}, 
  \mbox{F.~Leone$^{\ref{AFFIL::ItalyUCatania}}$\orcid{0000-0001-7626-3788}}, 
  \mbox{E.~Leonora$^{\ref{AFFIL::ItalyINFNCatania}}$\orcid{0000-0002-0536-3551}}, 
  \mbox{G.~Leto$^{\ref{AFFIL::ItalyOCatania}}$\orcid{0000-0002-0040-5011}}, 
  \mbox{E.~Lindfors$^{\ref{AFFIL::FinlandUTurku}}$\orcid{0000-0002-9155-6199}}, 
  \mbox{M.~Linhoff$^{\ref{AFFIL::GermanyUDortmundTU}}$\orcid{0000-0001-7993-8189}}, 
  \mbox{I.~Liodakis$^{\ref{AFFIL::FinlandFINCA}}$\orcid{0000-0001-9200-4006}}, 
  \mbox{A.~Lipniacka$^{\ref{AFFIL::NorwayUBergen}}$\orcid{0000-0002-8759-8564}}, 
  \mbox{S.~Lombardi$^{\ref{AFFIL::ItalyORoma}}$\orcid{0000-0002-6336-865X}}, 
  \mbox{F.~Longo$^{\ref{AFFIL::ItalyUandINFNTrieste}}$\orcid{0000-0003-2501-2270}}, 
  \mbox{R.~L\'opez-Coto$^{\ref{AFFIL::SpainIAACSIC}}$}, 
  \mbox{M.~L\'opez-Moya$^{\ref{AFFIL::SpainUCMAltasEnergias}}$\orcid{0000-0002-8791-7908}}, 
  \mbox{A.~L\'opez-Oramas$^{\ref{AFFIL::SpainIAC}}$\orcid{0000-0003-4603-1884}}, 
  \mbox{S.~Loporchio$^{\ref{AFFIL::ItalyPolitecnicoBari},\ref{AFFIL::ItalyINFNBari}}$}, 
  \mbox{J.~Lozano~Bahilo$^{\ref{AFFIL::SpainUAlcala}}$\orcid{0000-0003-0613-140X}}, 
  \mbox{P.~L.~Luque-Escamilla$^{\ref{AFFIL::SpainUJaen}}$\orcid{0000-0002-3306-9456}}, 
  \mbox{O.~Macias$^{\ref{AFFIL::NetherlandsUAmsterdam}}$\orcid{0000-0001-8867-2693}}, 
  \mbox{P.~Majumdar$^{\ref{AFFIL::IndiaSahaInstitute}}$\orcid{0000-0002-5481-5040}}, 
  \mbox{M.~Mallamaci$^{\ref{AFFIL::ItalyUPalermo},\ref{AFFIL::ItalyINFNCatania}}$\orcid{0000-0003-4068-0496}}, 
  \mbox{D.~Malyshev$^{\ref{AFFIL::GermanyIAAT}}$\orcid{0000-0001-9689-2194}}, 
  \mbox{D.~Mandat$^{\ref{AFFIL::CzechRepublicFZU}}$}, 
  \mbox{G.~Manic\`o$^{\ref{AFFIL::ItalyINFNCatania},\ref{AFFIL::ItalyUCatania}}$}, 
  \mbox{M.~Mariotti$^{\ref{AFFIL::ItalyUPadovaandINFN}}$\orcid{0000-0003-3297-4128}}, 
  \mbox{I.~M\'arquez$^{\ref{AFFIL::SpainIAACSIC}}$\orcid{0000-0003-2629-1945}}, 
  \mbox{P.~Marquez$^{\ref{AFFIL::SpainIFAEBIST}}$\orcid{0000-0002-9591-7967}}, 
  \mbox{G.~Marsella$^{\ref{AFFIL::ItalyUPalermo},\ref{AFFIL::ItalyINFNCatania}}$\orcid{0000-0002-3152-8874}}, 
  \mbox{J.~Mart{\'\i}$^{\ref{AFFIL::SpainUJaen}}$\orcid{0000-0001-5302-0660}}, 
  \mbox{G.~A.~Mart{\'\i}nez$^{\ref{AFFIL::SpainCIEMAT}}$\orcid{0000-0002-1061-8520}}, 
  \mbox{M.~Mart{\'\i}nez$^{\ref{AFFIL::SpainIFAEBIST}}$\orcid{0000-0002-9763-9155}}, 
  \mbox{O.~Martinez$^{\ref{AFFIL::SpainUCMElectronica},\ref{AFFIL::SpainUPCMadrid}}$\orcid{0000-0002-3353-7707}}, 
  \mbox{C.~Marty$^{\ref{AFFIL::FranceIRAPUToulouse}}$}, 
  \mbox{A.~Mas-Aguilar$^{\ref{AFFIL::SpainUCMAltasEnergias}}$\orcid{0000-0002-8893-9009}}, 
  \mbox{M.~Mastropietro$^{\ref{AFFIL::ItalyORoma}}$\orcid{0000-0002-6324-5713}}, 
  \mbox{D.~Mazin$^{\ref{AFFIL::JapanUTokyoICRR},\ref{AFFIL::GermanyMPP}}$}, 
  \mbox{S.~Menchiari$^{\ref{AFFIL::ItalyOArcetri}}$\orcid{0009-0006-6386-3702}}, 
  \mbox{E.~Mestre$^{\ref{AFFIL::SpainICECSIC}}$}, 
  \mbox{J.-L.~Meunier$^{\ref{AFFIL::FranceLPNHEUSorbonne}}$}, 
  \mbox{D.~M.-A.~Meyer$^{\ref{AFFIL::GermanyUPotsdam},\ref{AFFIL::SpainICECSIC}}$\orcid{0000-0001-8258-9813}}, 
  \mbox{M.~Meyer$^{\ref{AFFIL::GermanyUHamburg}}$\orcid{0000-0002-0738-7581}}, 
  \mbox{D.~Miceli$^{\ref{AFFIL::ItalyINFNPadova}}$\orcid{0000-0002-2686-0098}}, 
  \mbox{M.~Miceli$^{\ref{AFFIL::ItalyUPalermo},\ref{AFFIL::ItalyOPalermo}}$}, 
  \mbox{M.~Michailidis$^{\ref{AFFIL::GermanyIAAT}}$}, 
  \mbox{J.~Micha{\l}owski$^{\ref{AFFIL::PolandIFJ}}$}, 
  \mbox{T.~Miener$^{\ref{AFFIL::SwitzerlandUGenevaDPNC}}$\orcid{0000-0003-1821-7964}}, 
  \mbox{J.~M.~Miranda$^{\ref{AFFIL::SpainUCMElectronica},\ref{AFFIL::SpainIPARCOSInstitute}}$\orcid{0000-0002-1472-9690}}, 
  \mbox{A.~Mitchell$^{\ref{AFFIL::GermanyUErlangenECAP}}$\orcid{0000-0003-3631-5648}}, 
  \mbox{M.~Mizote$^{\ref{AFFIL::JapanUKonan}}$}, 
  \mbox{T.~Mizuno$^{\ref{AFFIL::JapanHASC}}$}, 
  \mbox{R.~Moderski$^{\ref{AFFIL::PolandNicolausCopernicusAstronomicalCenter}}$\orcid{0000-0002-8663-3882}}, 
  \mbox{M.~Molero$^{\ref{AFFIL::SpainIAC}}$\orcid{0000-0003-0967-715X}}, 
  \mbox{C.~Molfese$^{\ref{AFFIL::ItalyINAF}}$\orcid{0000-0002-2756-9075}}, 
  \mbox{E.~Molina$^{\ref{AFFIL::SpainIAC}}$\orcid{0000-0003-1204-5516}}, 
  \mbox{T.~Montaruli$^{\ref{AFFIL::SwitzerlandUGenevaDPNC}}$}, 
  \mbox{A.~Moralejo$^{\ref{AFFIL::SpainIFAEBIST}}$\orcid{0000-0002-1344-9080}}, 
  \mbox{D.~Morcuende$^{\ref{AFFIL::SpainIAACSIC}}$\orcid{0000-0001-9400-0922}}, 
  \mbox{A.~Morselli$^{\ref{AFFIL::ItalyINFNRomaTorVergata}}$\orcid{0000-0002-7704-9553}}, 
  \mbox{E.~Moulin$^{\ref{AFFIL::FranceCEAIRFUDPhP}}$\orcid{0000-0003-4007-0145}}, 
  \mbox{V.~Moya~Zamanillo$^{\ref{AFFIL::SpainUCMAltasEnergias}}$\orcid{0000-0001-9407-5545}}, 
  \mbox{K.~Munari$^{\ref{AFFIL::ItalyOCatania}}$}, 
  \mbox{T.~Murach$^{\ref{AFFIL::GermanyDESY}}$}, 
  \mbox{A.~Muraczewski$^{\ref{AFFIL::PolandNicolausCopernicusAstronomicalCenter}}$}, 
  \mbox{H.~Muraishi$^{\ref{AFFIL::JapanUKitasato}}$\orcid{0000-0003-3054-5725}}, 
  \mbox{T.~Nakamori$^{\ref{AFFIL::JapanUYamagata}}$\orcid{0000-0002-7308-2356}}, 
  \mbox{A.~Nayak$^{\ref{AFFIL::UnitedKingdomUDurham}}$}, 
  \mbox{R.~Nemmen$^{\ref{AFFIL::BrazilIAGUSaoPaulo},\ref{AFFIL::USAStanford}}$\orcid{0000-0003-3956-0331}}, 
  \mbox{J.~P.~Neto$^{\ref{AFFIL::BrazilURioGrandedoNortePhys},\ref{AFFIL::BrazilURioGrandedoNorteIIP}}$\orcid{0000-0002-8257-7369}}, 
  \mbox{L.~Nickel$^{\ref{AFFIL::GermanyUDortmundTU}}$\orcid{0000-0001-7110-0533}}, 
  \mbox{J.~Niemiec$^{\ref{AFFIL::PolandIFJ}}$\orcid{0000-0001-6036-8569}}, 
  \mbox{D.~Nieto$^{\ref{AFFIL::SpainUCMAltasEnergias}}$\orcid{0000-0003-3343-0755}}, 
  \mbox{M.~Nievas~Rosillo$^{\ref{AFFIL::SpainIAC}}$\orcid{0000-0002-8321-9168}}, 
  \mbox{M.~Niko{\l}ajuk$^{\ref{AFFIL::PolandUBiaystok}}$\orcid{0000-0003-4075-6745}}, 
  \mbox{L.~Nikoli\'c$^{\ref{AFFIL::ItalyUSienaandINFN}}$}, 
  \mbox{K.~Nishijima$^{\ref{AFFIL::JapanUTokai}}$\orcid{0000-0002-1830-4251}}, 
  \mbox{K.~Noda$^{\ref{AFFIL::JapanUChiba},\ref{AFFIL::JapanUTokyoICRR}}$\orcid{0000-0003-1397-6478}}, 
  \mbox{D.~Nosek$^{\ref{AFFIL::CzechRepublicUPrague}}$\orcid{0000-0001-6219-200X}}, 
  \mbox{V.~Novotny$^{\ref{AFFIL::CzechRepublicUPrague}}$\orcid{0000-0002-4319-4541}}, 
  \mbox{S.~Nozaki$^{\ref{AFFIL::GermanyMPP}}$\orcid{0000-0002-6246-2767}}, 
  \mbox{M.~Ohishi$^{\ref{AFFIL::JapanUTokyoICRR}}$\orcid{0000-0002-5056-0968}}, 
  \mbox{Y.~Ohtani$^{\ref{AFFIL::JapanUTokyoICRR}}$\orcid{0000-0001-7042-4958}}, 
  \mbox{A.~Okumura$^{\ref{AFFIL::JapanUNagoyaISEE},\ref{AFFIL::JapanUNagoyaKMI}}$\orcid{0000-0002-3055-7964}}, 
  \mbox{J.-F.~Olive$^{\ref{AFFIL::FranceIRAPUToulouse}}$}, 
  \mbox{R.~A.~Ong$^{\ref{AFFIL::USAUCLA}}$\orcid{0000-0002-4837-5253}}, 
  \mbox{M.~Orienti$^{\ref{AFFIL::ItalyRadioastronomiaINAF}}$\orcid{0000-0003-4470-7094}}, 
  \mbox{R.~Orito$^{\ref{AFFIL::JapanUTokushima}}$}, 
  \mbox{M.~Orlandini$^{\ref{AFFIL::ItalyOASBologna}}$\orcid{0000-0003-0946-3151}}, 
  \mbox{E.~Orlando$^{\ref{AFFIL::ItalyUandINFNTrieste}}$}, 
  \mbox{S.~Orlando$^{\ref{AFFIL::ItalyOPalermo}}$\orcid{0000-0003-2836-540X}}, 
  \mbox{M.~Ostrowski$^{\ref{AFFIL::PolandUJagiellonian}}$\orcid{0000-0002-9199-7031}}, 
  \mbox{J.~Otero-Santos$^{\ref{AFFIL::SpainIAACSIC}}$\orcid{0000-0002-4241-5875}}, 
  \mbox{I.~Oya$^{\ref{AFFIL::GermanyCTAOHeidelberg}}$\orcid{0000-0002-3881-9324}}, 
  \mbox{I.~Pagano$^{\ref{AFFIL::ItalyOCatania}}$\orcid{0000-0001-9573-4928}}, 
  \mbox{A.~Pagliaro$^{\ref{AFFIL::ItalyIASFPalermo}}$\orcid{0000-0002-6841-1362}}, 
  \mbox{M.~Palatiello$^{\ref{AFFIL::ItalyORoma}}$}, 
  \mbox{G.~Panebianco$^{\ref{AFFIL::ItalyOASBologna}}$\orcid{0000-0002-3410-8613}}, 
  \mbox{D.~Paneque$^{\ref{AFFIL::GermanyMPP}}$\orcid{0000-0002-2830-0502}}, 
  \mbox{F.~R.~Pantaleo$^{\ref{AFFIL::ItalyINFNBari},\ref{AFFIL::ItalyPolitecnicoBari}}$}, 
  \mbox{J.~M.~Paredes$^{\ref{AFFIL::SpainICCUB}}$\orcid{0000-0002-1566-9044}}, 
  \mbox{N.~Parmiggiani$^{\ref{AFFIL::ItalyOASBologna}}$\orcid{0000-0002-4535-5329}}, 
  \mbox{B.~Patricelli$^{\ref{AFFIL::ItalyORoma},\ref{AFFIL::ItalyUPisa}}$\orcid{0000-0001-6709-0969}}, 
  \mbox{A.~Pe'er$^{\ref{AFFIL::GermanyMPP}}$\orcid{0000-0001-8667-0889}}, 
  \mbox{M.~Pech$^{\ref{AFFIL::CzechRepublicFZU}}$\orcid{0000-0002-8421-0456}}, 
  \mbox{M.~Pecimotika$^{\ref{AFFIL::CroatiaURijeka},\ref{AFFIL::CroatiaIRB}}$\orcid{0000-0002-4699-1845}}, 
  \mbox{U.~Pensec$^{\ref{AFFIL::FranceLPNHEUSorbonne},\ref{AFFIL::FranceObservatoiredeParis}}$}, 
  \mbox{M.~Peresano$^{\ref{AFFIL::ItalyUTorino},\ref{AFFIL::ItalyINFNTorino}}$\orcid{0000-0002-7537-7334}}, 
  \mbox{J.~P\'erez-Romero$^{\ref{AFFIL::SloveniaUNovaGoricaCAC}}$\orcid{0000-0002-9408-3120}}, 
  \mbox{M.~Persic$^{\ref{AFFIL::ItalyOPadova},\ref{AFFIL::ItalyOandINFNTrieste}}$\orcid{0000-0003-1853-4900}}, 
  \mbox{K.~P.~Peters$^{\ref{AFFIL::GermanyUPotsdam}}$\orcid{0009-0000-4743-1463}}, 
  \mbox{O.~Petruk$^{\ref{AFFIL::UkraineIAPMMLviv},\ref{AFFIL::ItalyOPalermo}}$\orcid{0000-0003-3487-0349}}, 
  \mbox{G.~Piano$^{\ref{AFFIL::ItalyIAPS}}$\orcid{0000-0002-9332-5319}}, 
  \mbox{E.~Pierre$^{\ref{AFFIL::FranceLPNHEUSorbonne}}$}, 
  \mbox{E.~Pietropaolo$^{\ref{AFFIL::ItalyUandINFNAquila}}$\orcid{0000-0002-6633-9846}}, 
  \mbox{M.~Pihet$^{\ref{AFFIL::ItalyINFNPadova}}$\orcid{0009-0000-4691-3866}}, 
  \mbox{L.~Pinchbeck$^{\ref{AFFIL::AustraliaUMonash}}$\orcid{0009-0009-6802-2461}}, 
  \mbox{G.~Pirola$^{\ref{AFFIL::GermanyMPP}}$}, 
  \mbox{C.~Pittori$^{\ref{AFFIL::ItalyORoma}}$\orcid{0000-0001-6661-9779}}, 
  \mbox{C.~Plard$^{\ref{AFFIL::FranceLAPPUSavoieMontBlanc}}$\orcid{0000-0002-4061-3800}}, 
  \mbox{F.~Podobnik$^{\ref{AFFIL::ItalyUSienaandINFN}}$\orcid{0000-0001-6125-9487}}, 
  \mbox{M.~Pohl$^{\ref{AFFIL::GermanyUPotsdam},\ref{AFFIL::GermanyDESY}}$\orcid{0000-0001-7861-1707}}, 
  \mbox{V.~Pollet$^{\ref{AFFIL::FranceLAPPUSavoieMontBlanc}}$}, 
  \mbox{G.~Ponti$^{\ref{AFFIL::ItalyOBrera}}$\orcid{0000-0003-0293-3608}}, 
  \mbox{E.~Prandini$^{\ref{AFFIL::ItalyUPadovaandINFN}}$\orcid{0000-0003-4502-9053}}, 
  \mbox{G.~Principe$^{\ref{AFFIL::ItalyUandINFNTrieste}}$\orcid{0000-0003-0406-7387}}, 
  \mbox{C.~Priyadarshi$^{\ref{AFFIL::SpainIFAEBIST}}$\orcid{0000-0002-9160-9617}}, 
  \mbox{N.~Produit$^{\ref{AFFIL::SwitzerlandUGenevaISDC}}$\orcid{0000-0001-7138-7677}}, 
  \mbox{M.~Prouza$^{\ref{AFFIL::CzechRepublicFZU}}$\orcid{0000-0002-3238-9597}}, 
  \mbox{E.~Pueschel$^{\ref{AFFIL::GermanyUBochumPhysAst}}$\orcid{0000-0002-0529-1973}}, 
  \mbox{G.~P\"uhlhofer$^{\ref{AFFIL::GermanyIAAT}}$\orcid{0000-0003-4632-4644}}, 
  \mbox{M.~L.~Pumo$^{\ref{AFFIL::ItalyUCatania},\ref{AFFIL::ItalyINFNCatania}}$}, 
  \mbox{F.~Queiroz$^{\ref{AFFIL::BrazilURioGrandedoNorteIIP},\ref{AFFIL::BrazilURioGrandedoNortePhys}}$}, 
  \mbox{A.~Quirrenbach$^{\ref{AFFIL::GermanyLSW}}$}, 
  \mbox{S.~Rain\`o$^{\ref{AFFIL::ItalyUandINFNBari}}$\orcid{0000-0002-9181-0345}}, 
  \mbox{R.~Rando$^{\ref{AFFIL::ItalyUPadovaandINFN}}$\orcid{0000-0001-6992-818X}}, 
  \mbox{S.~Razzaque$^{\ref{AFFIL::SouthAfricaUJohannesburg},\ref{AFFIL::USAGWUWashingtonDC}}$\orcid{0000-0002-0130-2460}}, 
  \mbox{M.~Regeard$^{\ref{AFFIL::FranceAPCUParisCite}}$\orcid{0000-0002-3844-6003}}, 
  \mbox{A.~Reimer$^{\ref{AFFIL::AustriaUInnsbruck}}$\orcid{0000-0001-8604-7077}}, 
  \mbox{O.~Reimer$^{\ref{AFFIL::AustriaUInnsbruck}}$\orcid{0000-0001-6953-1385}}, 
  \mbox{A.~Reisenegger$^{\ref{AFFIL::ChileUPontificiaCatolicadeChile},\ref{AFFIL::ChileUMCE}}$\orcid{0000-0003-4059-6796}}, 
  \mbox{W.~Rhode$^{\ref{AFFIL::GermanyUDortmundTU}}$\orcid{0000-0003-2636-5000}}, 
  \mbox{D.~Ribeiro$^{\ref{AFFIL::USAUMinnesota}}$\orcid{0000-0002-7523-7366}}, 
  \mbox{M.~Rib\'o$^{\ref{AFFIL::SpainICCUB}}$\orcid{0000-0002-9931-4557}}, 
  \mbox{C.~Ricci$^{\ref{AFFIL::ChileUniversidadDiegoPortales}}$\orcid{0000-0001-5231-2645}}, 
  \mbox{T.~Richtler$^{\ref{AFFIL::ChileUdeConcepcion}}$}, 
  \mbox{J.~Rico$^{\ref{AFFIL::SpainIFAEBIST}}$\orcid{0000-0003-4137-1134}}, 
  \mbox{F.~Rieger$^{\ref{AFFIL::GermanyMPIK}}$}, 
  \mbox{L.~Riitano$^{\ref{AFFIL::USAUWisconsin}}$\orcid{0000-0003-2875-3066}}, 
  \mbox{V.~Rizi$^{\ref{AFFIL::ItalyUandINFNAquila}}$\orcid{0000-0002-5277-6527}}, 
  \mbox{E.~Roache$^{\ref{AFFIL::USACfAHarvardSmithsonian}}$}, 
  \mbox{G.~Rodriguez~Fernandez$^{\ref{AFFIL::ItalyINFNRomaTorVergata}}$\orcid{0000-0002-4683-230X}}, 
  \mbox{M.~D.~Rodr{\'\i}guez~Fr{\'\i}as$^{\ref{AFFIL::SpainUAlcala}}$\orcid{0000-0002-2550-4462}}, 
  \mbox{J.~J.~Rodr{\'\i}guez-V\'azquez$^{\ref{AFFIL::SpainCIEMAT}}$}, 
  \mbox{P.~Romano$^{\ref{AFFIL::ItalyOBrera}}$\orcid{0000-0003-0258-7469}}, 
  \mbox{G.~Romeo$^{\ref{AFFIL::ItalyOCatania}}$\orcid{0000-0003-3239-6057}}, 
  \mbox{J.~Rosado$^{\ref{AFFIL::SpainUCMAltasEnergias}}$\orcid{0000-0001-8208-9480}}, 
  \mbox{A.~Rosales~de~Leon$^{\ref{AFFIL::FranceLPNHEUSorbonne}}$}, 
  \mbox{G.~Rowell$^{\ref{AFFIL::AustraliaUAdelaide}}$\orcid{0000-0002-9516-1581}}, 
  \mbox{B.~Rudak$^{\ref{AFFIL::PolandNicolausCopernicusAstronomicalCenter}}$}, 
  \mbox{A.~J.~Ruiter$^{\ref{AFFIL::AustraliaUNewSouthWalesCanberra}}$\orcid{0000-0002-4794-6835}}, 
  \mbox{C.~B.~Rulten$^{\ref{AFFIL::UnitedKingdomUDurham}}$\orcid{0000-0001-7483-4348}}, 
  \mbox{I.~Sadeh$^{\ref{AFFIL::GermanyDESY}}$\orcid{0000-0003-1387-8915}}, 
  \mbox{L.~Saha$^{\ref{AFFIL::USACfAHarvardSmithsonian}}$\orcid{0000-0002-3171-5039}}, 
  \mbox{T.~Saito$^{\ref{AFFIL::JapanUTokyoICRR}}$\orcid{0000-0001-6201-3761}}, 
  \mbox{H.~Salzmann$^{\ref{AFFIL::GermanyIAAT}}$}, 
  \mbox{M.~S\'anchez-Conde$^{\ref{AFFIL::SpainIFTUAMCSIC}}$\orcid{0000-0002-3849-9164}}, 
  \mbox{H.~Sandaker$^{\ref{AFFIL::NorwayUOslo}}$}, 
  \mbox{P.~Sangiorgi$^{\ref{AFFIL::ItalyIASFPalermo}}$\orcid{0000-0001-8138-9289}}, 
  \mbox{H.~Sano$^{\ref{AFFIL::JapanUGifu},\ref{AFFIL::JapanUTokyoICRR}}$\orcid{0000-0003-2062-5692}}, 
  \mbox{M.~Santander$^{\ref{AFFIL::USAUAlabamaTuscaloosa}}$\orcid{0000-0001-7297-8217}}, 
  \mbox{R.~Santos-Lima$^{\ref{AFFIL::BrazilIAGUSaoPaulo}}$\orcid{0000-0001-6880-4468}}, 
  \mbox{V.~Sapienza$^{\ref{AFFIL::ItalyOPalermo},\ref{AFFIL::ItalyUPalermo}}$\orcid{0000-0002-6045-136X}}, 
  \mbox{T.~\v{S}ari\'c$^{\ref{AFFIL::CroatiaFESB}}$\orcid{0000-0001-8731-8369}}, 
  \mbox{A.~Sarkar$^{\ref{AFFIL::GermanyDESY}}$\orcid{0000-0002-7559-4339}}, 
  \mbox{S.~Sarkar$^{\ref{AFFIL::UnitedKingdomUOxford}}$\orcid{0000-0002-3542-858X}}, 
  \mbox{F.~G.~Saturni$^{\ref{AFFIL::ItalyORoma}}$\orcid{0000-0002-1946-7706}}, 
  \mbox{S.~Savarese$^{\ref{AFFIL::ItalyINAF}}$}, 
  \mbox{A.~Scherer$^{\ref{AFFIL::ChileUniversidaddeSantiagodeChile}}$}, 
  \mbox{F.~Schiavone$^{\ref{AFFIL::ItalyUandINFNBari}}$}, 
  \mbox{P.~Schipani$^{\ref{AFFIL::ItalyOCapodimonte}}$\orcid{0000-0003-0197-589X}}, 
  \mbox{B.~Schleicher$^{\ref{AFFIL::GermanyUWurzburg},\ref{AFFIL::SwitzerlandETHZurich}}$}, 
  \mbox{P.~Schovanek$^{\ref{AFFIL::CzechRepublicFZU}}$}, 
  \mbox{J.~L.~Schubert$^{\ref{AFFIL::GermanyUDortmundTU}}$}, 
  \mbox{U.~Schwanke$^{\ref{AFFIL::GermanyUBerlin}}$\orcid{0000-0002-1229-278X}}, 
  \mbox{M.~Seglar~Arroyo$^{\ref{AFFIL::SpainIFAEBIST}}$\orcid{0000-0001-8654-409X}}, 
  \mbox{I.~R.~Seitenzahl$^{\ref{AFFIL::AustraliaUNewSouthWalesCanberra}}$\orcid{0000-0002-5044-2988}}, 
  \mbox{O.~Sergijenko$^{\ref{AFFIL::UkraineAstObsofUKyiv},\ref{AFFIL::UkraineObsNASUkraine},\ref{AFFIL::PolandAGHCracowSTC}}$\orcid{0000-0002-9212-7118}}, 
  \mbox{M.~Servillat$^{\ref{AFFIL::FranceObservatoiredeParis}}$}, 
  \mbox{T.~Siegert$^{\ref{AFFIL::GermanyUWurzburg}}$}, 
  \mbox{H.~Siejkowski$^{\ref{AFFIL::PolandCYFRONETAGH}}$\orcid{0000-0003-1673-2145}}, 
  \mbox{C.~Siqueira$^{\ref{AFFIL::BrazilIFSCUSaoPaulo}}$\orcid{0000-0001-5684-3849}}, 
  \mbox{V.~Sliusar$^{\ref{AFFIL::SwitzerlandUGenevaISDC}}$}, 
  \mbox{A.~Slowikowska$^{\ref{AFFIL::PolandTorunInstituteofAstronomy}}$\orcid{0000-0003-4525-3178}}, 
  \mbox{H.~Sol$^{\ref{AFFIL::FranceObservatoiredeParis}}$}, 
  \mbox{S.~T.~Spencer$^{\ref{AFFIL::GermanyUErlangenECAP},\ref{AFFIL::UnitedKingdomUOxford}}$\orcid{0000-0001-5516-1205}}, 
  \mbox{D.~Spiga$^{\ref{AFFIL::ItalyOBrera}}$\orcid{0000-0003-1163-7843}}, 
  \mbox{A.~Stamerra$^{\ref{AFFIL::ItalyORoma},\ref{AFFIL::ItalyCTAOBologna}}$\orcid{0000-0002-9430-5264}}, 
  \mbox{S.~Stani\v{c}$^{\ref{AFFIL::SloveniaUNovaGoricaCAC}}$\orcid{0000-0003-3344-8381}}, 
  \mbox{T.~Starecki$^{\ref{AFFIL::PolandWUTElectronics}}$\orcid{0000-0002-4730-6803}}, 
  \mbox{R.~Starling$^{\ref{AFFIL::UnitedKingdomULeicester}}$\orcid{0000-0001-5803-2038}}, 
  \mbox{{\L}.~Stawarz$^{\ref{AFFIL::PolandUJagiellonian}}$}, 
  \mbox{C.~Steppa$^{\ref{AFFIL::GermanyUPotsdam}}$}, 
  \mbox{E.~S{\ae}ther~Hatlen$^{\ref{AFFIL::NorwayUOslo}}$}, 
  \mbox{T.~Stolarczyk$^{\ref{AFFIL::FranceCEAIRFUDAp}}$\orcid{0000-0002-0551-7581}}, 
  \mbox{J.~Stri\v{s}kovi\'c$^{\ref{AFFIL::CroatiaUOsijek}}$\orcid{0000-0003-2902-5044}}, 
  \mbox{Y.~Suda$^{\ref{AFFIL::JapanUHiroshima}}$\orcid{0000-0002-2692-5891}}, 
  \mbox{P.~\'Swierk$^{\ref{AFFIL::PolandIFJ}}$}, 
  \mbox{H.~Tajima$^{\ref{AFFIL::JapanUNagoyaISEE},\ref{AFFIL::JapanUNagoyaKMI}}$\orcid{0000-0002-1721-7252}}, 
  \mbox{D.~Tak$^{\ref{AFFIL::GermanyDESY}}$\orcid{0000-0002-9852-2469}}, 
  \mbox{M.~Takahashi$^{\ref{AFFIL::JapanUNagoyaISEE}}$\orcid{0000-0002-0574-6018}}, 
  \mbox{R.~Takeishi$^{\ref{AFFIL::JapanUTokyoICRR}}$\orcid{0000-0001-6335-5317}}, 
  \mbox{T.~Tavernier$^{\ref{AFFIL::CzechRepublicFZU}}$}, 
  \mbox{L.~A.~Tejedor$^{\ref{AFFIL::SpainUCMAltasEnergias}}$\orcid{0000-0003-1525-9085}}, 
  \mbox{K.~Terauchi$^{\ref{AFFIL::JapanUKyotoPhysicsandAstronomy}}$}, 
  \mbox{M.~Teshima$^{\ref{AFFIL::GermanyMPP}}$}, 
  \mbox{V.~Testa$^{\ref{AFFIL::ItalyORoma}}$\orcid{0000-0003-1033-1340}}, 
  \mbox{W.~W.~Tian$^{\ref{AFFIL::JapanUTokyoICRR}}$}, 
  \mbox{L.~Tibaldo$^{\ref{AFFIL::FranceIRAPUToulouse}}$\orcid{0000-0001-7523-570X}}, 
  \mbox{O.~Tibolla$^{\ref{AFFIL::UnitedKingdomUDurham}}$}, 
  \mbox{C.~J.~Todero~Peixoto$^{\ref{AFFIL::BrazilEELUSaoPaulo},\ref{AFFIL::BrazilIFSCUSaoPaulo}}$\orcid{0000-0003-3669-8212}}, 
  \mbox{F.~Torradeflot$^{\ref{AFFIL::SpainPIC},\ref{AFFIL::SpainCIEMAT}}$\orcid{0000-0003-1160-1517}}, 
  \mbox{D.~F.~Torres$^{\ref{AFFIL::SpainICECSIC}}$\orcid{0000-0002-1522-9065}}, 
  \mbox{G.~Tosti$^{\ref{AFFIL::ItalyOBrera},\ref{AFFIL::ItalyUPerugiaandINFN}}$\orcid{0000-0002-0839-4126}}, 
  \mbox{N.~Tothill$^{\ref{AFFIL::AustraliaUWesternSydney}}$\orcid{0000-0002-9931-5162}}, 
  \mbox{F.~Toussenel$^{\ref{AFFIL::FranceLPNHEUSorbonne}}$}, 
  \mbox{A.~Tramacere$^{\ref{AFFIL::SwitzerlandUGenevaISDC}}$\orcid{0000-0002-8186-3793}}, 
  \mbox{P.~Travnicek$^{\ref{AFFIL::CzechRepublicFZU}}$}, 
  \mbox{G.~Tripodo$^{\ref{AFFIL::ItalyUPalermo},\ref{AFFIL::ItalyINFNCatania}}$}, 
  \mbox{A.~Trois$^{\ref{AFFIL::ItalyINAFCagliari}}$\orcid{0000-0002-3180-6002}}, 
  \mbox{S.~Truzzi$^{\ref{AFFIL::ItalyUSienaandINFN}}$}, 
  \mbox{A.~Tutone$^{\ref{AFFIL::ItalyIASFPalermo}}$\orcid{0000-0002-2840-0001}}, 
  \mbox{L.~Vaclavek$^{\ref{AFFIL::CzechRepublicUOlomouc},\ref{AFFIL::CzechRepublicFZU}}$\orcid{0000-0002-0910-3415}}, 
  \mbox{M.~Vacula$^{\ref{AFFIL::CzechRepublicUOlomouc},\ref{AFFIL::CzechRepublicFZU}}$\orcid{0000-0003-4844-3962}}, 
  \mbox{P.~Vallania$^{\ref{AFFIL::ItalyINFNTorino},\ref{AFFIL::ItalyOTorino}}$\orcid{0000-0001-9089-7875}}, 
  \mbox{R.~Vall\'es$^{\ref{AFFIL::SpainICECSIC}}$\orcid{0000-0001-7701-2163}}, 
  \mbox{C.~van~Eldik$^{\ref{AFFIL::GermanyUErlangenECAP}}$\orcid{0000-0001-9669-645X}}, 
  \mbox{J.~van~Scherpenberg$^{\ref{AFFIL::GermanyMPP}}$\orcid{0000-0002-6173-867X}}, 
  \mbox{J.~Vandenbroucke$^{\ref{AFFIL::USAUWisconsin}}$\orcid{0000-0002-9867-6548}}, 
  \mbox{V.~Vassiliev$^{\ref{AFFIL::USAUCLA}}$}, 
  \mbox{M.~V\'azquez~Acosta$^{\ref{AFFIL::SpainIAC}}$\orcid{0000-0002-2409-9792}}, 
  \mbox{M.~Vecchi$^{\ref{AFFIL::NetherlandsUGroningen}}$\orcid{0000-0002-5338-6029}}, 
  \mbox{S.~Ventura$^{\ref{AFFIL::ItalyINFNPisa}}$\orcid{0000-0001-7065-5342}}, 
  \mbox{S.~Vercellone$^{\ref{AFFIL::ItalyOBrera}}$\orcid{0000-0003-1163-1396}}, 
  \mbox{G.~Verna$^{\ref{AFFIL::ItalyUSienaandINFN}}$\orcid{0000-0001-5916-9028}}, 
  \mbox{A.~Viana$^{\ref{AFFIL::BrazilIFSCUSaoPaulo}}$}, 
  \mbox{N.~Viaux$^{\ref{AFFIL::ChileDepFisUTecnicaFedericoSantaMaria}}$\orcid{0000-0002-5102-9140}}, 
  \mbox{A.~Vigliano$^{\ref{AFFIL::ItalyUUdineandINFNTrieste}}$\orcid{0009-0001-3508-4019}}, 
  \mbox{J.~Vignatti$^{\ref{AFFIL::ChileDepFisUTecnicaFedericoSantaMaria}}$\orcid{0000-0002-1494-9562}}, 
  \mbox{C.~F.~Vigorito$^{\ref{AFFIL::ItalyINFNTorino},\ref{AFFIL::ItalyUTorino}}$\orcid{0000-0002-0069-9195}}, 
  \mbox{J.~Villanueva$^{\ref{AFFIL::ChileUdeValparaiso}}$}, 
  \mbox{E.~Visentin$^{\ref{AFFIL::ItalyINFNTorino},\ref{AFFIL::ItalyUTorino}}$}, 
  \mbox{V.~Vitale$^{\ref{AFFIL::ItalyINFNRomaTorVergata}}$}, 
  \mbox{V.~Vodeb$^{\ref{AFFIL::SloveniaUNovaGoricaCAC}}$}, 
  \mbox{V.~Voisin$^{\ref{AFFIL::FranceLPNHEUSorbonne}}$}, 
  \mbox{V.~Voitsekhovskyi$^{\ref{AFFIL::SwitzerlandUGenevaDPNC}}$\orcid{0000-0002-3906-4840}}, 
  \mbox{S.~Vorobiov$^{\ref{AFFIL::SloveniaUNovaGoricaCAC}}$\orcid{0000-0001-8679-3424}}, 
  \mbox{G.~Voutsinas$^{\ref{AFFIL::SwitzerlandUGenevaDPNC}}$}, 
  \mbox{I.~Vovk$^{\ref{AFFIL::JapanUTokyoICRR}}$\orcid{0000-0003-3444-3830}}, 
  \mbox{T.~Vuillaume$^{\ref{AFFIL::FranceLAPPUSavoieMontBlanc}}$\orcid{0000-0002-5686-2078}}, 
  \mbox{S.~J.~Wagner$^{\ref{AFFIL::GermanyLSW}}$}, 
  \mbox{R.~Walter$^{\ref{AFFIL::SwitzerlandUGenevaISDC}}$\orcid{0000-0003-2362-4433}}, 
  \mbox{M.~White$^{\ref{AFFIL::AustraliaUAdelaide}}$}, 
  \mbox{R.~White$^{\ref{AFFIL::GermanyMPIK}}$}, 
  \mbox{A.~Wierzcholska$^{\ref{AFFIL::PolandIFJ}}$\orcid{0000-0003-4472-7204}}, 
  \mbox{M.~Will$^{\ref{AFFIL::GermanyMPP}}$\orcid{0000-0002-7504-2083}}, 
  \mbox{D.~A.~Williams$^{\ref{AFFIL::USASCIPP}}$\orcid{0000-0003-2740-9714}}, 
  \mbox{F.~Wohlleben$^{\ref{AFFIL::GermanyMPIK}}$\orcid{0000-0002-6451-4188}}, 
  \mbox{A.~Wolter$^{\ref{AFFIL::ItalyOBrera}}$\orcid{0000-0001-5840-9835}}, 
  \mbox{T.~Yamamoto$^{\ref{AFFIL::JapanUKonan}}$}, 
  \mbox{L.~Yang$^{\ref{AFFIL::SouthAfricaUJohannesburg},\ref{AFFIL::ChinaUSunYatsen}}$\orcid{0000-0001-7416-7434}}, 
  \mbox{T.~Yoshida$^{\ref{AFFIL::JapanUIbaraki}}$}, 
  \mbox{T.~Yoshikoshi$^{\ref{AFFIL::JapanUTokyoICRR}}$\orcid{0000-0002-6045-9839}}, 
  \mbox{G.~Zaharijas$^{\ref{AFFIL::SloveniaUNovaGoricaCAC}}$}, 
  \mbox{L.~Zampieri$^{\ref{AFFIL::ItalyOPadova}}$\orcid{0000-0002-6516-1329}}, 
  \mbox{R.~Zanmar~Sanchez$^{\ref{AFFIL::ItalyOCapodimonte},\ref{AFFIL::ItalyOCatania}}$\orcid{0000-0002-6997-0887}}, 
  \mbox{D.~Zavrtanik$^{\ref{AFFIL::SloveniaUNovaGoricaCAC}}$\orcid{0000-0002-4596-1521}}, 
  \mbox{M.~Zavrtanik$^{\ref{AFFIL::SloveniaUNovaGoricaCAC}}$}, 
  \mbox{A.~A.~Zdziarski$^{\ref{AFFIL::PolandNicolausCopernicusAstronomicalCenter}}$}, 
  \mbox{A.~Zech$^{\ref{AFFIL::FranceObservatoiredeParis}}$\orcid{0000-0002-4388-5625}}, 
  \mbox{W.~Zhang$^{\ref{AFFIL::SpainICECSIC}}$\orcid{0000-0003-2839-1325}}, 
  \mbox{V.~I.~Zhdanov$^{\ref{AFFIL::UkraineAstObsofUKyiv}}$\orcid{0000-0003-3690-483X}}, 
  \mbox{K.~Zi\k{e}tara$^{\ref{AFFIL::PolandUJagiellonian}}$}, 
  \mbox{M.~\v{Z}ivec$^{\ref{AFFIL::SloveniaUNovaGoricaCAC}}$\orcid{0009-0003-8528-1453}}, 
  \mbox{J.~Zuriaga-Puig$^{\ref{AFFIL::SpainIFTUAMCSIC}}$\orcid{0000-0003-0652-6700}}
}

\affiliation{
\begin{enumerate}
\item Institute for Cosmic Ray Research, University of Tokyo, 5-1-5, Kashiwa-no-ha, Kashiwa, Chiba 277-8582, Japan\label{AFFIL::JapanUTokyoICRR}
\item ETH Z\"urich, Institute for Particle Physics and Astrophysics, Otto-Stern-Weg 5, 8093 Z\"urich, Switzerland\label{AFFIL::SwitzerlandETHZurich}
\item INFN and Universit\`a degli Studi di Siena, Dipartimento di Scienze Fisiche, della Terra e dell'Ambiente (DSFTA), Sezione di Fisica, Via Roma 56, 53100 Siena, Italy\label{AFFIL::ItalyUSienaandINFN}
\item Universit\'e Paris-Saclay, Universit\'e Paris Cit\'e, CEA, CNRS, AIM, F-91191 Gif-sur-Yvette Cedex, France\label{AFFIL::FranceCEAIRFUDAp}
\item FSLAC IRL 2009, CNRS/IAC, La Laguna, Tenerife, Spain\label{AFFIL::SpainFSLACIRLCNRSIAC}
\item University of Alabama, Tuscaloosa, Department of Physics and Astronomy, Gallalee Hall, Box 870324 Tuscaloosa, AL 35487-0324, USA\label{AFFIL::USAUAlabamaTuscaloosa}
\item Universit\'e C\^ote d'Azur, Observatoire de la C\^ote d'Azur, CNRS, Laboratoire Lagrange, France\label{AFFIL::FranceOCotedAzur}
\item Laboratoire Leprince-Ringuet, CNRS/IN2P3, \'Ecole polytechnique, Institut Polytechnique de Paris, 91120 Palaiseau, France\label{AFFIL::FranceLLREcolePolytechnique}
\item Departament de F{\'\i}sica Qu\`antica i Astrof{\'\i}sica, Institut de Ci\`encies del Cosmos, Universitat de Barcelona, IEEC-UB, Mart{\'\i} i Franqu\`es, 1, 08028, Barcelona, Spain\label{AFFIL::SpainICCUB}
\item Instituto de Astrof{\'\i}sica de Andaluc{\'\i}a-CSIC, Glorieta de la Astronom{\'\i}a s/n, 18008, Granada, Spain\label{AFFIL::SpainIAACSIC}
\item Institute for Computational Cosmology and Department of Physics, Durham University, South Road, Durham DH1 3LE, United Kingdom\label{AFFIL::UnitedKingdomICCUDurham}
\item Pontificia Universidad Cat\'olica de Chile, Av. Libertador Bernardo O'Higgins 340, Santiago, Chile\label{AFFIL::ChileUPontificiaCatolicadeChile}
\item Universidad Nacional Aut\'onoma de M\'exico, Delegaci\'on Coyoac\'an, 04510 Ciudad de M\'exico, Mexico\label{AFFIL::MexicoUNAMMexico}
\item IPARCOS-UCM, Instituto de F{\'\i}sica de Part{\'\i}culas y del Cosmos, and EMFTEL Department, Universidad Complutense de Madrid, E-28040 Madrid, Spain\label{AFFIL::SpainUCMAltasEnergias}
\item Instituto de F{\'\i}sica Te\'orica UAM/CSIC and Departamento de F{\'\i}sica Te\'orica, Universidad Aut\'onoma de Madrid, c/ Nicol\'as Cabrera 13-15, Campus de Cantoblanco UAM, 28049 Madrid, Spain\label{AFFIL::SpainIFTUAMCSIC}
\item LUTH, GEPI and LERMA, Observatoire de Paris, Universit\'e PSL, Universit\'e Paris Cit\'e, CNRS, 5 place Jules Janssen, 92190, Meudon, France\label{AFFIL::FranceObservatoiredeParis}
\item INAF - Osservatorio Astrofisico di Arcetri, Largo E. Fermi, 5 - 50125 Firenze, Italy\label{AFFIL::ItalyOArcetri}
\item INFN Sezione di Perugia and Universit\`a degli Studi di Perugia, Via A. Pascoli, 06123 Perugia, Italy\label{AFFIL::ItalyUPerugiaandINFN}
\item International Institute of Physics, Universidade Federal do Rio Grande do Norte, 59078-970, Natal, RN, Brasil\label{AFFIL::BrazilURioGrandedoNorteIIP}
\item Departamento de F{\'\i}sica, Universidade Federal do Rio Grande do Norte, 59078-970, Natal, RN, Brasil\label{AFFIL::BrazilURioGrandedoNortePhys}
\item INFN Sezione di Napoli, Via Cintia, ed. G, 80126 Napoli, Italy\label{AFFIL::ItalyINFNNapoli}
\item INFN Sezione di Padova, Via Marzolo 8, 35131 Padova, Italy\label{AFFIL::ItalyINFNPadova}
\item Instituto de Astrof{\'\i}sica de Canarias and Departamento de Astrof{\'\i}sica, Universidad de La Laguna, La Laguna, Tenerife, Spain\label{AFFIL::SpainIAC}
\item Laboratoire Univers et Particules de Montpellier, Universit\'e de Montpellier, CNRS/IN2P3, CC 72, Place Eug\`ene Bataillon, F-34095 Montpellier Cedex 5, France\label{AFFIL::FranceLUPMUMontpellier}
\item Kapteyn Astronomical Institute, University of Groningen, Landleven 12, 9747 AD, Groningen, The Netherlands\label{AFFIL::NetherlandsUGroningen}
\item Department of Physics, Chemistry \& Material Science, University of Namibia, Private Bag 13301, Windhoek, Namibia\label{AFFIL::NamibiaUNamibia}
\item Centre for Space Research, North-West University, Potchefstroom, 2520, South Africa\label{AFFIL::SouthAfricaNWU}
\item Universit\"at Hamburg, Institut f\"ur Experimentalphysik, Luruper Chaussee 149, 22761 Hamburg, Germany\label{AFFIL::GermanyUHamburg}
\item School of Physics and Astronomy, Monash University, Melbourne, Victoria 3800, Australia\label{AFFIL::AustraliaUMonash}
\item D\'epartement de physique nucl\'eaire et corpusculaire, University de Gen\`eve,  Facult\'e de Sciences, 1205 Gen\`eve, Switzerland\label{AFFIL::SwitzerlandUGenevaDPNC}
\item Faculty of Science and Technology, Universidad del Azuay, Cuenca, Ecuador.\label{AFFIL::EcuadorUAzuay}
\item Deutsches Elektronen-Synchrotron, Platanenallee 6, 15738 Zeuthen, Germany\label{AFFIL::GermanyDESY}
\item Centro Brasileiro de Pesquisas F{\'\i}sicas, Rua Xavier Sigaud 150, RJ 22290-180, Rio de Janeiro, Brazil\label{AFFIL::BrazilCBPF}
\item Instituto de Astronomia, Geof{\'\i}sica e Ci\^encias Atmosf\'ericas - Universidade de S\~ao Paulo, Cidade Universit\'aria, R. do Mat\~ao, 1226, CEP 05508-090, S\~ao Paulo, SP, Brazil\label{AFFIL::BrazilIAGUSaoPaulo}
\item INFN Sezione di Padova and Universit\`a degli Studi di Padova, Via Marzolo 8, 35131 Padova, Italy\label{AFFIL::ItalyUPadovaandINFN}
\item Institut f\"ur Physik \& Astronomie, Universit\"at Potsdam, Karl-Liebknecht-Strasse 24/25, 14476 Potsdam, Germany\label{AFFIL::GermanyUPotsdam}
\item University of the Witwatersrand, 1 Jan Smuts Avenue, Braamfontein, 2000 Johannesburg, South Africa\label{AFFIL::SouthAfricaUWitwatersrand}
\item Center for Astrophysics | Harvard \& Smithsonian, 60 Garden St, Cambridge, MA 02138, USA\label{AFFIL::USACfAHarvardSmithsonian}
\item Department of Physics, Humboldt University Berlin, Newtonstr. 15, 12489 Berlin, Germany\label{AFFIL::GermanyUBerlin}
\item CIEMAT, Avda. Complutense 40, 28040 Madrid, Spain\label{AFFIL::SpainCIEMAT}
\item Max-Planck-Institut f\"ur Kernphysik, Saupfercheckweg 1, 69117 Heidelberg, Germany\label{AFFIL::GermanyMPIK}
\item Max-Planck-Institut f\"ur Physik, Boltzmannstr. 8, 85748 Garching, Germany\label{AFFIL::GermanyMPP}
\item Univ. Savoie Mont Blanc, CNRS, Laboratoire d'Annecy de Physique des Particules - IN2P3, 74000 Annecy, France\label{AFFIL::FranceLAPPUSavoieMontBlanc}
\item Center for Astrophysics and Cosmology (CAC), University of Nova Gorica, Nova Gorica, Slovenia\label{AFFIL::SloveniaUNovaGoricaCAC}
\item INAF - Osservatorio Astronomico di Roma, Via di Frascati 33, 00078, Monteporzio Catone, Italy\label{AFFIL::ItalyORoma}
\item Politecnico di Bari, via Orabona 4, 70124 Bari, Italy\label{AFFIL::ItalyPolitecnicoBari}
\item INFN Sezione di Bari, via Orabona 4, 70126 Bari, Italy\label{AFFIL::ItalyINFNBari}
\item Universit\'e Paris-Saclay, CNRS/IN2P3, IJCLab, 91405 Orsay, France\label{AFFIL::FranceIJCLab}
\item Institut universitaire de France (IUF)\label{AFFIL::FranceIUFInstitutuniversitairedeFrance}
\item Institut de Fisica d'Altes Energies (IFAE), The Barcelona Institute of Science and Technology, Campus UAB, 08193 Bellaterra (Barcelona), Spain\label{AFFIL::SpainIFAEBIST}
\item FZU - Institute of Physics of the Czech Academy of Sciences, Na Slovance 1999/2, 182 00 Praha 8, Czech Republic\label{AFFIL::CzechRepublicFZU}
\item INAF - Osservatorio Astronomico di Palermo {\textquotedblleft}G.S. Vaiana{\textquotedblright}, Piazza del Parlamento 1, 90134 Palermo, Italy\label{AFFIL::ItalyOPalermo}
\item Sorbonne Universit\'e, CNRS/IN2P3, Laboratoire de Physique Nucl\'eaire et de Hautes Energies, LPNHE, 4 place Jussieu, 75005 Paris, France\label{AFFIL::FranceLPNHEUSorbonne}
\item INAF - Osservatorio Astronomico di Brera, Via Brera 28, 20121 Milano, Italy\label{AFFIL::ItalyOBrera}
\item INFN Sezione di Pisa, Edificio C {\textendash} Polo Fibonacci, Largo Bruno Pontecorvo 3, 56127 Pisa\label{AFFIL::ItalyINFNPisa}
\item University School for Advanced Studies IUSS Pavia, Palazzo del Broletto, Piazza della Vittoria 15, 27100 Pavia, Italy\label{AFFIL::ItalyIUSSPaviaINAF}
\item Universit\`a degli Studi di Trento, Via Calepina, 14, 38122 Trento, Italy\label{AFFIL::ItalyUTrento}
\item University of Zagreb, Faculty of electrical engineering and computing, Unska 3, 10000 Zagreb, Croatia\label{AFFIL::CroatiaUZagreb}
\item University of Oslo, Department of Physics, Sem Saelandsvei 24 - PO Box 1048 Blindern, N-0316 Oslo, Norway\label{AFFIL::NorwayUOslo}
\item INAF - Osservatorio di Astrofisica e Scienza dello spazio di Bologna, Via Piero Gobetti 93/3, 40129  Bologna, Italy\label{AFFIL::ItalyOASBologna}
\item Dublin City University, Glasnevin, Dublin 9, Ireland\label{AFFIL::IrelandDCU}
\item Dublin Institute for Advanced Studies, 31 Fitzwilliam Place, Dublin 2, Ireland\label{AFFIL::IrelandDIAS}
\item Centre for Advanced Instrumentation, Department of Physics, Durham University, South Road, Durham, DH1 3LE, United Kingdom\label{AFFIL::UnitedKingdomUDurham}
\item Astronomical Observatory, Department of Physics, University of Warsaw, Aleje Ujazdowskie 4, 00478 Warsaw, Poland\label{AFFIL::PolandUWarsawPhysics}
\item INFN Sezione di Trieste and Universit\`a degli Studi di Udine, Via delle Scienze 208, 33100 Udine, Italy\label{AFFIL::ItalyUUdineandINFNTrieste}
\item Armagh Observatory and Planetarium, College Hill, Armagh BT61 9DB, United Kingdom\label{AFFIL::UnitedKingdomArmaghObservatoryandPlanetarium}
\item School of Physics, University of New South Wales, Sydney NSW 2052, Australia\label{AFFIL::AustraliaUNewSouthWales}
\item INFN Sezione di Catania, Via S. Sofia 64, 95123 Catania, Italy\label{AFFIL::ItalyINFNCatania}
\item Unitat de F{\'\i}sica de les Radiacions, Departament de F{\'\i}sica, and CERES-IEEC, Universitat Aut\`onoma de Barcelona, Edifici C3, Campus UAB, 08193 Bellaterra, Spain\label{AFFIL::SpainUABandCERESIEEC}
\item Department of Physics, Faculty of Science, Kasetsart University, 50 Ngam Wong Wan Rd., Lat Yao, Chatuchak, Bangkok, 10900, Thailand\label{AFFIL::ThailandUKasetsart}
\item National Astronomical Research Institute of Thailand, 191 Huay Kaew Rd., Suthep, Muang, Chiang Mai, 50200, Thailand\label{AFFIL::ThailandNARIT}
\item INAF - Osservatorio Astronomico di Capodimonte, Via Salita Moiariello 16, 80131 Napoli, Italy\label{AFFIL::ItalyOCapodimonte}
\item Universidade Cidade de S\~ao Paulo, N\'ucleo de Astrof{\'\i}sica, R. Galv\~ao Bueno 868, Liberdade, S\~ao Paulo, SP, 01506-000, Brazil\label{AFFIL::BrazilUCidadeSPaulo}
\item Dep. of Physics, Sapienza, University of Roma, Piazzale A. Moro 5, 00185, Roma, Italy \label{AFFIL::ItalyURomaSapienza}
\item INAF - Istituto di Astrofisica Spaziale e Fisica Cosmica di Milano, Via A. Corti 12, 20133 Milano, Italy\label{AFFIL::ItalyIASFMilano}
\item CCTVal, Universidad T\'ecnica Federico Santa Mar{\'\i}a, Avenida Espa\~na 1680, Valpara{\'\i}so, Chile\label{AFFIL::ChileUTecnicaFedericoSantaMaria}
\item Aix Marseille Univ, CNRS/IN2P3, CPPM, Marseille, France\label{AFFIL::FranceCPPMUAixMarseille}
\item Universidad de Alcal\'a - Space \& Astroparticle group, Facultad de Ciencias, Campus Universitario Ctra. Madrid-Barcelona, Km. 33.600 28871 Alcal\'a de Henares (Madrid), Spain\label{AFFIL::SpainUAlcala}
\item INFN Sezione di Bari and Universit\`a degli Studi di Bari, via Orabona 4, 70124 Bari, Italy\label{AFFIL::ItalyUandINFNBari}
\item Universit\'e Paris Cit\'e, CNRS, Astroparticule et Cosmologie, F-75013 Paris, France\label{AFFIL::FranceAPCUParisCite}
\item INFN Sezione di Torino, Via P. Giuria 1, 10125 Torino, Italy\label{AFFIL::ItalyINFNTorino}
\item Dipartimento di Fisica - Universit\`a degli Studi di Torino, Via Pietro Giuria 1 - 10125 Torino, Italy\label{AFFIL::ItalyUTorino}
\item Dipartimento di Fisica e Chimica {\textquotedblleft}E. Segr\`e{\textquotedblright}, Universit\`a degli Studi di Palermo, Via Archirafi 36, 90123, Palermo, Italy\label{AFFIL::ItalyUPalermo}
\item Universidade Federal Do Paran\'a - Setor Palotina, Departamento de Engenharias e Exatas, Rua Pioneiro, 2153, Jardim Dallas, CEP: 85950-000 Palotina, Paran\'a, Brazil\label{AFFIL::BrazilUFPR}
\item INAF - Osservatorio Astrofisico di Catania, Via S. Sofia, 78, 95123 Catania, Italy\label{AFFIL::ItalyOCatania}
\item University of Oxford, Department of Physics, Clarendon Laboratory, Parks Road, Oxford, OX1 3PU, United Kingdom\label{AFFIL::UnitedKingdomUOxford}
\item Universidad de Valpara{\'\i}so, Blanco 951, Valparaiso, Chile\label{AFFIL::ChileUdeValparaiso}
\item University of Wisconsin, Madison, 500 Lincoln Drive, Madison, WI, 53706, USA\label{AFFIL::USAUWisconsin}
\item Department of Physics and Technology, University of Bergen, Museplass 1, 5007 Bergen, Norway\label{AFFIL::NorwayUBergen}
\item INAF - Istituto di Radioastronomia, Via Gobetti 101, 40129 Bologna, Italy\label{AFFIL::ItalyRadioastronomiaINAF}
\item Western Sydney University, Locked Bag 1797, Penrith, NSW 2751, Australia\label{AFFIL::AustraliaUWesternSydney}
\item INAF - Istituto Nazionale di Astrofisica, Viale del Parco Mellini 84, 00136 Rome, Italy\label{AFFIL::ItalyINAF}
\item IRFU, CEA, Universit\'e Paris-Saclay, B\^at 141, 91191 Gif-sur-Yvette, France\label{AFFIL::FranceCEAIRFUDPhP}
\item Instituto de F{\'\i}sica de S\~ao Carlos, Universidade de S\~ao Paulo, Av. Trabalhador S\~ao-carlense, 400 - CEP 13566-590, S\~ao Carlos, SP, Brazil\label{AFFIL::BrazilIFSCUSaoPaulo}
\item Port d'Informaci\'o Cient{\'\i}fica, Edifici D, Carrer de l'Albareda, 08193 Bellaterrra (Cerdanyola del Vall\`es), Spain\label{AFFIL::SpainPIC}
\item Universit\`a degli Studi di Napoli {\textquotedblleft}Federico II{\textquotedblright} - Dipartimento di Fisica {\textquotedblleft}E. Pancini{\textquotedblright}, Complesso Universitario di Monte Sant'Angelo, Via Cintia - 80126 Napoli, Italy\label{AFFIL::ItalyUNapoli}
\item Universit\`a degli Studi di Modena e Reggio Emilia, Dipartimento di Ingegneria ''Enzo Ferrari'', via Pietro Vivarelli 10, 41125, Modena, Italy\label{AFFIL::ItalyUModena}
\item Institut f\"ur Astronomie und Astrophysik, Universit\"at T\"ubingen, Sand 1, 72076 T\"ubingen, Germany\label{AFFIL::GermanyIAAT}
\item Astroparticle Physics, Department of Physics, TU Dortmund University, Otto-Hahn-Str. 4a, 44227 Dortmund, Germany\label{AFFIL::GermanyUDortmundTU}
\item University of Rijeka, Faculty of Physics, Radmile Matejcic 2, 51000 Rijeka, Croatia\label{AFFIL::CroatiaURijeka}
\item Institute for Theoretical Physics and Astrophysics, Universit\"at W\"urzburg, Campus Hubland Nord, Emil-Fischer-Str. 31, 97074 W\"urzburg, Germany\label{AFFIL::GermanyUWurzburg}
\item Institut f\"ur Theoretische Physik, Lehrstuhl IV: Plasma-Astroteilchenphysik, Ruhr-Universit\"at Bochum, Universit\"atsstra{\ss}e 150, 44801 Bochum, Germany\label{AFFIL::GermanyUBochum}
\item Friedrich-Alexander-Universit\"at Erlangen-N\"urnberg, Erlangen Centre for Astroparticle Physics, Nikolaus-Fiebiger-Str. 2, 91058 Erlangen, Germany\label{AFFIL::GermanyUErlangenECAP}
\item Department of Astronomy and Astrophysics, University of Chicago, 5640 S Ellis Ave, Chicago, Illinois, 60637, USA\label{AFFIL::USAUChicagoDAA}
\item LAPTh, CNRS, USMB, F-74940 Annecy, France\label{AFFIL::FranceLAPTh}
\item School of Physics, Chemistry and Earth Sciences, University of Adelaide, Adelaide SA 5005, Australia\label{AFFIL::AustraliaUAdelaide}
\item Department of Physics, Washington University, St. Louis, MO 63130, USA\label{AFFIL::USAWashingtonU}
\item Escuela de Ingenier{\'\i}a El\'ectrica, Facultad de Ingenier{\'\i}a, Pontificia Universidad Cat\'olica de Valpara{\'\i}so, Avenida Brasil 2147, Valpara{\'\i}so, Chile\label{AFFIL::ChileEscIngElec}
\item Santa Cruz Institute for Particle Physics and Department of Physics, University of California, Santa Cruz, 1156 High Street, Santa Cruz, CA 95064, USA\label{AFFIL::USASCIPP}
\item Escola de Artes, Ci\^encias e Humanidades, Universidade de S\~ao Paulo, Rua Arlindo Bettio, CEP 03828-000, 1000 S\~ao Paulo, Brazil\label{AFFIL::BrazilEACHUSaoPaulo}
\item Astronomical Observatory of Taras Shevchenko National University of Kyiv, 3 Observatorna Street, Kyiv, 04053, Ukraine\label{AFFIL::UkraineAstObsofUKyiv}
\item Department of Physics and Astronomy, University of Utah, Salt Lake City, UT 84112-0830, USA\label{AFFIL::USAUUtah}
\item The University of Manitoba, Dept of Physics and Astronomy, Winnipeg, Manitoba R3T 2N2, Canada\label{AFFIL::CanadaUManitoba}
\item RIKEN, Institute of Physical and Chemical Research, 2-1 Hirosawa, Wako, Saitama, 351-0198, Japan\label{AFFIL::JapanRIKEN}
\item INFN Sezione di Roma La Sapienza, P.le Aldo Moro, 2 - 00185 Roma, Italy\label{AFFIL::ItalyINFNRomaLaSapienza}
\item INAF - Osservatorio Astronomico di Padova, Vicolo dell'Osservatorio 5, 35122 Padova, Italy\label{AFFIL::ItalyOPadova}
\item INAF - Istituto di Astrofisica e Planetologia Spaziali (IAPS), Via del Fosso del Cavaliere 100, 00133 Roma, Italy\label{AFFIL::ItalyIAPS}
\item Physics Program, Graduate School of Advanced Science and Engineering, Hiroshima University, 739-8526 Hiroshima, Japan\label{AFFIL::JapanUHiroshima}
\item Department of Physics, Nagoya University, Chikusa-ku, Nagoya, 464-8602, Japan\label{AFFIL::JapanUNagoya}
\item Department of Information Technology, Escuela Polit\'ecnica Superior, Universidad San Pablo-CEU, CEU Universities, Campus Montepr{\'\i}ncipe, Boadilla del Monte, Madrid 28668, Spain\label{AFFIL::SpainUniversidadSanPabloCEU}
\item INFN Sezione di Roma Tor Vergata, Via della Ricerca Scientifica 1, 00133 Rome, Italy\label{AFFIL::ItalyINFNRomaTorVergata}
\item Alikhanyan National Science Laboratory, Yerevan Physics Institute, 2 Alikhanyan Brothers St., 0036, Yerevan, Armenia\label{AFFIL::ArmeniaNSLAlikhanyan}
\item Universit\'e Paris Cit\'e, CNRS, CEA, Astroparticule et Cosmologie, F-75013 Paris, France\label{AFFIL::FranceAPCUParisCiteCEAaffiliatedpersonnel}
\item Universidad Andr\'es Bello, Av. Fern\'andez Concha 700, Las Condes, Santiago, Chile\label{AFFIL::ChileUAndresBello}
\item N\'ucleo de Astrof{\'\i}sica e Cosmologia (Cosmo-ufes) \& Departamento de F{\'\i}sica, Universidade Federal do Esp{\'\i}rito Santo (UFES), Av. Fernando Ferrari, 514. 29065-910. Vit\'oria-ES, Brazil\label{AFFIL::BrazilUFES}
\item Astrophysics Research Center of the Open University (ARCO), The Open University of Israel, P.O. Box 808, Ra{\textquoteright}anana 4353701, Israel\label{AFFIL::IsraelOpenUniversityofIsrael}
\item Department of Physics, The George Washington University, Washington, DC 20052, USA\label{AFFIL::USAGWUWashingtonDC}
\item Cherenkov Telescope Array Observatory gGmbH, Via Gobetti, Bologna, Italy\label{AFFIL::ItalyCTAOBologna}
\item Learning and Education Development Center, Yamanashi-Gakuin University, Kofu, Yamanashi 400-8575, Japan\label{AFFIL::JapanUYamanashiGakuin}
\item Sendai College, National Institute of Technology, Natori, Miyagi 981-1239, Japan\label{AFFIL::JapanNITSendaiNatori}
\item Universit\"at Innsbruck, Institut f\"ur Astro- und Teilchenphysik, Technikerstr. 25/8, 6020 Innsbruck, Austria\label{AFFIL::AustriaUInnsbruck}
\item University of Toyama, Department of Physics, 3190 Gofuku, Toyama 930-8555, Japan \label{AFFIL::JapanUToyama}
\item Department of Physics and Astronomy and the Bartol Research Institute, University of Delaware, Newark, DE 19716, USA\label{AFFIL::USAUDelaware}
\item Palack\'y University Olomouc, Faculty of Science, Joint Laboratory of Optics of Palack\'y University and Institute of Physics of the Czech Academy of Sciences, 17. listopadu 1192/12, 779 00 Olomouc, Czech Republic\label{AFFIL::CzechRepublicUOlomouc}
\item Finnish Centre for Astronomy with ESO, University of Turku, Finland, FI-20014 University of Turku, Finland\label{AFFIL::FinlandFINCA}
\item Aalto University, Mets\"ahovi Radio Observatory, Mets\"ahovintie 114, FI-02540 Kylm\"al\"a, Finland\label{AFFIL::FinlandUAalto}
\item Josip Juraj Strossmayer University of Osijek, Trg Ljudevita Gaja 6, 31000 Osijek, Croatia\label{AFFIL::CroatiaUOsijek}
\item CETEMPS Dipartimento di Scienze Fisiche e Chimiche, Universit\`a degli Studi dell{\textquoteright}Aquila and GSGC-LNGS-INFN, Via Vetoio 1, L{\textquoteright}Aquila, 67100, Italy\label{AFFIL::ItalyCETEMPSandUandINFNAquila}
\item Chiba University, 1-33, Yayoicho, Inage-ku, Chiba-shi, Chiba, 263-8522 Japan\label{AFFIL::JapanUChiba}
\item Department of Earth and Space Science, Graduate School of Science, Osaka University, Toyonaka 560-0043, Japan\label{AFFIL::JapanUOsaka}
\item Institute of Astronomy, Faculty of Physics, Astronomy and Informatics, Nicolaus Copernicus University in Toru\'n, ul. Grudzi\k{a}dzka 5, 87-100 Toru\'n, Poland\label{AFFIL::PolandTorunInstituteofAstronomy}
\item Astronomical Observatory, Jagiellonian University, ul. Orla 171, 30-244 Cracow, Poland\label{AFFIL::PolandUJagiellonian}
\item Landessternwarte, Zentrum f\"ur Astronomie  der Universit\"at Heidelberg, K\"onigstuhl 12, 69117 Heidelberg, Germany\label{AFFIL::GermanyLSW}
\item IRAP, Universit\'e de Toulouse, CNRS, CNES, UPS, 9 avenue Colonel Roche, 31028 Toulouse, Cedex 4, France\label{AFFIL::FranceIRAPUToulouse}
\item Department of Physics and Astronomy, University of California, Los Angeles, CA 90095, USA\label{AFFIL::USAUCLA}
\item Sendai College, National Institute of Technology, 4-16-1 Ayashi-Chuo, Aoba-ku, Sendai city, Miyagi 989-3128, Japan\label{AFFIL::JapanNITSendaiHirose}
\item Astronomical Institute of the Czech Academy of Sciences, Bocni II 1401 - 14100 Prague, Czech Republic\label{AFFIL::CzechRepublicASU}
\item Faculty of Science, Ibaraki University, Mito, Ibaraki, 310-8512, Japan\label{AFFIL::JapanUIbaraki}
\item Faculty of Science and Engineering, Waseda University, Shinjuku, Tokyo 169-8555, Japan\label{AFFIL::JapanUWaseda}
\item School of Physics, Aristotle University, Thessaloniki, 54124 Thessaloniki, Greece\label{AFFIL::GreeceUThessaloniki}
\item National Astronomical Observatory of Japan (NAOJ), Division of Science, 2-21-1, Osawa, Mitaka, Tokyo 181-8588, Japan\label{AFFIL::JapanNAOJ}
\item Institute of Particle and Nuclear Studies,  KEK (High Energy Accelerator Research Organization), 1-1 Oho, Tsukuba, 305-0801, Japan\label{AFFIL::JapanKEK}
\item Department of Physics, Tokai University, 4-1-1, Kita-Kaname, Hiratsuka, Kanagawa 259-1292, Japan\label{AFFIL::JapanUTokai}
\item INAF - Istituto di Astrofisica Spaziale e Fisica Cosmica di Palermo, Via U. La Malfa 153, 90146 Palermo, Italy\label{AFFIL::ItalyIASFPalermo}
\item School of Physics and Astronomy, University of Leicester, Leicester, LE1 7RH, United Kingdom\label{AFFIL::UnitedKingdomULeicester}
\item Universit\'e Bordeaux, CNRS, LP2I Bordeaux, UMR 5797, 19 Chemin du Solarium, F-33170 Gradignan, France\label{AFFIL::FranceLP2IUBordeaux}
\item Universit\`a degli studi di Catania, Dipartimento di Fisica e Astronomia {\textquotedblleft}Ettore Majorana{\textquotedblright}, Via S. Sofia 64, 95123 Catania, Italy\label{AFFIL::ItalyUCatania}
\item Department of Physics and Astronomy, University of Turku, Finland, FI-20014 University of Turku, Finland\label{AFFIL::FinlandUTurku}
\item INFN Sezione di Trieste and Universit\`a degli Studi di Trieste, Via Valerio 2 I, 34127 Trieste, Italy\label{AFFIL::ItalyUandINFNTrieste}
\item Escuela Polit\'ecnica Superior de Ja\'en, Universidad de Ja\'en, Campus Las Lagunillas s/n, Edif. A3, 23071 Ja\'en, Spain\label{AFFIL::SpainUJaen}
\item Anton Pannekoek Institute/GRAPPA, University of Amsterdam, Science Park 904 1098 XH Amsterdam, The Netherlands\label{AFFIL::NetherlandsUAmsterdam}
\item Saha Institute of Nuclear Physics, A CI of Homi Bhabha National Institute, Kolkata 700064, West Bengal, India\label{AFFIL::IndiaSahaInstitute}
\item UCM-ELEC group, EMFTEL Department, University Complutense of Madrid, 28040 Madrid, Spain\label{AFFIL::SpainUCMElectronica}
\item Departamento de Ingenier{\'\i}a El\'ectrica, Universidad Pontificia Comillas - ICAI, 28015 Madrid\label{AFFIL::SpainUPCMadrid}
\item Institute of Space Sciences (ICE, CSIC), and Institut d'Estudis Espacials de Catalunya (IEEC), and Instituci\'o Catalana de Recerca I Estudis Avan\c{c}ats (ICREA), Campus UAB, Carrer de Can Magrans, s/n 08193 Cerdanyola del Vall\'es, Spain\label{AFFIL::SpainICECSIC}
\item The Henryk Niewodnicza\'nski Institute of Nuclear Physics, Polish Academy of Sciences, ul. Radzikowskiego 152, 31-342 Cracow, Poland\label{AFFIL::PolandIFJ}
\item IPARCOS Institute, Faculty of Physics (UCM), 28040 Madrid, Spain\label{AFFIL::SpainIPARCOSInstitute}
\item Department of Physics, Konan University, Kobe, Hyogo, 658-8501, Japan\label{AFFIL::JapanUKonan}
\item Hiroshima Astrophysical Science Center, Hiroshima University, Higashi-Hiroshima, Hiroshima 739-8526, Japan\label{AFFIL::JapanHASC}
\item Nicolaus Copernicus Astronomical Center, Polish Academy of Sciences, ul. Bartycka 18, 00-716 Warsaw, Poland\label{AFFIL::PolandNicolausCopernicusAstronomicalCenter}
\item School of Allied Health Sciences, Kitasato University, Sagamihara, Kanagawa 228-8555, Japan\label{AFFIL::JapanUKitasato}
\item Department of Physics, Yamagata University, Yamagata, Yamagata 990-8560, Japan\label{AFFIL::JapanUYamagata}
\item Kavli Institute for Particle Astrophysics and Cosmology, Stanford University, Stanford, CA 94305, USA\label{AFFIL::USAStanford}
\item University of Bia{\l}ystok, Faculty of Physics, ul. K. Cio{\l}kowskiego 1L, 15-245 Bia{\l}ystok, Poland\label{AFFIL::PolandUBiaystok}
\item Charles University, Institute of Particle \& Nuclear Physics, V Hole\v{s}ovi\v{c}k\'ach 2, 180 00 Prague 8, Czech Republic\label{AFFIL::CzechRepublicUPrague}
\item Institute for Space{\textemdash}Earth Environmental Research, Nagoya University, Furo-cho, Chikusa-ku, Nagoya 464-8601, Japan\label{AFFIL::JapanUNagoyaISEE}
\item Kobayashi{\textemdash}Maskawa Institute for the Origin of Particles and the Universe, Nagoya University, Furo-cho, Chikusa-ku, Nagoya 464-8602, Japan\label{AFFIL::JapanUNagoyaKMI}
\item Graduate School of Technology, Industrial and Social Sciences, Tokushima University, Tokushima 770-8506, Japan\label{AFFIL::JapanUTokushima}
\item Cherenkov Telescope Array Observatory, Saupfercheckweg 1, 69117 Heidelberg, Germany\label{AFFIL::GermanyCTAOHeidelberg}
\item University of Pisa, Largo B. Pontecorvo 3, 56127 Pisa, Italy \label{AFFIL::ItalyUPisa}
\item Rudjer Boskovic Institute, Bijenicka 54, 10 000 Zagreb, Croatia\label{AFFIL::CroatiaIRB}
\item INAF - Osservatorio Astronomico di Padova and INFN Sezione di Trieste, gr. coll. Udine, Via delle Scienze 208 I-33100 Udine, Italy\label{AFFIL::ItalyOandINFNTrieste}
\item Pidstryhach Institute for Applied Problems in Mechanics and Mathematics NASU, 3B Naukova Street, Lviv, 79060, Ukraine\label{AFFIL::UkraineIAPMMLviv}
\item Dipartimento di Scienze Fisiche e Chimiche, Universit\`a degli Studi dell'Aquila and GSGC-LNGS-INFN, Via Vetoio 1, L'Aquila, 67100, Italy\label{AFFIL::ItalyUandINFNAquila}
\item Department of Astronomy, University of Geneva, Chemin d'Ecogia 16, CH-1290 Versoix, Switzerland\label{AFFIL::SwitzerlandUGenevaISDC}
\item Ruhr University Bochum, Faculty of Physics and Astronomy, Astronomical Institute (AIRUB), Universit\"atsstra{\ss}e 150, 44801 Bochum, Germany\label{AFFIL::GermanyUBochumPhysAst}
\item Centre for Astro-Particle Physics (CAPP) and Department of Physics, University of Johannesburg, PO Box 524, Auckland Park 2006, South Africa\label{AFFIL::SouthAfricaUJohannesburg}
\item Departamento de F{\'\i}sica, Facultad de Ciencias B\'asicas, Universidad Metropolitana de Ciencias de la Educaci\'on, Avenida Jos\'e Pedro Alessandri 774, \~Nu\~noa, Santiago, Chile\label{AFFIL::ChileUMCE}
\item School of Physics and Astronomy, University of Minnesota, 116 Church Street S.E. Minneapolis, Minnesota 55455-0112, USA\label{AFFIL::USAUMinnesota}
\item Instituto de Estudios Astrof{\'\i}sicos, Facultad de Ingenier{\'\i}a y Ciencias, Universidad Diego Portales, Av. Ej\'ercito Libertador 441, 8370191 Santiago, Chile\label{AFFIL::ChileUniversidadDiegoPortales}
\item Departamento de Astronom{\'\i}a, Universidad de Concepci\'on, Barrio Universitario S/N, Concepci\'on, Chile\label{AFFIL::ChileUdeConcepcion}
\item University of New South Wales, School of Science, Australian Defence Force Academy, Canberra, ACT 2600, Australia \label{AFFIL::AustraliaUNewSouthWalesCanberra}
\item Gifu University, Faculty of Engineering, 1-1 Yanagido, Gifu 501-1193, Japan \label{AFFIL::JapanUGifu}
\item University of Split  - FESB, R. Boskovica 32, 21 000 Split, Croatia\label{AFFIL::CroatiaFESB}
\item Departamento de F{\'\i}sica, Universidad de Santiago de Chile (USACH), Av. Victor Jara 3493, Estaci\'on Central, Santiago, Chile\label{AFFIL::ChileUniversidaddeSantiagodeChile}
\item Main Astronomical Observatory of the National Academy of Sciences of Ukraine, Zabolotnoho str., 27, 03143, Kyiv, Ukraine\label{AFFIL::UkraineObsNASUkraine}
\item Space Technology Centre, AGH University of Krakow, Aleja Mickiewicza 30, Krak\'ow 30-059, Poland\label{AFFIL::PolandAGHCracowSTC}
\item Academic Computer Centre CYFRONET AGH, ul. Nawojki 11, 30-950, Krak\'ow, Poland\label{AFFIL::PolandCYFRONETAGH}
\item Warsaw University of Technology, Faculty of Electronics and Information Technology, Institute of Electronic Systems, Nowowiejska 15/19, 00-665 Warsaw, Poland\label{AFFIL::PolandWUTElectronics}
\item Division of Physics and Astronomy, Graduate School of Science, Kyoto University, Sakyo-ku, Kyoto, 606-8502, Japan\label{AFFIL::JapanUKyotoPhysicsandAstronomy}
\item Escola de Engenharia de Lorena, Universidade de S\~ao Paulo, \'Area I - Estrada Municipal do Campinho, s/n{\textdegree}, CEP 12602-810, Pte. Nova, Lorena, Brazil\label{AFFIL::BrazilEELUSaoPaulo}
\item INAF - Osservatorio Astronomico di Cagliari, Via della Scienza 5, I-09047 Selargius (CA), Italy\label{AFFIL::ItalyINAFCagliari}
\item INAF - Osservatorio Astrofisico di Torino, Strada Osservatorio 20, 10025  Pino Torinese (TO), Italy\label{AFFIL::ItalyOTorino}
\item Departamento de F{\'\i}sica, Universidad T\'ecnica Federico Santa Mar{\'\i}a, Avenida Espa\~na, 1680 Valpara{\'\i}so, Chile\label{AFFIL::ChileDepFisUTecnicaFedericoSantaMaria}
\item School of Physics and Astronomy, Sun Yat-sen University, Zhuhai, China\label{AFFIL::ChinaUSunYatsen}
\end{enumerate}
}

\affiliation{\mbox{ }\newline\textbf{Corresponding authors}: T.~Bringmann (\url{torsten.bringmann@fys.uio.no}),}
\affiliation{\hspace*{131pt} E.~S{\ae}ther~Hatlen (\url{e.s.hatlen@fys.uio.no}),}
\affiliation{\hspace*{131pt} G.~Zaharijas (\url{gabrijela.zaharijas@ung.si})}

\abstract{%
Monochromatic gamma-ray signals constitute a potential smoking gun signature for annihilating or decaying
dark matter particles that could relatively easily be distinguished from astrophysical or instrumental backgrounds. 
We provide an updated assessment of the sensitivity of the Cherenkov Telescope Array (CTA)
to such signals, based on observations of the Galactic centre region as well as of selected dwarf spheroidal galaxies. 
We find that current limits and detection prospects for dark matter masses 
above 300\,GeV will be significantly improved, by up to an order of magnitude in the multi-TeV range. 
This demonstrates that CTA will set a new standard for gamma-ray astronomy also in this respect, as the world's largest and 
most sensitive high-energy gamma-ray observatory, in particular due to its exquisite energy resolution at TeV energies
and the adopted observational strategy focussing on regions with large dark matter densities. 
Throughout our analysis, we use up-to-date instrument response functions, 
and we 
thoroughly model the effect of instrumental systematic uncertainties in our statistical treatment. We further present
results for other potential signatures with sharp spectral features, e.g.~box-shaped spectra, that would
likewise very clearly point to a particle dark matter origin.
}

\begin{document}
\maketitle

\section{Introduction}

The nature of the cosmological dark matter (DM), contributing about $26\,\%$ to the total energy content of 
the universe~\cite{Aghanim:2018eyx}, remains unknown. The most often discussed explanation is that of a
hypothetical elementary particle, and a plethora of viable DM candidates of this type has been 
suggested in the literature~\cite{Jungman:1995df,Bertone:2004pz,Feng:2010gw}. 
Gamma rays produced from the annihilation or decay of these 
particles may provide a promising way to test the particle hypothesis of 
DM~\cite{Bringmann:2012ez}.

The Cherenkov Telecope Array Observatory (CTAO)~\cite{CTAweb}, whose construction is starting, 
will be in an excellent position to perform such an indirect search for DM. One of the reasons is 
the estimated unprecedented angular resolution and sensitivity of this observatory, for gamma-ray 
energies from below 100\,GeV to at least several tens of TeV.
As recently demonstrated~\cite{CTA:2020qlo}, in particular, these properties imply the exciting prospect 
that the Cherenkov Telecope Array (CTA)
 may be able to robustly probe thermally produced weakly interacting massive particles (WIMPs), 
i.e.~the most prominently discussed  type of DM candidates
(for earlier work arriving at similar conclusions, see also 
Refs.~\cite{Doro:2012xx,Silverwood:2014yza,Pierre:2014tra,Carr:2015hta,Lefranc:2015pza}). 
Here we focus instead on 
a different property of CTAO, namely its very good {\it energy resolution}.
As we show here, this may help to single out characteristic spectral features expected in several DM models
-- which, in the case of a detection, would allow a much more robust {signal} claim because the 
discrimination against astrophysical and instrumental backgrounds would be significantly easier than for 
the generic WIMP signals studied in Ref.~\cite{CTA:2020qlo}.

Examples for such {\it smoking gun} signatures of DM include monochromatic gamma-ray 
`lines'~\cite{Srednicki:1985sf,Bergstrom:1988fp,Bergstrom:1997fj}, 
box-shaped signals~\cite{Ibarra:2012dw} and other strongly enhanced spectral features at energies close to the
DM particle's mass~\cite{Bringmann:2007nk}. 
In fact, the details of the spectrum allow to not only discriminate DM from
background components, but can also provide valuable insights about the underlying particle 
physics model~\cite{Bringmann:2012ez}.
On the other hand, such features in the gamma-ray spectra from DM typically appear at smaller rates
than the generic spectra expected from the simplest WIMP models (though, as discussed explicitly 
further down, prominent counterexamples exist). 
In this sense, those generic spectra typically have a significantly better DM {\it constraining} potential, 
while distinct spectral features provide a very promising {\it discovery} 
channel (for DM models that exhibit such spectra).

This difference is also reflected in the analysis methods that are most suitable to identify a potential DM signal.
For the continuum signals expected from generic WIMP models the spectral information is less important 
than the angular information, motivating the use of detailed spatial templates for the DM and the 
various background components~\cite{CTA:2020qlo}. Clearly, this approach is limited by the precision
to which in particular the different background components can be modelled. For (almost) monochromatic signals,
on the other hand, the exact knowledge of the spatial morphology of the background is less crucial. 
In fact, the analysis also becomes to some degree independent of the energy dependence of the background,
as long as it varies much less strongly with energy than the signal. It is worth noting that this generic property of 
spectral `line searches' has been successfully employed not only in the context of DM 
searches~\cite{Bringmann:2012vr,Weniger:2012tx,HESS:2013rld,Fermi-LAT:2015kyq,HESS:2018kom,
HAWC:2019jvm,MAGIC:2022acl} 
but also, e.g., in the discovery of the standard model Higgs boson~\cite{Aad:2012tfa,Chatrchyan:2012ufa}.

In this article we complement the DM analysis of Ref.~\cite{CTA:2020qlo} by estimating the sensitivity 
of CTAO to monochromatic and similar `smoking gun' signals, highly localized in energy. 
We adopt up-to-date background models and the current best estimates for the expected instrument 
performance, using a binned profile likelihood ratio test inside a sliding energy 
window in the range from 200\,GeV to 30\,TeV. For this analysis approach, we pay special attention 
to quantify the impact of systematic uncertainties in the event reconstruction. 
We discuss prospects both for observations of the Galactic Centre (GC) region, where the DM density and 
hence the signal strength is expected to be largest, and for combining observations of dwarf spheroidal galaxies
(dSPhs) where astrophysical gamma-ray backgrounds can largely be neglected at the energies of interest 
here.
For previous work estimating the CTA prospects to observe sharp spectral features, see 
Refs.~\cite{Weniger:2011ik,Ibarra:2015tya,CTAConsortium:2017dvg,Lefranc:2016fgn,Hryczuk:2019nql}.

\medskip
This article is organized as follows. In Section~\ref{section:cta} we give a brief introduction to CTAO
and its expected performance. 
Section~\ref{section:spectral_signatures} introduces in more detail the characteristic spectral features
that we focus our analysis on,  along with a motivation from the underlying DM models.
We discuss the specifics of the target regions of this sensitivity analysis in Section~\ref{sec:targets}, 
both with respect to the modelling of the astrophysical emission components and with respect to the 
expected DM distribution.
In Section~\ref{section:analysis} we provide details about the analysis techniques adopted in this work.
We present our results in Section~\ref{section:results}, and discuss them further in Section~\ref{section:discussion}.
Our final conclusions are given in Section~\ref{section:conclusion}.
In Appendix \ref{app} we provide further details about the statistical analysis method that we adopted.
\section{The Cherenkov Telescope Array Observatory}
\label{section:cta}

Ground-based gamma-ray astronomy started  in the 1980s when the Whipple telescope~\cite{Weekes:1989tc}
demonstrated the feasibility of  the imaging atmospheric Cherenkov light technique. The field of  
ground-based observations  of very high-energy gamma rays then quickly grew to one of the main 
contributors to modern-day astroparticle physics,  
expanding to include also water Cherenkov techniques
(as pioneered, starting from 1999, by Milagro~\cite{Atkins:2004yb}).

Imaging Atmospheric Cherenkov Telescopes (IACTs) operate by detecting extended showers of
Cherenkov light that are produced in the atmosphere due to cascades of relativistic particles
resulting from incident high-energy cosmic ray (CR) particles and 
gamma rays~\cite{DiSciascio:2019ieq}. 
Due to telescope and camera architecture, the field of view (FoV)  of current IACTs is generally limited to 
several degrees.
Currently operating IACT systems are H.E.S.S (5 telescopes, Namibia)~\cite{HESSweb}, VERITAS (4 telescopes, 
Arizona)~\cite{VERITASweb}, and MAGIC (2 telescopes, La Palma)~\cite{MAGICweb}. Having a larger number of 
telescopes is beneficial, as it allows tracking the shower from multiple angles, and therefore improving the 
reconstruction of the arrival direction and energy of the event.  The discrimination between CR proton and 
gamma-ray induced events is possible via the image shape, based on Monte Carlo (MC) simulations, 
which however cannot discriminate electrons and gamma rays.
Since CRs arriving at the top of the atmosphere  are dominated by protons,  with gamma rays only 
making up a tiny fraction (e.g.~$10^{-4}$  of the proton flux at 1\,TeV), 
 large backgrounds due to misidentified charged CRs often present an unavoidable consequence for 
 ground-based detection. 
Next generation water Cherenkov facilities like SWGO may have comparable 
sensitivity in the multi TeV range~\cite{Viana:2019ucn,Conceicao:2023tfb}; 
their expectedly worse energy resolution, however, makes them less competitive  
to search for the kind of monochromatic spectral features that we will focus on in our analysis.

\begin{figure}[t]
\centering
\includegraphics[width=0.48\textwidth]{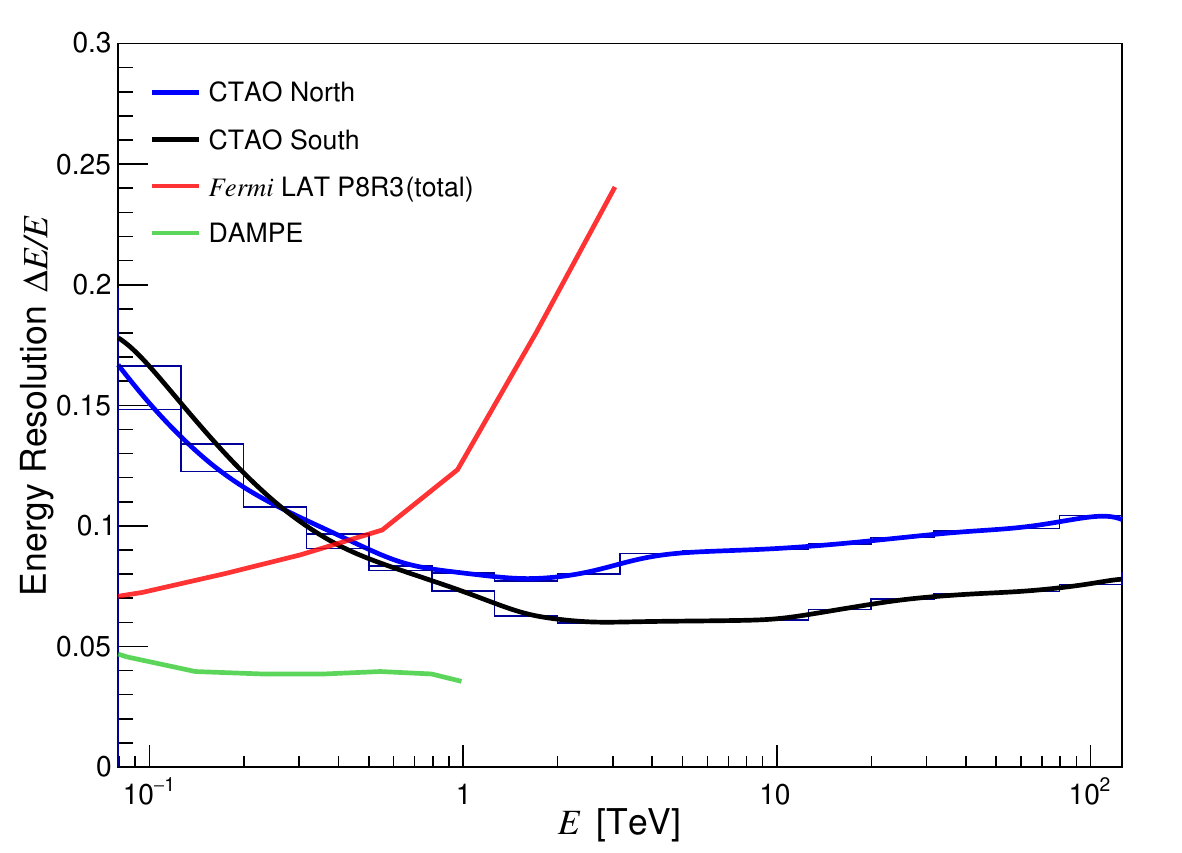}
~~~
\includegraphics[width=0.48\textwidth]{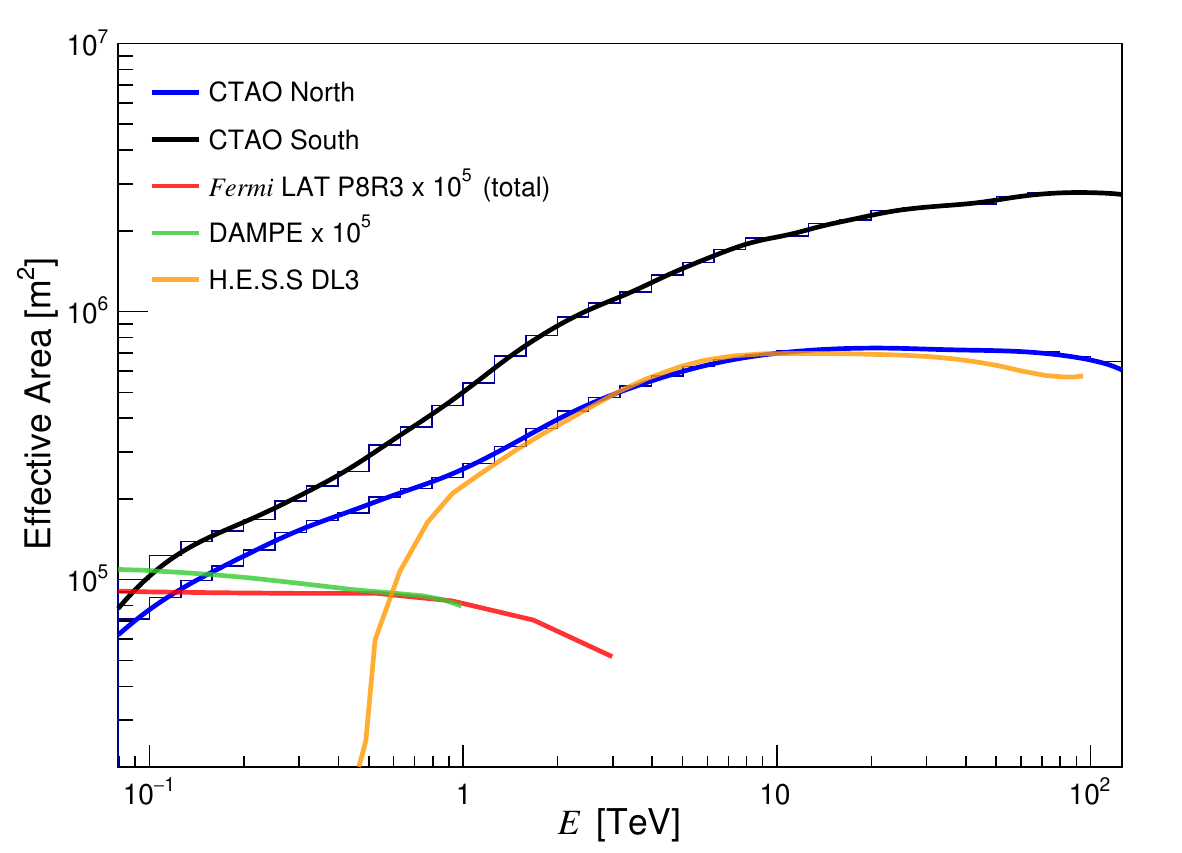}
\vspace{-1ex}
\caption{\textit{Left panel.} The expected energy resolution of CTAO as a function of (true) energy for the 
northern (blue) and southern (black) array, obtained as linear interpolation of the histograms provided with the IRF
(indicated with thinner lines; cf.~footnote \ref{foot_IRF}).
Here the energy resolution $\Delta E$ is defined such that $68\%$ of the reconstructed gamma-rays will have 
a true energy within $\Delta E$.
For comparison, we also show in red the energy resolution for  {\it Fermi}-LAT Pass 8 Release 3 SOURCE V3 
(total)~\cite{Bruel:2018lac} and in green that for DAMPE~\cite{DAMPE:2021hsz}.
\textit{Right panel.} 
Effective area of the two site locations as a function of energy.
The thick solid lines are based on a Gaussian smoothing of width $\Delta E$, as used in our analysis.
In addition, we show the effective areas for  Fermi (red), DAMPE (green) and
H.E.S.S.~Data Level 3 (DL3)~\cite{HESS:2018zix} (orange).
}
\label{fig:cta}
\end{figure}

CTAO~\cite{Acharya:2013sxa} is the next-generation ground-based gamma-ray instrument facility. 
Its construction is already starting,  and large-scale telescope production is expected to begin in 2025. 
The goal of CTA (for the so-called `Omega' configuration) is to build about 100 IACTs of three different sizes
and distribute them among two locations, one for each hemisphere: Paranal in Chile for the 
southern hemisphere, and La Palma in Spain for the northern. The southern hemisphere array will 
consist of 
telescopes covering the entire energy range of CTAO; LSTs (Large-Sized 
Telescopes) for the $20\!-\!150$\,GeV range, MSTs (Medium-Sized Telescopes) for the $150$\,GeV 
to $5$\,TeV  range and finally SSTs (Small-Sized Telescope) for energies from $5$\,TeV to $300$\,TeV 
and more.
The northern hemisphere array will instead be more limited in size, and will focus on energies from 
20\,GeV to 20\,TeV. 
In a first stage of CTAO construction, the so-called `Alpha' configuration will be built -- which is the configuration
we will focus on in this work. It will consists of 4 LSTs and 9 MSTs in the
Northern Array, and 14 MSTs and 37 SSTs in the southern array. 
CTAO will reach 
better sensitivities than current generation instruments by a factor of $5\!-\!10$~\cite{Hassan:2017paq}, 
reaching an energy resolution of order $\Delta E/E \sim \mathcal{O}(0.1)$ for TeV energies 
(\FIG\ref{fig:cta}, left panel). This makes CTAO an excellent instrument to search for exotic localized spectral 
features, e.g.~from DM, over several orders of magnitude in gamma-ray energies.

Satellite experiments -- like {\it Fermi} LAT~\cite{Atwood:2009ez},  AGILE~\cite{AGILE:2008nyq} or DAMPE~\cite{DAMPE:2021hsz} -- offer a complementary 
strategy to detect gamma rays, based on the direct detection of electron-positron pairs produced by the incoming gamma ray.  
As a result, satellite-borne gamma-ray telescopes typically have larger FoV and can cover lower energies
than ground-based observatories, but have a smaller effective area. 
More importantly for the
present study, IACTs have an excellent energy resolution at TeV energies, i.e.~higher than the reach of satellite-borne
experiments. For comparison, we 
also indicate in \FIG\ref{fig:cta} the energy resolution of {\it Fermi} LAT and DAMPE.

Key Science Projects discussed for CTA~\cite{CTAConsortium:2017dvg} include a range of surveys covering
extended portions of the sky that will surpass in ambition previous IACT attempts. 
Since the GC region is especially interesting for DM-related searches we will here focus on the 
GC survey (see Section \ref{section:galactic_centre} 
for details), along with traditional pointing 
observations of additional targets relevant for DM detection (dwarf spheroidal galaxies, dSphs, see Section \ref{section:dsph}). 
We study these observational strategies by benefitting from the latest instrument response functions (IRFs) for the Alpha configuration
provided by the CTA consortium, derived from detailed MC simulations.\footnote{%
\label{foot_IRF}
Concretely, we make use of  Prod5 v.~12.06 (Alpha configuration), based on an average of 50\,hr
observation time at $20^\circ$ zenith angle. 
All IRFs files are publicly available at the CTA website~\cite{CTAweb_IRF}.
} 
An important ingredient besides the energy resolution, in particular, is the effective area $A_{\rm eff}$ of CTAO.
 We show this in the right panel of \FIG\ref{fig:cta}, along with a smoothed version that we will adopt in our analysis 
 in order to avoid numerical binning artefacts. As visible in this figure, $A_{\rm eff}$ rises continuously with energy
 up to at least about 10\,TeV; the visible (beginning of a) sharper drop towards low energies at the southern array (black line) is due to 
 the absence  of LSTs at this site.

\section{Spectral signatures from dark matter}
\label{section:spectral_signatures}

For the energies of interest to this analysis, gamma rays propagate without significant interactions through
the Galaxy. This makes it straightforward to calculate the signal expected from DM based on 
its density distribution $\rho_\chi(\mathbf{r})$ and the {\it in situ} energy injection rate 
(see e.g.~Ref.~\cite{Bringmann:2012ez}). For the case of annihilating DM particles $\chi$, e.g., the
differential gamma-ray flux per unit energy and solid angle is given by 
\be
\label{DMflux}
  \frac{d\Phi_{\gamma}}{d\Omega\, dE_\gamma} (E_\gamma,\psi) = 
  \frac{1}{4\pi} \int_\mathrm{l.o.s}
  d\ell(\psi) \rho_\chi^2(\mathbf{r}) 
  \left({\frac{\langle\sigma v\rangle_\mathrm{ann}}{2S_\chi m_{\chi}^2} 
  \frac{dN_\gamma}{dE_\gamma}}\right) \,,
\ee
where the integration is performed along the line of sight (l.o.s.) in the observing 
direction ($\psi$). 
The term inside the parenthesis depends on model-specific particle physics parameters.
Here $\langle\sigma v\rangle_\mathrm{ann}$ is the average velocity-weighted
annihilation cross section, $m_\chi$ is the DM mass, and the symmetry factor 
$S_\chi$ indicates whether the DM particle is its own antiparticle ($S_\chi=1$)
or not  ($S_\chi=2$).
The main focus of our analysis will be the photon {\it spectrum} produced by DM, $dN_\gamma/dE_\gamma$, which
in this case corresponds to the (differential) number of photons per annihilation.

It is typically assumed that the factor in parenthesis can be taken outside the line-of-sight and 
angular integrals.\footnote{%
More concretely, the flux given in Eq.~(\ref{DMflux}) fully factorizes into a part depending  on
particle physics (as described by the quantities in parenthesis) and a part depending on astrophysics
(encoded in what will be introduced as the $J$-factor) 
only if both $(\sigma v)_\mathrm{ann}$ and $dN_\gamma/dE_\gamma$
are sufficiently independent of the DM velocity. This is the case in many typical WIMP models -- though
notable exceptions exist not the least for the type of pronounced spectral features that this article
focusses on~\cite{Hisano:2003ec,Arina:2014fna}. A full analysis of these necessarily 
model-dependent effects, however, is beyond the scope of the present work.
}
Spatial and spectral information of the signal are then uncorrelated, and the flux from a given angular 
region $\Delta\Omega$ becomes directly proportional to the `$J$-factor'
\be
\label{Jfactor}
J_{\Delta\Omega}\equiv\int_{\Delta\Omega} d\Omega\int\!\!d\ell\,\rho_\chi^2\,.
\ee
The $J$ factor thus depends on the choice of target, and its DM distribution, 
which is discussed in Section \ref{sec:targets}.
While we will mostly refer to the case of annihilating DM, let us briefly mention
that it is straightforward to generalize our results to the case of decaying DM~\cite{Ibarra:2013cra}:
in the above expression for the DM-induced flux, one then simply has to replace
$J_{\Delta\Omega} \langle\sigma v\rangle_\mathrm{ann}/(2S_\chi m_\chi)$ by $D_{\Delta\Omega} \Gamma_\chi$, 
where $\Gamma_\chi$ is the total DM decay rate and the `$D$-factor' is defined in analogy to the $J$-factor as 
$D_{\Delta\Omega}\equiv\int_{\Delta\Omega} d\Omega\int\!\!d\ell\,\rho_\chi$.

\medskip

Let us now turn to a discussion of the signal shapes expected from DM annihilation. In generic WIMP
models, the dominant source of prompt gamma-ray emission often stems from the tree-level annihilation
to pairs of standard model particles. These particle then decay and fragment, producing a large  
multiplicity of photons in each of the annihilation channels $f$, mostly through the decay of 
neutral pions and final state radiation (FSR). The total yield
$dN_\gamma/dE_\gamma=\sum_f  B_f {dN_\gamma^{f}}/{dE_\gamma}$, 
with $B_f$ the he branching ratio into final state $f$, 
then describes a photon spectrum 
with a rather universal form that lacks distinct features apart from a rather soft cutoff at the kinematical 
limit $E_\gamma=m_\chi$~\cite{Bringmann:2012ez}. Against typical instrumental and astrophysical backgrounds, 
these DM candidates would produce a broadly distributed excess (in energy), which means that
the identification of a subdominant signal would require an exquisite understanding of the background spectra.
In fact, a detailed template-based study of  the CTA sensitivity to a DM signal from the GC 
region~\cite{CTA:2020qlo} recently confirmed that
the {\it spatial} distribution of gamma rays becomes a much more powerful tool to distinguish 
signal and backgrounds in such cases. 

The goal of this work is to complement that analysis by assessing the prospects for CTA to detect 
`smoking gun' DM signals,
i.e~signal shapes that would clearly stick out against the typical backgrounds and hence, if detected, leave 
little doubt about their origin.\footnote{%
A possible exception to this statement may, perhaps, be cold pulsar winds that have been argued to produce
relatively narrow spectral features in certain, non-generic scenarios~\cite{Aharonian:2012cs}. Such pulsar winds 
would in any case 
be (quasi) point-like sources, and hence could easily be distinguished from annihilating DM  once the photon count is sufficiently 
high to infer spatial information about the signal. We will here not discus this possibility further.
} 
For concreteness, we will consider three classes of such narrow spectral features that are 
exemplary for the range of possibilities from a model-building perspective:
\begin{enumerate}
\item {\bf Line signals}. Monochromatic, or `line', spectra of the form (in units of photons per energy)
\be
\label{eq:line}
\frac{dN_\gamma}{dE_\gamma}=N_\gamma^0\,\delta(E_\gamma-E_0)
\ee
have early been pointed out as a DM signature that would be straight-forward to distinguish
from astrophysical backgrounds~\cite{Srednicki:1985sf,Bergstrom:1988fp,Bergstrom:1997fj}.
Concretely, such a contribution to the total spectrum is expected whenever DM annihilates to a pair of final states containing 
at least one photon, $\chi\bar\chi\to X\gamma$, where $X$ can either be a neutral boson of the
standard model ($X=\gamma,Z,H$) or  a new neutral state (like a $Z'$, or a `dark' photon).\footnote{%
Strictly speaking, the expected observable spectrum from such annihilations is a very narrow {\it Gaussian}
centered around $E_0$, with a width set by Doppler shift and hence the velocity dispersion of Galactic 
DM, $v_0/c\sim10^{-3}$. 
Radiative corrections will further somewhat distort the
spectrum~\cite{Ciafaloni:2010ti,Ovanesyan:2014fwa,Baumgart:2017nsr,Baumgart:2018yed,Beneke:2018ssm,
Beneke:2019vhz,Beneke:2019gtg, Bauer:2020jay,Beneke:2022eci}, which however is not 
completely model-independent.
For IACTs, usually, the signal shape is still to an excellent approximation given by Eq.~(\ref{eq:line}).
} 
The line energy is then given by $E_0=m_\chi(1-m^2_{X}/4m^2_\chi)$, and the total number of 
photons per annihilations $N_\gamma^0=1$ (unless
$X=\gamma$, in which case $N_\gamma^0=2$). It is worth noting that these processes are necessarily
loop-suppressed, parametrically by a factor of $\mathcal{O}(\alpha_{\rm em}^2)$, because DM cannot 
directly couple to photons, thus generically leading to correspondingly
low gamma-ray fluxes. There are, however, examples of well-motivated DM candidates where
particularly strong line signals are expected
in the energy range accessible to CTAO~\cite{Hisano:2003ec,Guo:2009aj,Mambrini:2009ad,Dudas:2012pb,Arina:2014fna}.

\item {\bf Virtual internal bremsstrahlung (VIB)}.
A single photon in the final state can also appear along with two charged particles (instead of one neutral
particle, as in the previous example). Such a process
is referred to as internal bremsstrahlung, and parametrically only suppressed by a factor of  
$\mathcal{O}(\alpha_{\rm em})$
with respect to the (tree-level) annihilation to the charged-particle pair. Just as in the case of line signals, 
furthermore, there are indeed cases in which internal bremsstrahlung constitutes the {\it dominant} contribution to the 
annihilation rate -- or at least to the photon yield at energies close to the kinematical endpoint at 
$E_\gamma=m_\chi$, giving rise to pronounced spectral 
signatures~\cite{Bergstrom:2004cy,Birkedal:2005ep,Bergstrom:2005ss,Bringmann:2007nk,Barger:2011jg,
Garcia-Cely:2013zga,Toma:2013bka,Giacchino:2013bta}.
A notable example that we will explicitly consider here is the case of neutralino DM, or any other Majorana DM 
candidate, annihilating to standard model fermions.
In this case `virtual' internal bremsstrahlung (VIB)\footnote{%
Here, `virtual' refers to the dominant contribution resulting from photons radiated off virtual sfermions.
Technically, VIB is the final state radiation (FSR) subtracted part of internal bremsstrahlung 
(see Ref.~\cite{Bringmann:2007nk} for a detailed discussion). 
}
 dominates, which in the limit of large DM masses and 
degenerate sfermions takes the form~\cite{Bergstrom:1989jr,Bringmann:2007nk}
\be
\label{eq:VIB}
\frac{dN_\gamma}{dE_\gamma}=A_\gamma^{\rm VIB}\,
\frac{x(x^3-4x^2+6x-4)-4(x-1)^2\log(1-x)}{(x-2)^3}\,,
\ee
with $x=E_\gamma/m_\chi$ and $A_\gamma^{\rm VIB}=6/(21-2\pi^2)\simeq4.76$. We note that a somewhat similar 
spectral shape
also arises for $W^+W^-\gamma$ final states~\cite{Bergstrom:2005ss}; this is, e.g., highly relevant for Wino DM, 
for which there has recently been a significant theoretical effort to model the exact shape of the kinematic endpoint
features of 
$dN_\gamma/dE_\gamma$~\cite{Ovanesyan:2016vkk,Beneke:2018ssm,Baumgart:2018yed,Beneke:2019vhz,Beneke:2019gtg}, 
as well as a dedicated analysis of the prospects to
detect such a feature with an instrument like CTAO~\cite{Rinchiuso:2020skh}.

\item {\bf Box signals}.
A third type of pronounced spectral signal, {\it not} necessarily suppressed with respect to the leading
annihilation rate, arises if the DM particles annihilate into a pair of new, long-lived neutral states $\phi$.
If these in turn decay dominantly into photons, $\phi\to\gamma\gamma$, the result is a `box-shaped' signal
of the form~\cite{Ibarra:2012dw}
\be
\label{eq:box}
\frac{dN_\gamma}{dE_\gamma}= \frac{4}{\Delta E}\times \theta\left(E_\gamma-\frac{m_\chi-\Delta E}2\right)\,\theta\left(\frac{m_\chi+\Delta E}2-E_\gamma\right).
\ee 
Here $\theta(x)$ is the Heaviside step function, and the width of the box
constitutes a free parameter that can be expressed in terms of the mass of the intermediate particle $\phi$
as $\Delta E=\sqrt{m_\chi^2-m_\phi^2}$. The above expression assumes DM annihilation to two identical 
states, $\chi\bar\chi\to \phi\phi$, which we will consider here. We note however that it is straight-forward to 
generalize the above expression to two different intermediate states, $\chi\bar\chi\to \phi_1\phi_2$, 
resulting in a linear superposition of 
box-spectra of the above type, with different central values and widths~\cite{Ibarra:2012dw,Workgroup:2017lvb}.
 
\end{enumerate}

\begin{figure}[t]
\centering
\centerline{
\includegraphics[width=0.49\textwidth]{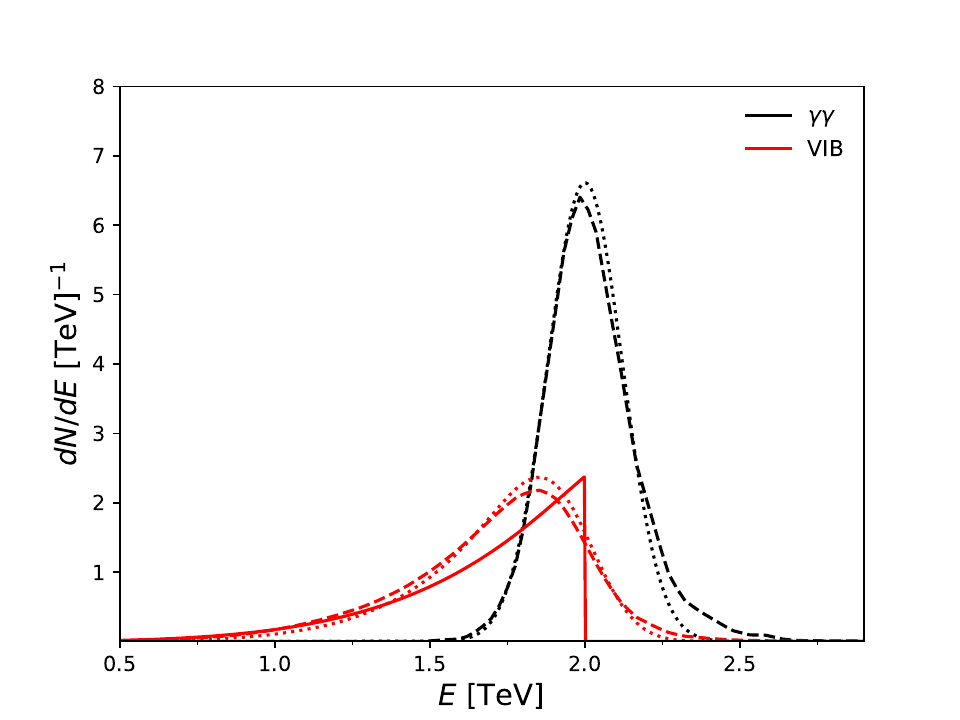}
\includegraphics[width=0.49\textwidth]{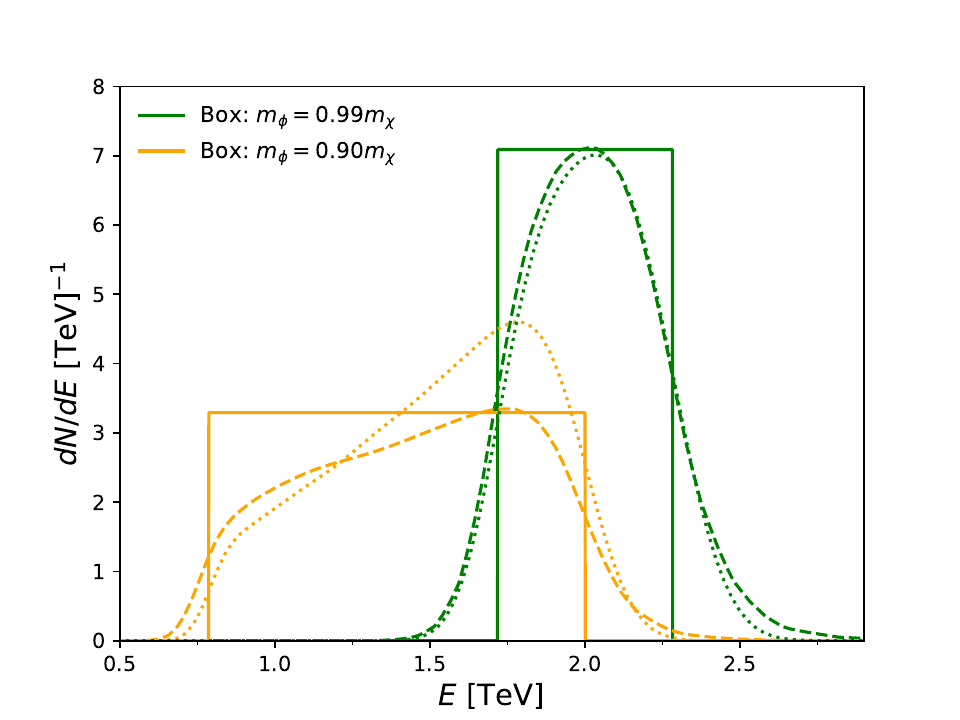}
}
\caption{The figures show characteristic DM signal spectra, $dN_\gamma/dE_\gamma$, of the type discussed
in Section~\ref{section:spectral_signatures}, featuring sharp endpoints at or around $E_0=2$\,TeV. 
Solid lines correspond to the physical, injected spectra, while
dashed lines show the observed signal spectra as modeled by including
the IRF of CTAO (see section \ref{sec:sliding}).
For comparison, dotted lines show the result of the physical spectrum convoluted
with a Gaussian of width equaling the energy resolution displayed in \FIG\ref{fig:cta}.
\textit{Left panel:}
Monochromatic line (black), Eq.~(\ref{eq:line}), and VIB (red), Eq.~(\ref{eq:VIB}).
The solid, monochromatic line at $E_\gamma=m_\chi=2$\,TeV is not shown explicitly.
\textit{Right panel:}
The signal spectrum for two different box scenarios, Eq.~(\ref{eq:box});
green (orange) curves show the case of the box width $\Delta E$ being
smaller (larger) than the energy window.
The DM mass for the narrow (wide) box shape in these examples is $m_\chi=4 \ (2.87$) TeV.
We note that the different areas under these curves directly reflect the different number of photons
per annihilation, namely~$N_\gamma=2$ for the line spectrum, $N_\gamma=1$
for VIB and  $N_\gamma=4$ for box-shaped spectra.
}
\label{fig:signalshapes}
\end{figure}

In \FIG\ref{fig:signalshapes} we provide concrete examples to illustrate these spectral shapes. 
As apparent from the above list, furthermore, the exact shape of the spectra we consider here strongly depends
on the details of the underlying particle model (in contrast to the spectra considered in 
Ref.~\cite{CTA:2020qlo}).
This implies that the detection of such a signal would not only provide smoking gun evidence for particle
DM, but immediately allow to reach far-reaching conclusions about the more general theory these DM particles
are embedded in~\cite{Bringmann:2012ez}. 

Eventually we will be interested in deriving CTA sensitivities in terms of projected upper limits
on the (velocity-weighted) DM annihilation cross section $\langle\sigma v\rangle_\mathrm{ann}$,
for a given spectral shape $dN_\gamma/dE_\gamma$.
Let us therefore close this section by briefly reflecting about the expected size of 
$\langle\sigma v\rangle_\mathrm{ann}$ for thermally produced DM. In particular, the total 
annihilation rate required to produce the observed DM relic abundance in the early universe is 
often referred to as the `thermal' annihilation rate, and numerically given by about  
$\langle \sigma v \rangle_{\rm therm}~\sim2.1\times10^{-26}~\rm{cm^3s^{-1}}$ for DM particles with 
$m_\chi\sim1$\,TeV~\cite{Gondolo:1990dk}. 
For {\it line signals}, it {\it is} in principle possible that $\chi\bar\chi\to X\gamma$ is the dominant annihilation 
channel -- e.g.~because DM only couples to heavier, charged states~\cite{Cline:2012nw} -- in which case the 
correct `benchmark' cross section is indeed $\langle \sigma v \rangle_{\rm therm}$. 
More generically, however, this channel will be suppressed by a
loop-factor of $(\alpha_{\rm em}/4\pi)^2$ with respect to the tree-level annihilations that are responsible for
setting the relic density, resulting in $\langle\sigma v\rangle_\mathrm{ann}\sim10^{-31}~\rm{cm^3s^{-1}}$
and lower; however, near-resonant annihilation can lead to line signals significantly
larger than this estimate~\cite{Guo:2009aj,Dudas:2012pb,Arina:2014fna}
and non-perturbative effects can even result in present-day annihilation cross sections 
{\it higher} than the `thermal' value responsible for setting the relic density in the early universe 
(prominent examples being Wino and Higgsino DM~\cite{Hisano:2003ec}).
{\it VIB signals}, on the other hand, are inevitably accompanied by tree-level processes (without the
additional photon in the final state) that set the relic density and hence generically suppressed only by 
a factor of $\sim{\alpha_{\rm em}}/\pi$ with respect to the `thermal' rate. For {\it box signals}, finally,
the relic density is often set by the same process that gives rise to the signal, namely 
$\chi\bar\chi\to\phi\phi$; in fact, the value of the relevant `thermal' cross section can easily be a factor
of a few higher because, for such an annihilation scenario, freeze-out would typically happen in 
a secluded dark sector (see Ref.~\cite{Bringmann:2020mgx} for how to determine the relic density 
in such cases).

\section{Target Regions}
\label{sec:targets}
In Section~\ref{section:spectral_signatures} we discussed spectral signatures of annihilating DM, 
related to the particle physics aspects of DM.
In this section we turn our attention to the expected spatial distribution of cold DM, largely independent of its
particle properties, and how this motivates our choice of target regions.
Generally speaking, as evident from Eq.~(\ref{DMflux}), close-by regions with a high DM density are good 
targets for observing DM annihilation 
signals.
The GC region has the largest $J$-factor, Eq.~(\ref{Jfactor}), among all possible targets,
making it arguably the best DM target from the point of view of the overall expected signal strength 
(even when taking into account that the uncertainty on the $J$-factor, $\Delta J$, is considerable). 
However, the GC hosts a rich environment of astrophysical gamma-ray emitters, resulting  in complex 
backgrounds for DM searches. 

Complementary targets to the GC are dwarf spheroidal galaxies, which have practically no astrophysical 
background in gamma rays~\cite{Winter:2016wmy}, but are farther away  and less massive, resulting in 
lower $J$-factors. Many dSphs are very faint in terms of visible  gravitational field tracers (stars and gas), 
thus leading to substantial uncertainties in the DM density distribution, and hence $J$, also for these targets.  

Below we discuss in more detail the GC target in Section~\ref{section:galactic_centre}, including astrophysical 
backgrounds, 
as well as dSphs in Section~\ref{section:dsph}.

\subsection{Galactic centre}
\label{section:galactic_centre}

\paragraph{Observational program.}
There is a large number of independent science drivers that motivates
an observational strategy for CTAO specifically targeting the GC region~\cite{CTAConsortium:2017dvg}. 
We follow the recommendation for the GC survey from that work and consider 
$9$ pointings centred at $l:\{\pm1^\circ,0^\circ\}$, $b: \{\pm1^\circ,0^\circ\}$, each with an observation time 
of $58.3$ hours. 
Effectively, this gives a total of $500$ hours of observation time of the GC with a roughly 
homogeneous exposure over the inner $4^\circ$ (see also Ref.~\cite{CTA:2020qlo} 
for further details, including full exposure maps).

We will base our analysis on this GC survey setup, but will optimize our region of interest (RoI) 
to comprise a region that is generally significantly smaller than the above mentioned $4^\circ$ (by 
maximizing the expected signal-to-noise ratio, see Section \ref{sec:ROI} for further details). 
Based on this observational (and analysis) strategy we simulate all signals and backgrounds 
using \ctools~\cite{Knodlseder:2016nnv}, a public software package developed for the scientific 
analysis of gamma-ray data.

\paragraph{Dark Matter Distribution.}
Numerical $N$-body simulations of collision-less cold DM clustering, neglecting the effect of baryons, 
have over the past decades consistently found that DM halos develop a universal density profile on all clustering scales 
\cite{Zavala:2019gpq}. While there are  differences  in the exact parametrization of such a profile, its salient feature is 
that it is `cuspy', i.e.~it follows a power law $\rho_\chi\propto r^{-n}$ with $n\gtrsim1$, at small (kpc) galactocentric distances $r$. 
Due to the limited resolution of  $N$-body simulations, as well as the fact that baryonic feedback is expected to 
become more relevant close to the halo centres, it is however unclear whether the extrapolation of such 
power laws remains valid to sub-kpc scales.  

From the purely observational side, stellar data and gas tracers of the gravitational potential are typically used to 
constrain the underlying DM density profile on Galactic scales (with gravitational lensing providing a competitive 
alternative on larger scales). While this method works well for large galactocentric distances, where DM dominates, 
the gravitational potential in the inner $\sim$kpc of the GC is dominated by baryons. 
DM density measurements  therefore 
remain inconclusive at small scales, being consistent with both cuspy and more shallow inner density profiles. The latter are,
in fact, also 
found in $N$-body simulations including baryons, indicating that cores of constant DM density can develop
due to baryonic feedback on the gravitational potential~\cite{DiCintio:2013qxa}. For 
example, a high concentration of baryons typically leads to a more vibrant star formation rate and hence
an enhanced supernova (SN) feedback due to the injection of significant amounts of energy on short timescales, 
effectively `heating' DM and dispersing the cusp. DM halos with active 
super-massive black holes can show a similar effect. These processes  are 
however not yet understood  in sufficient  detail. In fact, the presence of baryons could also have the opposite
effect, since the cooling of baryonic gas in the GC region may well lead to an adiabatic contraction and hence 
a {\it steepening} of the DM density profile with respect to the one found in DM-only  
simulations~\cite{Gnedin:2004cx}. 

For these reasons, we follow Ref.~\cite{CTA:2020qlo} (see also there for a more detailed
discussion) and adopt two bracketing DM density profiles in the main part
of our analysis: 
\textit{Einasto}~\cite{Einasto:1965czb}  as a representative of cuspy profiles and  \textit{cored Einasto}~\cite{DiCintio:2013qxa} to estimate a possible conservative lower bound for the expected limits on (and discovery potential of) a
DM signal:

\begin{equation}
\label{eq:Einasto}
\rho_{\mbox{\tiny{\rm{Einasto}}}} (r) \hspace{0.2cm} = \hspace{0.2cm} \rho_s \hspace{1mm} \mbox{e}^{-\bigl(\frac{2}{\alpha}\bigr)\bigl[\bigl(\frac{r}{r_s}\bigr)^\alpha -1\bigr]}
\end{equation}

\begin{equation}
\label{eq:cEinasto}
\rho_{\mbox{\tiny{\rm cored\ Einasto}}}\!\left(r\right)=
\begin{cases}
\rho_{\mbox{\tiny{\rm Einasto}}}\!\left(r_{c}\right) & \textrm{if}\;r\leq r_{c}\\
\rho_{\mbox{\tiny{\rm Einasto}}}\!\left(r\vphantom{r_{c}}\right) & \textrm{if}\;r>r_{c}
\end{cases}\,.
\end{equation}

\noindent Here $\rho_s$ is the characteristic density, normalized to an average DM density of
$\rho(r_{\rm{\odot}})= 0.4$ GeV$/$cm$^3$ at the same galactocentric distance as the sun 
($r_\odot = 8.5\ \mathrm{kpc}$),  $r_s = 20$\,kpc is the characteristic radius and $\alpha = 0.17$
is the Einasto shape parameter. 
The core radius is chosen as $r_c=1~\rm{kpc}$, which for this analysis essentially implies $\rho=const$ 
for the cored Einasto profile as we only focus on the inner few degrees of the GC. 
\TAB\ref{table:jfrac} lists the resulting \textit{J}-factor values for the inner $2^\circ$ of the GC, as 
computed with 
\darksusy~\cite{Bringmann:2018lay}  and cross-checked with \clumpy~\cite{Hutten:2018aix}. 
Here, we include for completeness also the often quoted Navarro-Frenk-White profile~\cite{Navarro:1995iw}, which is 
similarly cuspy to the Einasto profile, for the same choice of parameters as adopted in Ref.~\cite{CTA:2020qlo}.
For a more detailed discussion of how the choice of DM profile affects our results, we refer to Section 
\ref{sec:profile_discussion}.

\begin{table}[t]
\centering
\begin{tabular}{ |l ||c|c|c|c|}
 \hline
   && \multicolumn{3}{|c|}{}\\[-2.5ex] 
 &  \multirow{3}{*}{Angular Size [sr]} &\multicolumn{2}{|c|}{\textit{J}-factor [$\rm{GeV^{2} cm^{-5} }$]} &\\
   &&&& \\[-2.5ex] 
& & Einasto & cored Einasto & NFW  \\
 \hline\hline
   &&&&\\[-2.5ex] 
 $J_{0.5^\circ}$   
 &  $2.39\times 10^{-4}$ 
 & $ 3.48\times 10^{21}$
 & $1.93\times 10^{20}$ 
 & $2.65\times 10^{21}$\\
  $J_{1^\circ}$
  & $7.18\times 10^{-4}$
  & $ 5.14\times 10^{21}$
  & $5.55\times 10^{20}$
  &$2.69\times 10^{21}$ \\
  $J_{1.5^\circ}$
  & $1.20\times 10^{-3}$
  &  $ 5.53\times 10^{21}$
  & $9.38\times 10^{20}$
  & $2.67\times 10^{21}$\\
  $J_{2^\circ}$
  & $1.67\times 10^{-3}$
  & $ 5.41\times 10^{21}$
  &$1.29\times 10^{21}$
  &$2.56\times 10^{21}$\\
   $J_{2.5^\circ}$
   & $2.15\times 10^{-3}$ 
   &   $ 5.27\times 10^{21}$
   & $1.64\times 10^{21}$ 
   & $2.49\times 10^{21}$\\
   $J_{3^\circ}$
   & $2.63\times 10^{-3}$
   &$ 5.10\times 10^{21}$
   & $1.99\times 10^{21}$
   & $2.44\times 10^{21}$\\
   \hline
   &&&&\\[-2.5ex] 
    $\sum J_{\leq2^\circ}$ 
    & $3.83\times 10^{-3}$
    &$ 1.96\times 10^{22}$
    & $2.97\times 10^{21}$ 
    & $1.06\times 10^{22}$\\
 \hline
\end{tabular}
\caption{\textit{J}-factors $[\rm{GeV}^{2} \rm{cm}^{-5} ]$ for the benchmark DM profiles adopted in our GC analysis, 
as computed with {\sf DarkSUSY}. $J_{\theta}$ indicates the \textit{J}-factor for a 
concentric ring with outer radius $\theta$ and inner radius $\theta -0.5^\circ$, with a total angular size
as indicated in the 2nd column. The last row states the total $J$-factor from the inner 2 degrees. 
}
 \label{table:jfrac}
\end{table}

\paragraph{Background Components.}
The fact that CTAO effectively uses the atmosphere as a 
calorimeter implies an inevitable source of background from misidentified CRs, independent of the target that
is observed (in this sense, this could be called an `instrumental' background). 
CRs hitting the upper atmosphere consist mainly of protons and electrons, with fluxes that are 
(at $\sim100$ GeV) a factor of $10^4$ and $10^2$ times 
higher, respectively,  than the diffuse gamma-ray flux \cite{Abdo:2009zk,Fegan:1997db}. Though energy-dependent,
the proton rejection rate is typically better than $10^{-2}$ due to the different shape of proton-induced showers
compared to those induced by gamma rays. Electrons, on the other hand, produce almost identical shower 
shapes and are thus practically indistinguishable from gamma rays. 
The misidentified CR background has to be estimated based on detailed MC simulations of
the shower evolution and the response of the instrument. As detailed in \SEC\ref{sec:ROI}, we will use
\ctools\ for the generation of mock data, automatically including this component.

In terms of astrophysical emission, the GC region is an active environment, rich with non-thermal emitters such as radio 
filaments~\cite{YusefZadeh:2003qx}, 
young massive stellar clusters~\cite{Aharonian:2018oau},
a number of pulsars, SNR shells etc., in addition to the super massive black hole, 
Sagittarius A*~\cite{Genzel:2010zy}. Furthermore, 
the whole region is  embedded in the bright emission stemming from the Galactic CR population, 
producing gamma rays by interacting with magnetic fields,  interstellar light and gas.  This so-called Interstellar Emission (IE) 
extends to high latitudes at GeV energies~\cite{Ackermann:2012pya}, while  at TeV energies it  was so far only detected in the limited  
region of the GC Ridge~\cite{Abdalla:2017xja}. 
In order to model this component we take advantage of a recent study  \cite{Luque:2022buq} based on available GeV to PeV 
gamma-ray data (from {\it Fermi} LAT, Tibet AS$\gamma$, LHAASO and ARGO-YBJ), together with  local charged cosmic ray 
measurements (from AMS-02, DAMPE, CALET, ATIC-2, CREAM-III and NUCLEON).
Modelling the IE over such a wide energy range  is achieved via two complementary approaches to describe 
the diffusion of  CRs: in the so-called `Base' models the diffusion coefficient is assumed to be constant throughout the Galaxy,
while in the `Gamma' models it is allowed to vary radially. Both sets of models are 
further divided in MIN and MAX setups in order to reflect uncertainties of the CR proton and helium source spectra, 
see Ref.~\cite{Luque:2022buq} for more details. We choose Base MAX as our 
benchmark model, noting that current Gamma models were not tested in the vicinity  of the GC, 
where by  construction they should become increasingly brighter (and, likely, overshooting what can realistically be expected
in this region). On the other hand, the Base models might somewhat underestimate the emission in the innermost region of the  
GC Ridge~\cite{HESS:2016pst}. We explore these uncertainties in \SEC\ref{section:conclusion}, but note that due to the 
methodology of the line search, 
background modelling is expected to have a rather limited impact  on our results (as opposed to the case of continuum DM 
signals, cf.~Ref.~\cite{CTA:2020qlo}).

In addition to the IE, our RoI also includes localised sources such as
the point source associated with Sgr\,A*, \textit{HESS J1745-290}\cite{Collaboration:2009tm},  \textit{G0.9+01} and the
recently discovered, still unidentified faint source \textit{HESS J1741-302}\cite{hess302}. 
We take into account these sources in our simulations, as well as 
the two extended sources \textit{HESS J1741-303} and \textit{HESS J1741-308}.
Although highly uncertain at small latitudes, finally, we further include a template of the \textit{Fermi bubbles} (FBs)
based on a recent analysis from Ref.~\cite{Herold:2019pei}. \footnote{%
In view of recent limits from H.E.S.S.~\cite{Moulin:2021mug}, this template likely overestimates the actual flux
at multi-TeV energies. However, at these energies the FB contribution is negligible compared to other background components; 
our template thus leads to too conservative limits on an exotic signal -- but only very slightly so. 
}

When implementing the contribution from both point sources 
and FBs, we thus follow again the same modelling treatment as in Ref.~\cite{CTA:2020qlo}.
For a more detailed discussion of all background components we therefore also refer to that reference.

\begin{figure}[t]
\centering
	\includegraphics[width=0.8\linewidth]{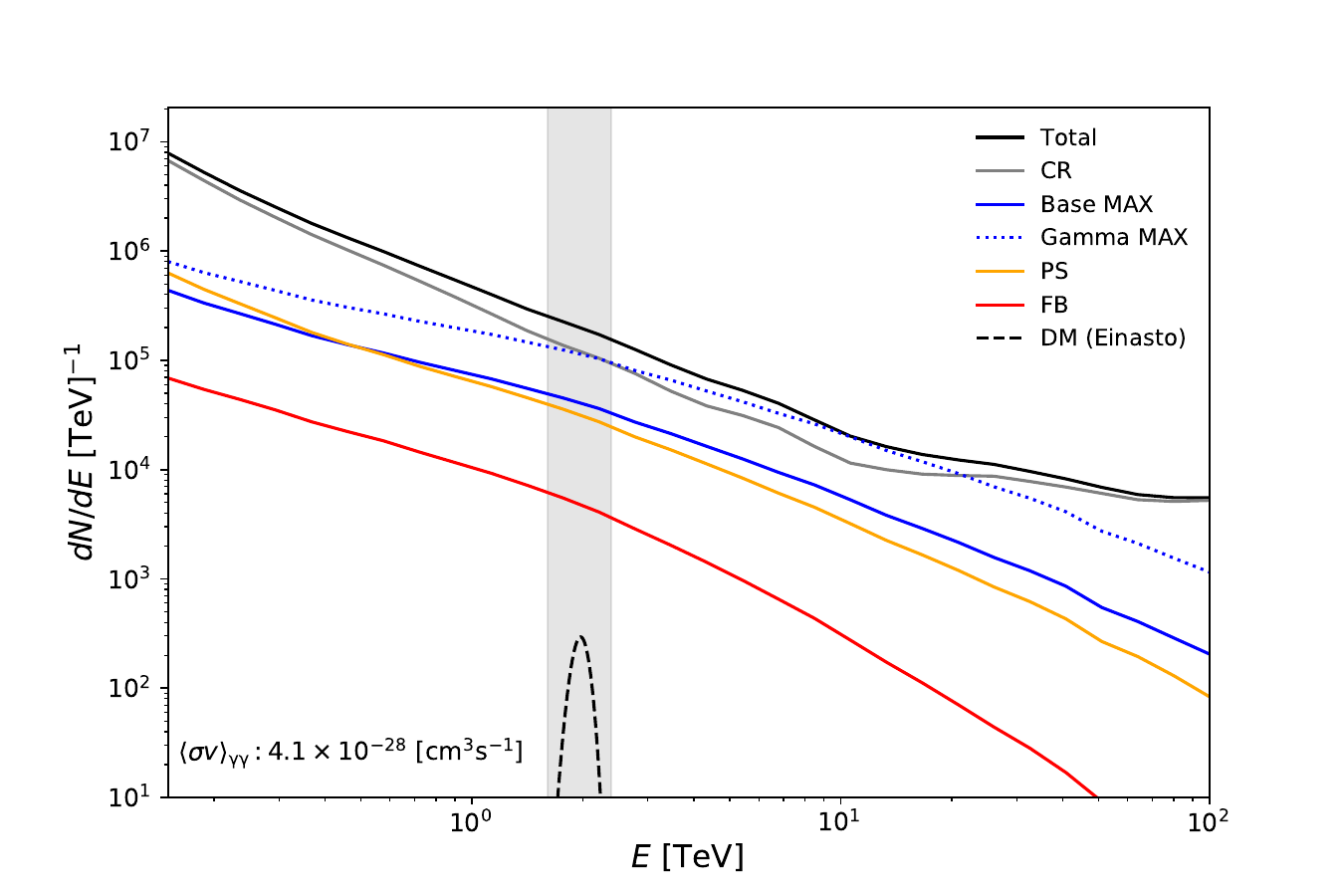}
\caption{%
The total expected photon count (black solid line) from the individual background components, for the inner $2^\circ$ of the 
considered GC observation of 500\,hr, that are included in the background 
simulations: Fermi Bubbles (\textit{FB}, red), combined point sources (\textit{PS}, orange), misidentified 
cosmic rays 
(\textit{CR}, gray) and the diffuse gamma-ray emission (\textit{IE}, blue).
The solid line shows the benchmark model, Base MAX, included in the total count, 
while the dotted line indicates the alternative Gamma MAX model;
see text for further details. 
For comparison, we also show a DM line signal (black dashed), assuming a DM mass $m_\chi = 2\ \mathrm{TeV}$ and 
an annihilation strength that would result in a $5\sigma$ discovery; the shaded region corresponds to the 
size of the energy window used in the analysis in that case. 
Simulations are performed with \ctools.
}
\label{fig:background_wider}
\end{figure}

We display the expected count spectrum from the inner $2^\circ$ of the GC region, 
broken down into individual components, in \FIG\ref{fig:background_wider}. While the expected counts
are clearly dominated by misidentified cosmic rays, the figure also illustrates that the astrophysical 
components discussed above can by no means be neglected for the analysis.
For comparison, we also include a DM line signal (black dashed line), for a DM particle with mass $m_\chi=2$\,TeV and 
annihilation cross section $\ov = 8.10\times10^{-28}\ \mathrm{cm^3s^{-1}}$ which would lead to a $5\sigma$ discovery (see \SEC\ref{sec:statistics}). 
The shaded region corresponds to the size of the `sliding' energy window used to analyse such signals. 
We will discuss this analysis technique in detail in \SEC\ref{section:analysis}, but note already here
that the total expected background count spectrum can be well described by a simple power law within the 
shaded region. As we will demonstrate, this observation makes it possible to robustly distinguish a sharply peaked 
DM signal, even if it is highly subdominant.

\subsection{Dwarf Spheroidal Galaxies}
\label{section:dsph}

The dSph satellites surrounding the Milky Way are old and DM-dominated systems. Due to their age and 
the lack of gas content, they are not expected to source any significant non-thermal emission. Consequently  they are 
considered  to be essentially background-free targets for DM signal searches~\cite{Winter:2016wmy}, such that the detection of 
a gamma-ray signal might in itself 
constitute a smoking gun for the presence of particle DM (see e.g.~Refs.~\cite{HESS:2018kom, H.E.S.S.:2020jez}).
It is not only the substantial DM content (e.g.~\cite{Simon:2007dq})
and their relative proximity that makes dSphs promising targets, but also the fact that they are distributed over a significant 
range of Galactic latitudes, including regions with low diffuse foreground emission.
As of today, no gamma-ray signal has been conclusively associated with dSphs, either individually or as a population, 
and the corresponding upper limits have been used to set competitive constraints on the DM annihilation strength
(summarised by e.g.~Ref.~\cite{Fermi-LAT:2016afa}). 

The statement that no dSph galaxy has been found to significantly emit gamma rays in the GeV or TeV band has 
recently been challenged by Crocker\;{\it et al.}~\cite{Crocker:2022aml}, who report evidence of extended 
gamma-ray emission from the Sagittarius dSph (Sgr II). This emission appears as a well-known substructure inside the 
rather uniform FBs, which also has been coined the Fermi Bubbles' cocoon region~\cite{Fermi-LAT:2014ryh}.
A possible explanation for such a signal from Sgr II would be a population of around 700 
millisecond pulsars (MSPs), based on a strong correlation between the distribution of old stars in the system and the measured 
gamma rays. Indeed, the expected number of  MSPs in dSphs only depends on the initial gas content (unlike in the case of
the much higher stellar densities in globular clusters, where not only direct formation of 
binaries~\cite{Gilfanov:2003th,Mirabal:2013rba,Gautam:2021wqn} but also formation 
in later stages via stellar encounters~\cite{1998MNRAS.301...15D,Hui:2010vt,deMenezes:2023mzx} plays a role). Based on this 
observation,  a classical dSph like Fornax may host up to  300 MSPs~\cite{Winter:2016wmy}; 
since Sgr II  contains about four times as many stars~\cite{2020MNRAS.497.4162V}, $\mathcal{O}(1000)$ MSPs 
appear fully possible.
On the other hand, the significance of Sagittarius' gamma-ray emission reported in Ref.~\cite{Crocker:2022aml} could also 
be the 
result of mis-modelling the diffuse Galactic gamma-ray foregrounds~\cite{Calore:2023prep} and hence remains the 
subject of a still ongoing debate. Let us in any case stress that a continuous background with a normalization as found in
Ref.~\cite{Crocker:2022aml} will not affect in any appreciable way searches for monochromatic features.
We tested this explicitly, conservatively allowing also for correspondingly re-scaled 
contributions from other dSphs, and found that
our results (presented in \SEC\ref{section:results_dsph}) are affected only at the sub-percent level.

In an accompanying paper~\cite{CTAdSphsInPrep} we defined the most promising dSphs targets based on 
an updated analysis of stellar kinematic data and CTA observational strategy.
While Ref.~\cite{CTAdSphsInPrep} is concerned about continuum spectra from DM annihilation and decay,  
our discussion of line searches here represents an extension of that work 
and follows the target selection and observational strategy considered there.
Concretely, it is argued that the optimal strategy for CTA, given the relatively limited FoV, is not to observe 
as many targets as possible, but rather to focus on a limited number of dSphs with  the highest chance of detection. 
The recommendation is to observe one classical and two ultra-faint 
dwarfs per hemisphere, namely  Coma Berenices, Draco~I and Willman~1 in the Northern hemisphere, as well as  Reticulum~II, 
the Sgr dSph and Sculptor in the South. In \TAB\ref{table:jfracDsph_cta} we show the corresponding $J$-factors derived 
in Ref.~\cite{CTAdSphsInPrep},  cf.~Eq.~(\ref{Jfactor}), thereby updating the results from Ref.~\cite{Bonnivard:2015xpq}.
It should be noted that the observational strategy of CTA on one or more dSphs is not yet fully decided, 
but it was proposed~\cite{CTAConsortium:2017dvg} to dedicate 100\,hr per target per year and per CTAO site, 
for a total of about 500-600\,hr for both sites. Ref.~\cite{CTAdSphsInPrep} explores different strategies to optimally use 
an assumed total observing time of 600\,hr. Here we will focus on the `conservative' strategy,  in terms of mitigating 
the impact of underestimated uncertainties of $J$-factor calculations, based on the observation
of each of the six proposed candidates shown in \TAB\ref{table:jfracDsph_cta} for 100\,hr.
Let us also stress that the uncertainties in the $J$-factors quoted in  \TAB\ref{table:jfracDsph_cta}
are observationally driven (through the analysis of kinematic data) and much smaller than the $J$-factor 
uncertainties for the GC (which are driven by extrapolation of idealized numerical simulations).
As detailed in \SEC\ref{sec:statistics}, this warrants a different statistical treatment of these cases.

\begin{table}[t]
\centering
\begin{tabular}{ |l||llllll| }
 \hline
 \multicolumn{7}{|c|}{$\log_{10}J(0.5^\circ)$ [$\rm{GeV}^{2} \rm{cm}^{-5} $] } \\
 \hline \hline
dSph &  CBe &  DraI  & Wil1 & RetII & Scl  & SgrII  \\
 \hline
CTA Group~\cite{CTAdSphsInPrep}
& $ 19.5^{+0.9}_{-0.7} $
& $ 18.7^{+0.3}_{-0.1} $
& $ 19.1^{+0.6}_{-0.5} $
& $ 18.9^{+0.9}_{-0.6} $
& $ 18.4^{+0.1}_{-0.1} $
& $ 18.9^{+1.8}_{-0.9} $
\\
\mbox{Bonnivar {\it et al.}~\cite{Bonnivard:2015xpq}}
& $ 19.6^{+0.8}_{-0.8} $ 
& $ 19.5^{+0.4}_{-0.2} $ 
& $ 19.5^{+1.2}_{-0.6}  $ 
& $ 19.6^{+1.7}_{-0.7}  $ 
& $ 18.5^{+0.1}_{-0.1}  $ 
& $ -$
\\
 \hline
\end{tabular}
\caption{$J$-factors with mean standard deviations for a selection of dSph galaxies, as defined in Eq.~(\ref{Jfactor}), 
averaged over an RoI with radius $0.5^\circ$. 
Following Ref.~\cite{CTAdSphsInPrep}, we include in our analysis the dSphs Coma Berenices (CBe), Draco I (DraI), 
Willman I (Wil1), Reticulum II (RetII), Sculptor (Scl) and the Sgr dSph (SgrII). For comparison, we also show the 
corresponding $J$-factors from an older compilation~\cite{Bonnivard:2015xpq}. 
}
 \label{table:jfracDsph_cta}
\end{table}

Traditionally, dSphs were only considered in the context of generic DM annihilation or decay spectra, not in the context
of searches for pronounced spectral signatures (see, however, Ref.~\cite{HESS:2018kom} for an exception).
The latter searches, see also below in \SEC\ref{section:analysis} for a detailed description,  are by construction 
less limited by the presence of astrophysical backgrounds. This implies that it is in general favourable to focus on 
the region with the highest $J$-factor, namely the GC. 
However, given that CTA is anyway expected to dedicate substantial observation time to dSphs, we will
also perform a sensitivity study for these targets here, based on the observational strategy discussed above. 
As it turns out, the CTA spectral line sensitivities from dSphs might in fact (almost) become comparable to those from the 
GC, in case the DM density profile in the Milky Way is cored rather than cuspy (i.e.~a GC $J$-factor that
is unfavourably small, combined with optimistic assumptions about the largest $J$-factors in dSphs).

\section{Analysis}
\label{section:analysis}

In the past, different strategies have been followed to search for DM signals with 
sharp spectral features. The most recent such analysis of the H.E.S.S.~collaboration~\cite{HESS:2018cbt}, e.g.,
adopted a fully data-driven approach based on two spatially distinct `ON' and `OFF' regions, respectively.
Here, both regions are modelled as containing the same astrophysical and instrumental background;
the `OFF' region is assumed to contain no further emission components, 
such that any potential excess in the `ON' region can be attributed to a DM signal. For current gamma-ray 
telescopes, this approach has proven highly successful also in searches for exotic signals with a 
broader energy distribution~\cite{HESS:2016mib}. 
Given the increased DM sensitivity of CTA, the bright large-scale interstellar emission in the GC region can 
no longer be ignored~\cite{Neronov:2020zhd,CTA:2020qlo}. 
This would make this specific ON/OFF technique more challenging to use.

An alternative avenue is to model the astrophysical background components explicitly. 
The \textit{sliding energy window technique}
-- as e.g.~adopted by the {\it Fermi}-LAT collaboration~\cite{Abdo:2010nc,Ackermann:2012qk,Fermi-LAT:2015kyq}, 
but also in earlier IACT studies~\cite{HEGRA:2003wfz,Bringmann:2011ye,Bergstrom:2012vd,HESS:2018kom} --
aims to implement this approach in an as data-driven and model-independent way as feasible.
Realizing that the specific types of signals we are interested in here
vary much faster with energy than any of the expected background components, the basic 
analysis idea is to divide the total energy domain into overlapping narrow energy windows, each window covering only a few times the instrumental energy resolution. 
This allows 
remaining agnostic about the nature of the background, and to model the {\it cumulative}
 (instrumental and astrophysical) background as a simple parametric function with parameters fit 
 directly to the counts inside this narrow energy range.
For our default analysis we follow this approach, modelling the total counts locally as a power law
in energy.

A somewhat more sophisticated method of the background estimation is to separate the astrophysical and 
instrumental background components, noting that information about the latter is already contained 
in the IRFs. 
Indeed, these IRFs are based on a CR spectrum at the top of the atmosphere that is not, unlike the gamma-ray 
component, partially unknown but in fact well measured up to at 
least 100\,TeV~\cite{Chang:2008aa,CALET:2019bmh,DAMPE:2019gys} (with
percent-level precision up to 1\,TeV~\cite{Aguilar:2021tos}). This would motivate to use an interpolation of the 
misidentified CRs as provided by the IRF; only the \textit{intrinsic} astrophysical background would then be
locally modelled as a power law, convoluted with the IRF.
As a result, the overall background description and sensitivity to DM improves over the simple fit directly on the 
counts, as described above; on the other hand, this approach is more
dependent on explicit assumptions about the instrument performance (which will be more accurately known 
once the instrument is fully operational). Following this alternative approach can thus be used as an indication
of how much potential gain in sensitivity one may eventually hope for, compared to the more conservative 
pre-construction sensitivity derived with our default analysis procedure.

In the following, we describe our benchmark analysis procedure in terms of the generation of mock data for the chosen
RoI
 (\SEC\ref{sec:ROI}), explain in more detail how we model background and signal components inside the sliding energy window (\SEC\ref{sec:sliding}) and lay out the general analysis pipeline to derive  exclusion limits and discovery sensitivities 
(\SEC\ref{sec:statistics}).
Later, in \SEC\ref{section:discussion}, we will explicitly discuss how modifying the assumptions underlying 
the benchmark analysis settings defined here would impact our results (presented in \SEC\ref{section:results}).
 
\subsection{Data generation and analysis regions}
\label{sec:ROI}
Based on the observational strategies and expected signal and astrophysical background 
components outlined in \SEC\ref{sec:targets}, we generate mock data  using \ctools.\footnote{%
\label{foot_data_generation}
Concretely, we use {\sf ctobssim} to produce an event list (in the form of a .fits file) containing MC realisations of the data. 
The effective area and energy dispersion for CTAO are provided as histograms in the IRF \textit{.root} files, for which we 
use the official instrument response file \rm{Prod5-South-20deg-AverageAz-14MSTs37SSTs.180000s-v0.1.root}~\cite{prod5}.
} 
The exact definition of the analysis RoIs, and the masking that we adopt, depends on the target region:

\begin{figure}[t!]
\centering
\centerline{
\includegraphics[trim=20 3 49 0,clip,width=0.515\textwidth]{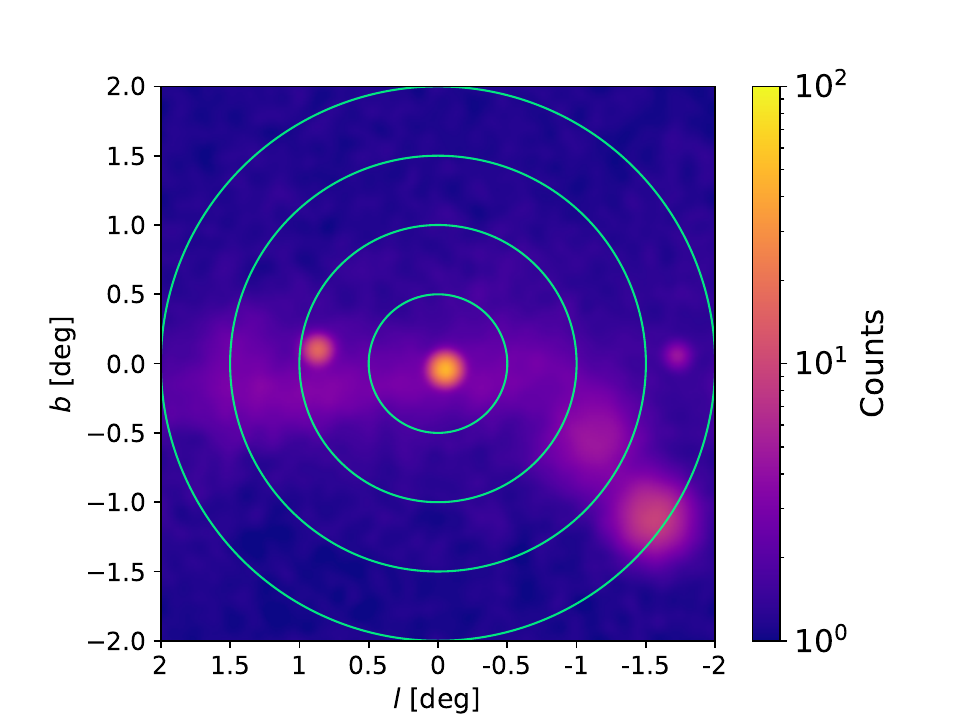}~
\includegraphics[clip,width=0.43\textwidth]{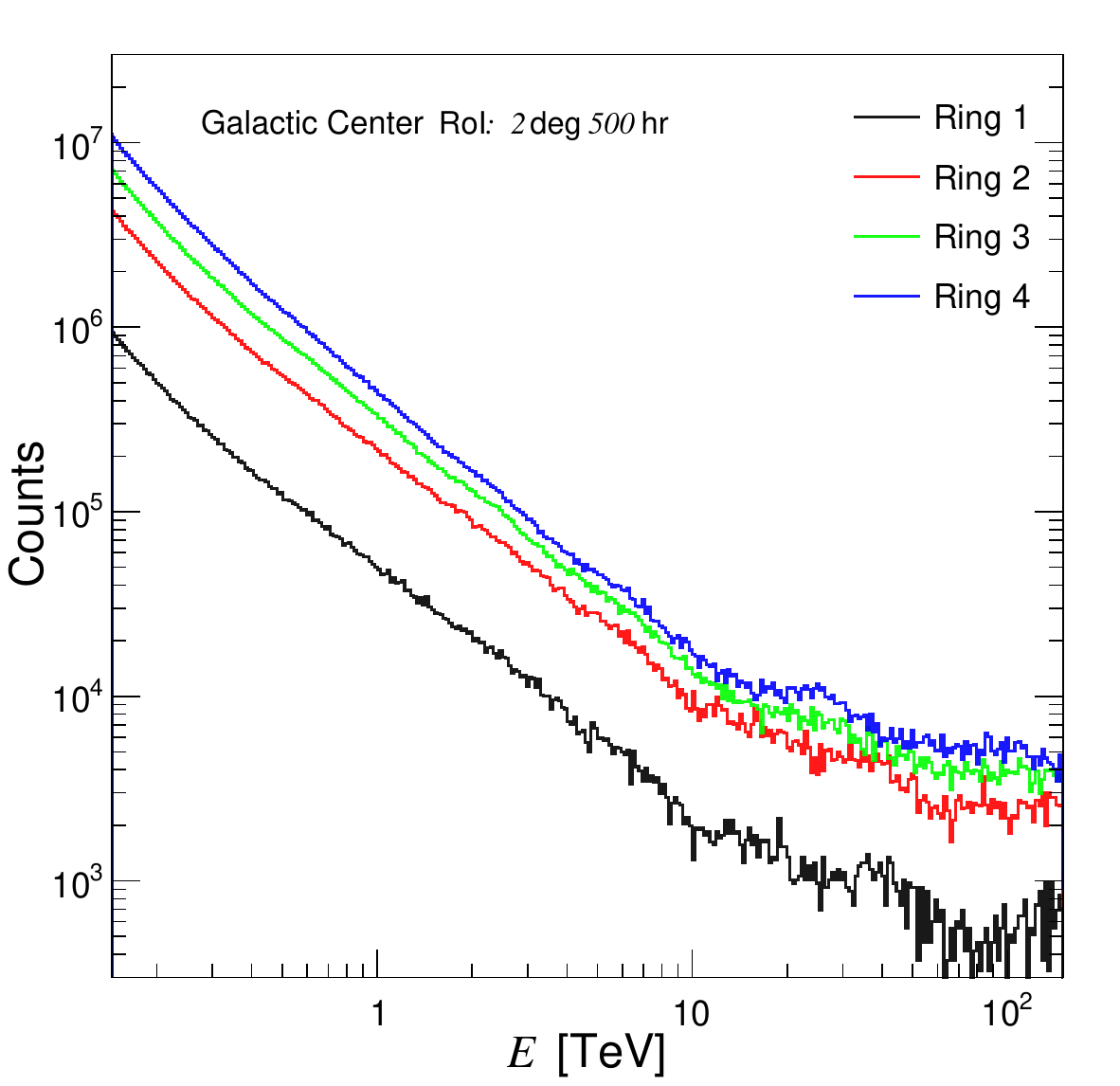}
}
\caption{ {\it Left.} Visualisation of the spatial binning geometry over a skymap of the GC background simulation
described in \SEC\ref{section:galactic_centre}. 
In Galactic coordinates, the figure shows the 
region $(b,l)=(-2^\circ..2^\circ,-2^\circ..2^\circ)$. The color scheme represent the counts 
in the energy range $[1.55,2.51]$ TeV with a pixel of size $(0.05^\circ)^2$ and a Gaussian smoothing with the same size.
 {\it Right.} Integrated background photon 
count for each spatial bin, for a specific MC realization, 
where Ring$\ 1$ refers to the innermost and Ring$\ 4$ to the outermost region.
The sum of the four histograms shown here can thus directly be compared to the (on average) expected photon 
count displayed as black line in \FIG\ref{fig:background_wider}.
The histogram has a log-even binning of $100$ bins per decade, similar to the  width used in our analysis.
}
\label{fig:background_spatial_bin}
\end{figure}

\paragraph{Galactic Centre.}
The GC survey will result in an almost isotropic exposure of the inner few degrees of the GC. 
We restrict our analysis to the inner $2^\circ$ of this survey, as motivated below,
and divide this RoI into four spatial bins consisting of concentric rings of 
width $0.5^\circ$ 
(\TAB\ref{table:jfrac} lists the corresponding angular sizes and \textit{J}-factor values). 
\FIG\ref{fig:background_spatial_bin} shows a skymap of the whole RoI illustrating this 
spatial binning configuration (left panel) and a realisation of the total photon count -- including misidentified 
CRs, point sources, the default ({\it base-max}) interstellar emission model (IEM) and Fermi bubbles --
for each of the spatial bins (right panel).  
In the left panel, the three point sources \textit{HESS J1745-290} (centre), \textit{G0.9+01} (centre left) and \textit{HESS J1741-302} (right) are clearly 
distinguishable by eye, as well as the IE (concentrated along the Galactic plane). 
The photon count in the outer
parts of the RoI (Ring 4), on the other hand, is dominated by misidentified CRs.
Features in the spectrum between about $5$ and $10$\,TeV reflect different spectral cuts in the 
transition region between the MSTs and SSTs;
still, as visible in the right panel of the figure, a power law {\it locally} provides a reasonable description of the 
spectra across the entire energy range.
It also becomes clear that up to energies of a few TeV, the photon count 
is so large that one would expect DM limits to be affected by the accuracy to which CTAO's energy resolution 
and effective area are known;
beyond multi-TeV energies, on the other hand, the limiting factor will be Poisson noise. We will return
to this observation in \SEC\ref{sec:syst_discussion}. 

Dividing the RoI in the GC region into several spatial bins is a relatively common procedure and motivated by
the different morphologies of signal and background components, see, e.g., 
Refs.~\cite{Calore:2014,Lefranc:2015pza,Lefranc:2016dgx}. 
In \SEC\ref{sec_mask:discussion} and Appendix~\ref{app:roi} we will discuss alternatives
to our default analysis setting illustrated in \FIG\ref{fig:background_spatial_bin}, and show that the final DM limits
and discovery prospects are rather robust with respect to the exact choice of the RoI and binning scheme.
In particular, concentric ring binning gives the highest statistical power to discriminate a DM signal 
among the binning geometries
that we checked explicitly, while providing an  equivalent $\chi^2$ score of the background fit.

\paragraph{Dwarf Spheroidal galaxies.}
We model the DM content of dSph galaxies ($J$-factor and its uncertainties, assuming a log-normal distribution) 
as stated in \TAB\ref{table:jfracDsph_cta},
based on the recent work developed within CTA~\cite{CTAdSphsInPrep}. 
We also follow the suggested observational strategy, i.e.~we assume 100\,hr for each of the targets shown in the table.
Note that here we use $J$-factors calculated within 0.5 degrees of the centre of each dSph, 
in order to optimize the expected DM signal. 
Further increasing the size of the disk would not significantly enhance the sensitivity, see also
Appendix \ref{app:roi} for a related discussion about how to choose the RoI in the context of the GC.
For the purpose of constructing the likelihood, see further down, we choose only one spatial bin per dSph; 
this is a simplification given the angular resolution of 
CTAO~\cite{CTAConsortium:2017dvg}, but justified for our analysis which emphasizes spectral shapes over 
morphology. Given that all selected dSphs are located at high latitudes, finally,
 we neglect any potential IEM emission and model only the (misidentified) CR backgrounds.

\subsection{Component modelling inside sliding energy window}
\label{sec:sliding}
\mbox{ }\\[-4ex]
As explained above, the mock data are {\it generated} based on a realistic implementation, as of current knowledge, 
of all relevant astrophysical (and signal) components in the respective RoIs. 
For the {\it analysis} of the data, on the other hand,
we adopt a much simpler, parametric description of all components related to the `background' (i.e.~everything but the
DM signal with its characteristic spectral shape). In particular, we will explore two strategies:

\begin{enumerate}
\item {\bf Power law on counts.} As our benchmark analysis strategy, we aim to remain fully agnostic about the 
`background' processes, other than
assuming that they lead to a spectrum much less localized in energy than the DM signal.
We therefore 
model the sum of the total  {\it counts} 
 (astrophysical  and instrumental) as a power law,
\begin{equation}
\label{eq:powerlaw_counts}
\mu^\mathrm{bg}_{ij} =  b_j \int_{\Delta E_i}dE\,E^{-\gamma_j}\,,
\end{equation}
Here, $j$ denotes spatial bins and $i$ energy bins, and $b_j$ and $\gamma_j$ describe
normalization and spectral index of the power law, respectively.
With this ansatz, any assumption about the instrument performance is removed from the analysis step (but of course not from the 
generation of mock data).

\item {\bf Power law on gamma-ray flux.} As an alternative analysis strategy we estimate the misidentified
CR component in the total counts directly from the IRF, using {\sf ctools}' \rm{ctmodel}, 
as given by the grey line in \FIG\ref{fig:background_wider}.
We note that, once the instrument is fully operational, an alternative to determine 
this component would be an auxiliary measurement from an empty area on the sky. 
For the astrophysical background component, on the other hand, we assume that a simple power law locally provides a satisfactory description of
the gamma-ray {\it flux}. We then estimate the contribution to the observed counts by 
convoluting this ansatz with the effective area shown in \FIG\ref{fig:cta}. 
The combined background model for the counts, including CRs and astrophysical gamma rays, is thus
\begin{equation}
\label{eq:aeff_treatment}
\mu^\mathrm{bg}_{ij} =  N_{ij}^{\rm CR} + b_j\int_{\Delta E_i}dE\, A_{\rm{eff}}(E) E^{-\gamma_j}\,,
\end{equation}
where $N_{ij}^{\rm CR}$ is the expected number of counts due to
unidentified cosmic rays; $b_j$ and $\gamma_j$ describe
normalization and spectral index, respectively, of {\it only} the gamma-ray component. 
Here, the effective area in this simplified form, neglecting the PSF and energy dispersion, 
is introduced exclusively to improve the (numerical) performance of the analysis. We checked explicitly that this 
description reproduces the results from a full {\sf ctools}  implementation (with a source spectrum following a power law) 
to sufficient accuracy.
\end{enumerate}
In Appendix \ref{app:corr}, cf.~\FIG\ref{fig:tsdistribution}, we will get back to the question of how well these two 
background descriptions fit the actual (mock) data.

As far as the {\it DM component} is concerned, we are interested in the detailed shape of the signal and
simply convolving the intrinsic annihilation spectrum $dN_\gamma/dE_\gamma$ with the effective area is no longer sufficient. 
Instead, we fully model the instrument response using {\sf ctools}.
For a line, VIB and box signal, 
cf.~Eqs.~(\ref{eq:line}, \ref{eq:VIB}, \ref{eq:box}),%
\footnote{%
Technically, we approximate the Dirac Delta function by using {\sf ctmodel}  with a narrow Gaussian, with 
an intrinsic width $\sigma_\chi \ll \sigma_\mathrm{res}$, and explicitly setting the flag {\sf edisp=yes}. 
} 
this results in the count spectra shown in \FIG\ref{fig:signalshapes}. 
We thus model the signal component as 
\begin{equation}
\mu^\chi_{ij} = \nu_j\int_{\Delta E_i} dE\, \zeta(E),
\end{equation}
where $\nu$ is the signal normalization and $\zeta$ the photon count of the signal spectrum convolved with the IRF (as
displayed in \FIG\ref{fig:signalshapes}).
The normalization of $\nu_j$ is fixed by Eq.~(\ref{DMflux}). 
In practice, 
we use this equation to calculate the total signal count rate only once, leading to some value of $\nu_0$ 
for the whole RoI (or, for the case of dSphs, the sum of all targets) and a reference cross section $\ov,_0$ and 
DM mass $m_{0,\chi}$. For a fixed value of 
the DM mass, $m_\chi$, $\nu_j$ is then directly related to the annihilation rate that is to be constrained
as $\nu_j/\nu_0= (m_{0,\chi}/m_\chi)^2 (\ov/\ov,_0) (J_j/J_{\rm tot})$, where $J_j$ ($J_{\rm tot}$) is the $J$-factor 
associated to the spatial bin $j$ (the total RoI).

The final task is to optimize the analysis region
-- the sliding energy window -- such that it is small enough for the effective description of the background model 
to hold, but at the same time large enough to give sufficient statistical power to test the DM signal.
The benchmark setting that we adopt in our analysis is a sliding energy window 
of width $\Delta = 8\sigma_\mathrm{res}(E_0)$, centred
on the putative DM signal localized
at $E_0$ 
(for a wide box, with width  $\Delta E>\Delta$, we choose the energy window instead to be centred on the upper edge of the box spectrum, cf.~the right panel of \FIG\ref{fig:signalshapes}).
Here, $\sigma_\mathrm{res}$ is the energy resolution of CTAO, as depicted in \FIG\ref{fig:cta}.
As detailed in Appendix~\ref{app:energy-window}, this choice of $\Delta$ is motivated by increasing the window size
until the signal significance begins to converge while at the same time ensuring that the background model
(described above) still gives a good fit to the data.
We use an energy binning of three energy bins per $\sigma_\mathrm{res}$,
i.e.~we are in some sense effectively working in the limit of an unbinned analysis (in energy).
Given the instrumental count normalization, our setup guarantees more than 
10 photons per bin even at the highest energies considered in the analysis.

\begin{figure}
\centering
\includegraphics[width=0.8\textwidth]{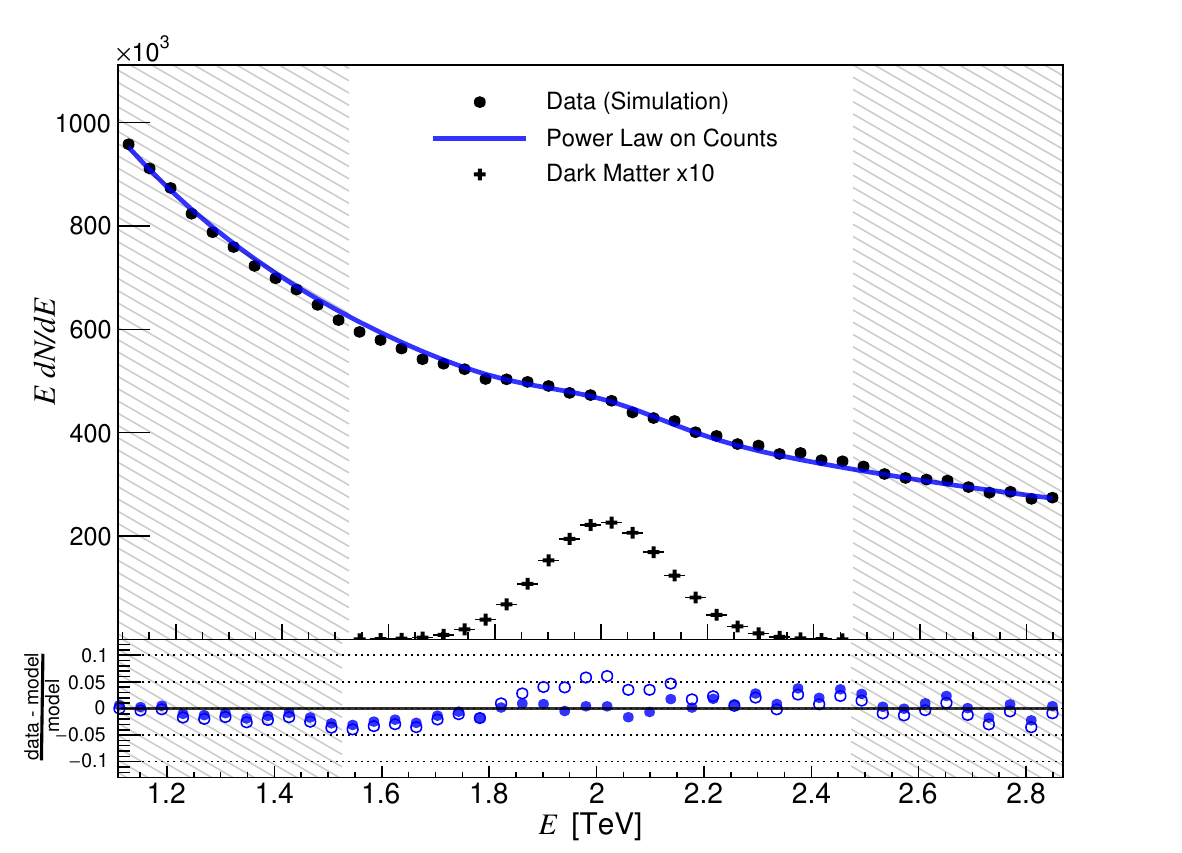}
\caption{%
Illustration of the sliding energy window technique to identify signals with sharp spectral features.
Mock data points (black dots) are based on the full background model, for the GC survey, and a 
monochromatic signal component at $E_0=2$\,TeV with a normalization that would allow a 
$5\sigma$ discovery. The white area that is not hatched corresponds to the sliding energy
window, of width $8\times\sigma_{\mathrm{res}}(E_0)$, within which the analysis is performed. 
Black crosses show the expected signal component (multiplied by a factor of 10 for 
better visualization). The blue solid line is the result of fitting the data with a monochromatic
signal component on top of a simple power law.
The lower panel shows residuals with (solid circles) and without (empty circles) 
including the signal component in the fit.
}
\label{fig:linesearch}
\end{figure}

In \FIG\ref{fig:linesearch}  we illustrate the analysis procedure by showing an explicit 
example of a monochromatic DM line injected into the data, which is then fitted by the assumed signal 
component and a simple power law on the `background counts'.
The region between the shaded areas is the sliding energy window inside which the analysis is performed.
We indicate the true signal with black crosses, scaled by a factor of 10 for better visibility, 
and the best-fit model (power law plus signal) with a solid blue line.
The residual plot in the lower panel gives a good visual impression of how well the power law fits the background
inside the analysis region -- even though it does not necessarily do so for a larger energy range. In what follows
we detail how this observation can be used to derive (expected) sensitivity limits for such line signals.
 
\subsection{Statistical procedure}
\label{sec:statistics}
Within each sliding energy window we implement a binned likelihood based on Poisson statistics,
$P\left[\left.{n}_{ij}\right|{\mu}_{ij}\right]=\prod_{i,j} e^{-\mu_{ij}} {\mu_{ij}^{n_{ij}}}/{\left(n_{ij}\right)!}$,
where $\bm{\mu}=\{\mu_{ij}\}$ denotes the model prediction and $\bm{n}=\{n_{ij}\}$ the (mock) 
data counts. The energy bins (indicated by an index $i$) are taken to be much smaller than the 
instrument's resolution, thus effectively implementing an unbinned approach; the spatial
bins, indicated by an index $j$, refer to the RoIs defined in \FIG\ref{fig:background_spatial_bin} 
(for the GC analysis) or
the individual galaxies stated in \TAB\ref{table:jfracDsph_cta} (for the combined dSph analysis), respectively.
The model prediction depends on the signal normalization $\nu$, and various background model 
and other nuisance parameters which we collectively denote as
$\bm{\theta}$.

\paragraph{Treatment of systematic uncertainties.}

\label{section:systematics}
Clearly, instrumental systematic uncertainties are challenging to model for a telescope still 
under construction. 
Even if the underlying event counts are uncorrelated, as assumed here, the finite energy resolution of CTA 
will correlate noise deriving from systematic deviations between the true and assumed IRFs. 
Here we take a parametric approach to estimate such systematic noise 
by introducing additional nuisance parameters $\eta_i$, one for each energy bin $i$, 
to rescale counts expected from the model prediction as $\mu_i\rightarrow \eta_i \mu_i$.
We model the covariance of these nuisance parameters by assuming multivariate normal distributions 
with means $\langle \eta_i\rangle=1$ and a covariance matrix $\Sigma$ with variance $\sigma$.
The off-diagonal part of the covariance matrix is thus modelled as 
\begin{equation}
\label{eq:defsigma}
 \Sigma_{ii'} = \sigma^2\exp{\left[ -\frac{(E_{i}-E_{i'})^2}{2(\lambda\, \Delta E)^2} \right]}\,,
\end{equation} 
where $\lambda$ denotes the correlation length and $\Delta E\equiv\sigma_\mathrm{eres} (E_0)$, with
$E_0$ being the energy at the center of the analysis window.
We find that this functional form describes the results of dedicated MC simulations very well when adopting 
a characteristic length scale $\lambda\simeq 1.5$, see Appendix~\ref{app:corr} for further details.
For the variance we choose $\sigma=0.025$ as a fiducial value which, at face value, is significantly larger than the
$\sim$1\% design goal of CTAO~\cite{CTAConsortium:2017dvg}. This choice avoids artificially strong limits due to an overfitting of the 
specific numerical IRF (and/or IEM) model realization that is used in our analysis.
See also \SEC\ref{sec:syst_discussion} for a discussion of how
the treatment of systematic uncertainties, and in particular the exact choice of $\sigma$,  impacts our final results.

\paragraph{Construction of likelihoods.}
Following the description above, the total likelihood that we adopt for the GC analysis is
given by 
\begin{equation}
\label{eq:likelihoodSyst}
\mathcal{L}(\nu, \bm{\theta}) \hspace{0.2cm} 
\propto \hspace{0.2cm}  
\prod_{i}
\prod_{j}
P[n_{ij}|\eta_{ij}\mu_{ij}]\, 
\exp\left[{-(1-\eta_{ij})\Sigma_{ii'}^{-1}(1-\eta_{i'j})}\right]\,,
\end{equation}
where the indices $i$ ($j$) run over all energy (spatial) bins within the sliding energy window,
and a summation over the energy bins $i'$ in the covariance part is implicit.
We recall that our model description is given by $\mu_{ij}= \mu^\chi_{ij}(\nu)+\mu^{\mathrm{bg}}_{ij}(\bm{\theta}_{\rm bg})$,
with $\nu$ being the signal normalization and $\bm{\theta}_{\rm bg}=\{b_j,\gamma_j\}$ describing the
background normalizations and slopes of every spatial bin that is considered (per energy window);
the full list of nuisance parameters for the GC likelihood 
is thus given by $\bm{\theta}=\bm{\theta}_{\rm bg}\cup \{\eta_i\}$.

The likelihood for dSphs is constructed by multiplying (sometimes referred to as {\it stacking} in this context) the individual likelihoods 
for each separate dSph observation, taking into account their respective $J$-factors and
associated uncertainties. 
For each dSph galaxy we model the likelihood for the true 
$J$-factor to follow a log-normal distribution $\mathrm{Log}\mathcal{N}$ around the mean observed value
(following, e.g., Ref.~\cite{Fermi-LAT:2015att}), 
$\bar{J}_{j}$, with the standard deviation $\sigma_{J,j}$ of $\ln J$ fitted to the mean absolute 
deviation stated in \TAB\ref{table:jfracDsph_cta}.
Since the DM flux is directly proportional to the $J$-factor, we thus arrive at the total
likelihood (see also Ref.~\cite{Fermi-LAT:2015att,H.E.S.S.:2020jez,Rico:2020vlg})
\bea
\label{eq:likelihoodDSPH}
\mathcal{L}(\nu, \bm{\theta}) \hspace{0.2cm} 
&\propto& 
\prod_{j}^{\rm dSph}
\mathrm{Log}\mathcal{N}
\left[ \log_{10} (J_j)|\log_{10}(\bar{J}_{j}),\sigma_{J,j}\right]\\
&&
\times\prod_{i}
P[n_{ij}|\eta_{ij}\mu_{ij}]\, 
\exp\left[{-(1-\eta_{ij})\Sigma_{ii'}^{-1}(1-\eta_{i'j})}\right]\,.\nonumber
\eea
Denoting with $\nu$ the signal normalization that would correspond to
a putative target with $J_{\rm eff}\equiv \sum_j \bar{J}_{j}$, the model description is now
given as $\mu_{ij}= \mu^\chi_{ij}(\alpha_j \nu)+\mu^{\mathrm{bg}}_{ij}(\bm{\theta}_{\rm bg})$, with $\alpha_j\equiv J_j/J_{\rm eff}$,
and the complete list of nuisance parameters is $\bm{\theta}=\{\log_{10} (J_j),\eta_i,b_j,\gamma_j\}$.

\paragraph{Expected limits and discovery prospects.}
Exclusion limits must correctly account for statistical downward fluctuations in the photon count, 
for a given signal strength, while discovery limits  should avoid falsely rejecting the 
background-only hypothesis 
in the presence of upward fluctuations of the background.
In order to distinguish the hypotheses of signal plus background and background only, respectively,
we estimate both types of limits by implementing a standard likelihood ratio test~\cite{Rolke:2004mj},
based on the test statistic (TS)
\begin{equation}
\label{eq:MLR}
{\rm TS}(\nu) \equiv -2\log{\frac{ \mathcal{L}(\nu, 
\hat{\hat{\boldsymbol{\theta}}})}{ \mathcal{L}(\hat{\nu}, {\boldsymbol{\hat{\theta}}})}}\,.
\end{equation} 
Here, $\hat{\hat{\boldsymbol{\theta}}}$ 
is the conditional estimate (best fit) for $\boldsymbol{\theta}$ under the hypothesis $\nu\ge0$. 
The best-fit estimates for the signal normalization and nuisance parameters are given by
$\hat{\nu}$ and $\hat{\boldsymbol{\theta}}$, respectively. 
We use  the Migrad algorithm~\cite{James:1975dr,migradweb} 
in  {\sf ROOT's MINUIT} package to maximize (profile over) the likelihoods given in 
Eqs.~(\ref{eq:likelihoodSyst},  \ref{eq:likelihoodDSPH}) to obtain these quantities.

In order to produce sensitivity curves for expected exclusion limits, one must generate mock data 
without a signal 
component. Taking into account that the signal normalization is non-negative, one-sided $95\%$ {\it upper 
exclusion limits} (U.L.) are found by increasing the signal normalisation, $\nu$, until
\begin{equation}
\label{eq:upperlimit}
{\rm TS}_{\mathrm{U.L.}}(\nu)= 2.71\,.
\end{equation}
In order to derive the {\it sensitivity for discovery}, on the other hand, one has to 
generate mock data including a signal with some normalization $\nu'$.
A $5\,\sigma$ discovery, corresponding to a p-value of 
$5.74\times10^{-7}$,  can be claimed when the test statistics for the background only hypothesis
($\nu=0$) on this data set evaluates to\footnote{%
The exact condition results from the fact that, for nested hypotheses with non-negative signal, $q(0)$ 
follows $\frac{1}{2}\chi^2_1 \equiv \frac{1}{2}\delta(q)+\chi^2_1$ under the background-only hypothesis, 
where $\chi^2_1$ is a chi-squared distribution with one degree of freedom,
cf.~Appendix \ref{App:Asimov}.
}
\begin{equation}
\label{eq:discovery}
{\rm TS}_\mathrm{discovery}\equiv TS(0)
 = 23.75\,.
\end{equation} 
In practice, this involves gradually increasing $\nu'$ until the best-fit signal normalization $\hat \nu$ 
satisfies the above condition. We note that, for the energies and analysis window considered here, a
signal discovery will always correspond to significantly more than 10 signal photons.
We further note that Eq.~(\ref{eq:discovery}) corresponds to the {\it local} significance for a $5\sigma$ discovery 
-- which formally reduces to a {\it global} significance of about $4.1\sigma$ for an assumed very conservative
trial factor of $80$ (based on how many lines naively `fit' into the analysis region) or $4.3\sigma$ 
when taking into account statistical correlations, based on a rough estimate following~Ref.~\cite{Gross:2010qma}.
For such a highly significant signal, however, TS is in any case a very steep function of the required signal normalization $\nu$.
The distinction between global and local significance has therefore 
only very limited impact on the reported $5\sigma$ discovery reach. Concretely, we find that a $\sim 10\%$ larger normalization
would raise the {\it global} significance of the signal to the $5\sigma$ level.

Since the likelihood is a function of the (mock) data, limits derived from 
Eqs.~(\ref{eq:upperlimit}, \ref{eq:discovery}) will necessarily be subject to statistical fluctuations. 
Rather than creating a large number of mock datasets to derive the {\it median} limits, and
their variances, we will here adopt the \emph{Asimov dataset} method~\cite{cowan:asimov}. 
This method allows to extract both results from a single, fiducial dataset that is defined by 
the observed photon counts in each bin being exactly equal to their expectation values.
 For further details on the construction of the Asimov dataset, including explicit validation checks
 with MC simulations, see  Appendix~\ref{App:Asimov}.

\section{Results}
\label{section:results}

\mbox{ }
\begin{table}[t]
\centering
\begin{tabular}{ |l||p{5cm}|p{5cm}| }
  \hline
  & Galactic Centre & dSphs \\
  \hline\hline
Exposure time  &  500\,hr & 100\,hr per target \\
   \hline
DM density profile &   Einasto [\ref{sec:profile_discussion}] & $J$-factors in \TAB\ref{table:jfracDsph_cta}\\
   \hline
RoI and binning  &  
$4$ rings of width $0.5^\circ$deg [\ref{app:roi}] & Single RoI per dSphs, $0.5^\circ$ \\
   \hline
Mask  &  none [\ref{sec_mask:discussion}] & none \\
 \hline
IEM &  Base MAX [\ref{sec_IEM_discussion}] &  none \\
 \hline
   \hline
Analysis  method & \multicolumn{2}{|c|} {Sliding energy window, PL assumption on counts} \\
  \hline
Window size & \multicolumn{2}{|c|} {$8\sigma_\mathrm{res}(E_0)$ [\ref{app:energy-window}]}\\ 
  \hline
Systematic uncertainty  & \multicolumn{2}{|c|} {$2.5\%$, per energy bin [\ref{sec:syst_discussion}]} \\
 \hline
\end{tabular}
 \caption{Summary of benchmark settings and assumptions for the analysis performed in this work. All our 
 main results, presented in \SEC\ref{section:results}, are exclusively based on these settings.   Numbers in 
 parentheses link to the subsections where we assess the impact of varying the respective assumption or 
 analysis setting on our results.
 \label{table:benchmark}}
 \end{table}
All results in this section will assume our set  of benchmark assumptions, summarised in 
\TAB\ref{table:benchmark}.
In particular, in \SEC\ref{section:results_galactic_centre} we present the sensitivity for exclusion and 
discovery of DM 
self-annihilating to a pair of monochromatic gamma rays from the GC,
 and in \SEC\ref{section:results_dsph} the sensitivity resulting from a combined 
analysis of six dSphs. Finally, in \SEC\ref{section:results_linelike}, we provide results for the case of other 
sharp spectral features that can originate from DM annihilation, focussing on box-shaped and VIB-like signals.

\subsection{Galactic Centre}
\label{section:results_galactic_centre}

\begin{figure}[t!]
\centering
\includegraphics[width=0.8\textwidth]{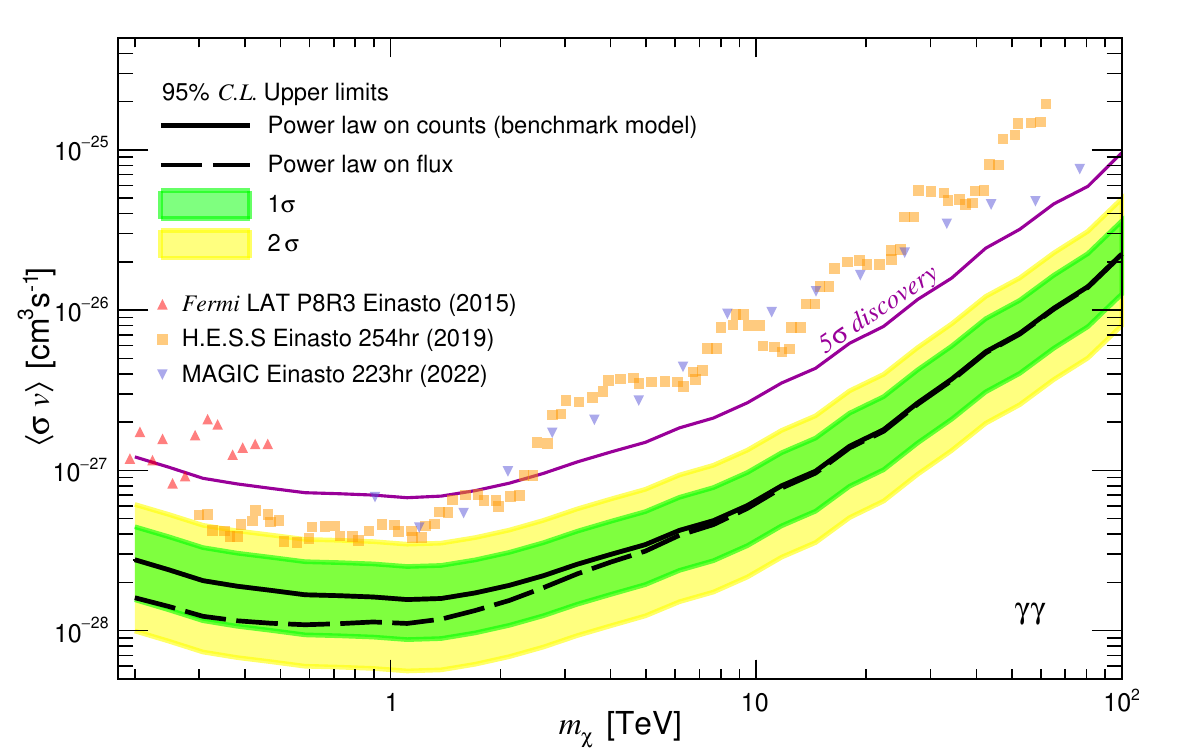}
\caption{
Median of expected $95\%~C.L.$ upper limits on the annihilation of DM into a pair of gamma-ray photons (black) 
as well as the $5\sigma$ discovery potential (purple), 
as a function of the DM mass $m_\chi$. The green and yellow bands show the expected variance of the 
median upper limits, as indicated,
and data points summarize $95\%~C.L.$ limits previously obtained by {\it Fermi} LAT~\cite{Fermi-LAT:2015kyq},  
H.E.S.S.~\cite{HESS:2018cbt} and MAGIC~\cite{MAGIC:2022acl}.  (Note that a significant scatter
between mass bins is expected for a limit on actual data, as opposed to the median of limits derived from
many MC realizations; the treatment of systematic uncertainties, furthermore, partially differs from the analysis adopted in this work).
The limits projected for CTA are based on the assumption of an Einasto DM profile and $500$ hours of 
observation of the inner GC, adopting our benchmark modelling of the background component in the 
analysis (solid lines); for comparison, we also indicate (with dashed black lines) the mean upper limits resulting 
from the more aggressive analysis method that relies on modelling the astrophysical gamma-ray flux -- rather 
than the total counts -- as a power law. 
\label{fig:limits_main}}
\end{figure}

In \FIG\ref{fig:limits_main}, we show the expected median $95\%~C.L.$ upper limits (black) and the $5\sigma$ discovery 
potential (purple) of the DM line signature.
While solid lines are the 
result of our default analysis strategy (power-law background on the measured counts), dashed lines show the alternative 
approach, where the power-law assumption is made on the gamma-ray fluxes instead. 
As stressed in \SEC\ref{sec:sliding}, the default approach neglects our knowledge of the 
IRFs and therefore results in more conservative estimates of the sensitivity. The inner (green) and outer 
(yellow) bands show the 1\,$\sigma$ 
and 2\,$\sigma$ confidence level of our sensitivity estimate, respectively, 
as derived from the Asimov dataset (for further discussion, see Appendix.~\ref{App:Asimov}). 
The lower DM mass threshold in this figure is set to 200 GeV,  from the requirement of the lower edge of the
sliding energy window to 
not  fall below 100 GeV. We prefer  to not use the lowest bins at this stage because
the effective area of CTAO drops rapidly when going below 100 GeV, 
cf.~\FIG\ref{fig:cta}, causing the current IRF estimate to be more uncertain. 

As demonstrated in the figure, the projected  CTA sensitivity to spectral line signatures improves upon current limits by 
ground-based experiments (notably HESS~\cite{HESS:2018cbt}) by a factor of $\sim$2 at 1\,TeV, and by up to one order of 
magnitude in the multi-TeV range. 
Such an improvement is in rough agreement with what one may expect from an increase of exposure alone, 
as a consequence of doubling the observation time and a larger effective area (cf.~right panel of \FIG\ref{fig:cta}).
Below about 300 GeV, the CTA sensitivity is expected to become worse than limits reported by 
the {\it Fermi} LAT~\cite{Fermi-LAT:2015kyq}. It is also intriguing to compare the current bounds to the CTA discovery potential.
The fact that CTA would potentially allow the robust discovery of a line signal above around 3\,TeV,
without being in tension with any known limits, offers exciting prospects for detecting heavy DM candidates. 
For example, this corresponds to the upper mass range of thermally produced Wino-like 
DM~\cite{Hisano:2006nn,Hryczuk:2010zi}.  
Let us stress that the results obtained in \FIG\ref{fig:limits_main} were obtained with the initially targeted `Alpha'
configuration of the instrument; we find that a fiducial `Omega' configuration corresponding to a later 
construction stage would result in a further improvement of the reported limits by about a factor of two.

\begin{figure}[t]
\centering
	\includegraphics[width=1.05\textwidth]{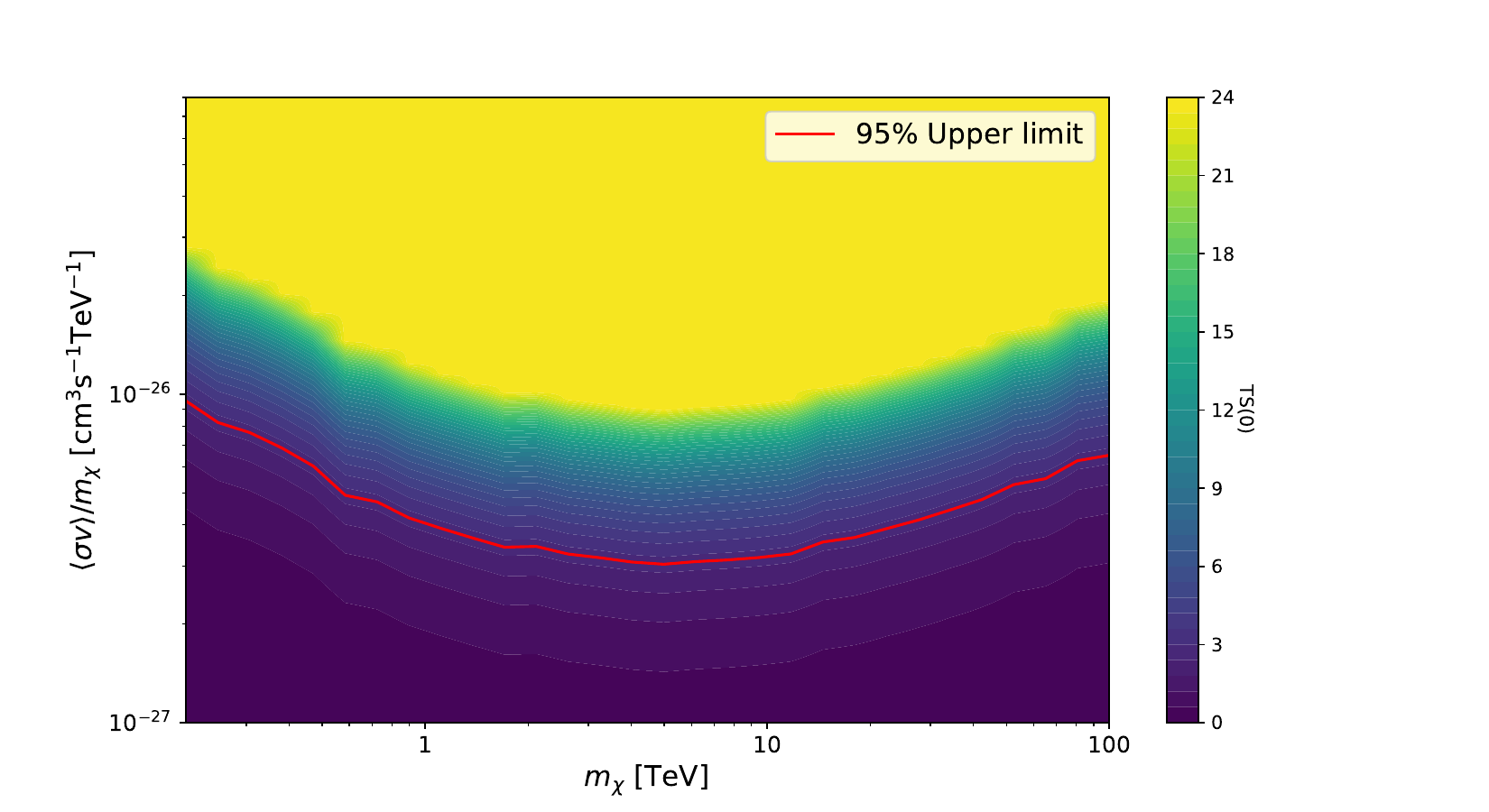}
\caption{Contour plot of the local test statistic for a monochromatic line signal from DM annihilations 
$\chi\chi\to\gamma\gamma$, as a function of $\langle \sigma v\rangle/m_\chi$ 
and the dark matter mass $m_\chi$.
To guide the eye, we apply a cap of TS\,$=23.75$ in this figure, corresponding to a 5\,$\sigma$ {\it discovery}.
The $95\%$ C.L.~{\it limit}, corresponding to the black solid line in \FIG\ref{fig:limits_main}, 
is indicated with a red line for comparison. 
The full likelihood tables for both limits and discovery potential, also for other DM profiles, are available 
for download at zenodo~\cite{zenodo}.
\label{fig:likelihood-scan}
}
\end{figure}

Consequently, CTAO data will likely also have a decisive impact on global fits of theories beyond the standard 
model that contain multi-TeV DM candidates (see, e.g., 
Refs.~\cite{Bechtle:2012zk,GAMBIT:2017snp,Bagnaschi:2017tru}). 
To facilitate such parameter scans we provide in \FIG\ref{fig:likelihood-scan} the full binned TS, 
from which the likelihood, up to an overall normalization, follows from Eq.~(\ref{eq:MLR}).
Note that, for plotting reasons, we choose here $\langle \sigma v\rangle/m_\chi$ rather than 
$\langle \sigma v\rangle$ for the $y$-axis.
 This figure complements the limits at a given confidence level shown in \FIG\ref{fig:limits_main}, and 
 illustrates how quickly it becomes impossible  to reject the signal hypothesis once the intrinsic signal 
 strength reaches a certain value (while at low signal strengths the test statistic, and hence the likelihood, 
 remains rather flat). We provide a tabulated version of the likelihood at zenodo~\cite{zenodo}.

\subsection{Dwarf Spheroidal Galaxies }
\label{section:results_dsph}

\begin{figure}[t!]
\centering
\includegraphics[width=0.8\textwidth]{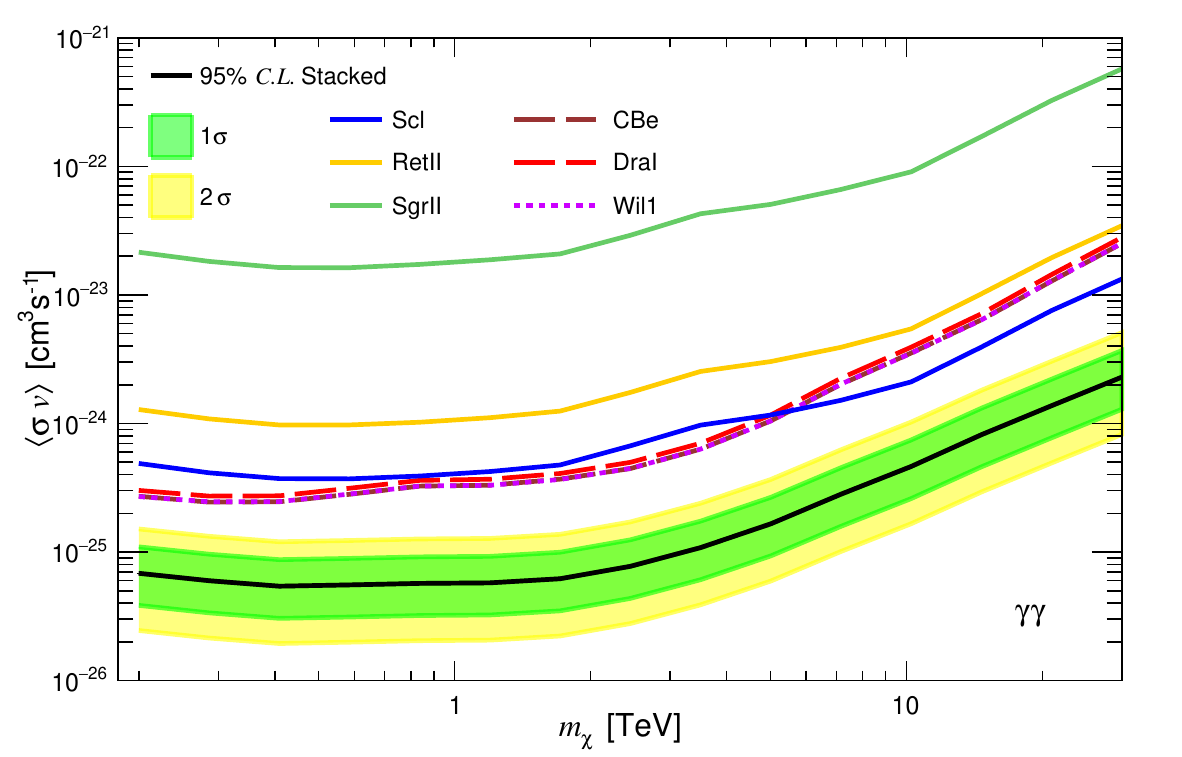}
\caption{CTA sensitivity limits of a dark matter line signal from $\chi\chi\to\gamma\gamma$, 
assuming $100$ hour observations of the individual (colored) 
and combined (black) dSphs. The green and yellow bands show the expected variance of the 
stacked limits at the 1\,$\sigma$ and 2\,$\sigma$ level, respectively.
For the individual objects, as indicated in the legend, solid lines styles are used for objects targeted 
by the southern array, while other lines styles are used for objects targeted by the northern array.
}
\label{fig:limits_dsph}
\end{figure}

We extend the DM line search to a combined analysis of the most promising dSphs for DM 
indirect detection, as described in Section \ref{section:dsph}. The result for the median expected limits on such a signal
is shown as a solid black line in \FIG\ref{fig:limits_dsph}, along with the expected variance of these 
limits at the 1\,$\sigma$ and 2\,$\sigma$ level (green and yellow bands, respectively). 
As expected, the sensitivity resulting from the observation of dSphs is significantly worse, by more than two
orders of magnitude, than the sensitivity shown in \FIG\ref{fig:limits_main} for the GC case. On the other hand, 
the DM distribution close to the GC is much more uncertain than the $J$-factor determination of dSphs.
This may reduce the GC sensitivity by a factor of 10 with respect to the default assumption of an 
Einasto density profile,
see the discussion in Section \ref{sec:profile_discussion} below, which could in fact make line limits obtained through
dSph observations (marginally) competitive.
Concerning discovery, the above discussion also makes clear that identifying a line(-like) signal in at least one dSph 
would be an extremely strong case in favour of a DM interpretation if -- and in fact only if -- an identical spectral shape is
seen from the direction of the GC.

Let us stress that the sensitivities shown in  \FIG\ref{fig:limits_dsph} crucially depend not only on the mean value
and standard deviations of the $J$-factors, as stated in \TAB\ref{table:jfracDsph_cta}, but in principle on their entire probability 
distribution. When eventually inferring limits from actual data taken by CTAO, it is thus important to 
include the full likelihoods from state-of-the-art kinematical analyses rather than just derived values for mean and standard
deviation of the $J$-factors. Incorrectly modelling the $J$-factor distribution beyond their first two moments may, in fact,
easily affect overall DM limits by a factor of a few.

In \FIG\ref{fig:limits_dsph} we also present, for comparison, the $95\%$ exclusion limits for the individual targets. 
In the limit of negligible $J$-factor uncertainties, these limits could simply be scaled with the square root of the 
observation time in order to estimate the effect of implementing different observational strategies.
Notably, the actual limit that we obtain 
from the combined analysis is somewhat stronger than just naively adding (and then squaring) 
the individual limits. This demonstrates the power of the
statistical analysis method to combine (`stack') several targets with intrinsically identical DM annihilation strengths, thereby
effectively reducing the overall $J$-factor uncertainty.
From the figure one can see that sensitivities derived from individual observations of Coma Berenices, Draco and Willman 1 
are comparable, and  that the combined
limit improves the best individual limit by about a factor of three. 
Indeed, these results might suggest that for the specific
case of line searches 
a slightly better observational strategy could be to focus the entire 600\,hr of available observation time on the three dSphs visible 
with the Northern array  (for a general and more detailed discussion of optimizing dSph observations for DM searches, we refer 
to Ref.~\cite{CTAdSphsInPrep}).

\subsection{General Signal Shapes}
\label{section:results_linelike}

\begin{figure}[t]
\centering
\includegraphics[width=0.8\textwidth]{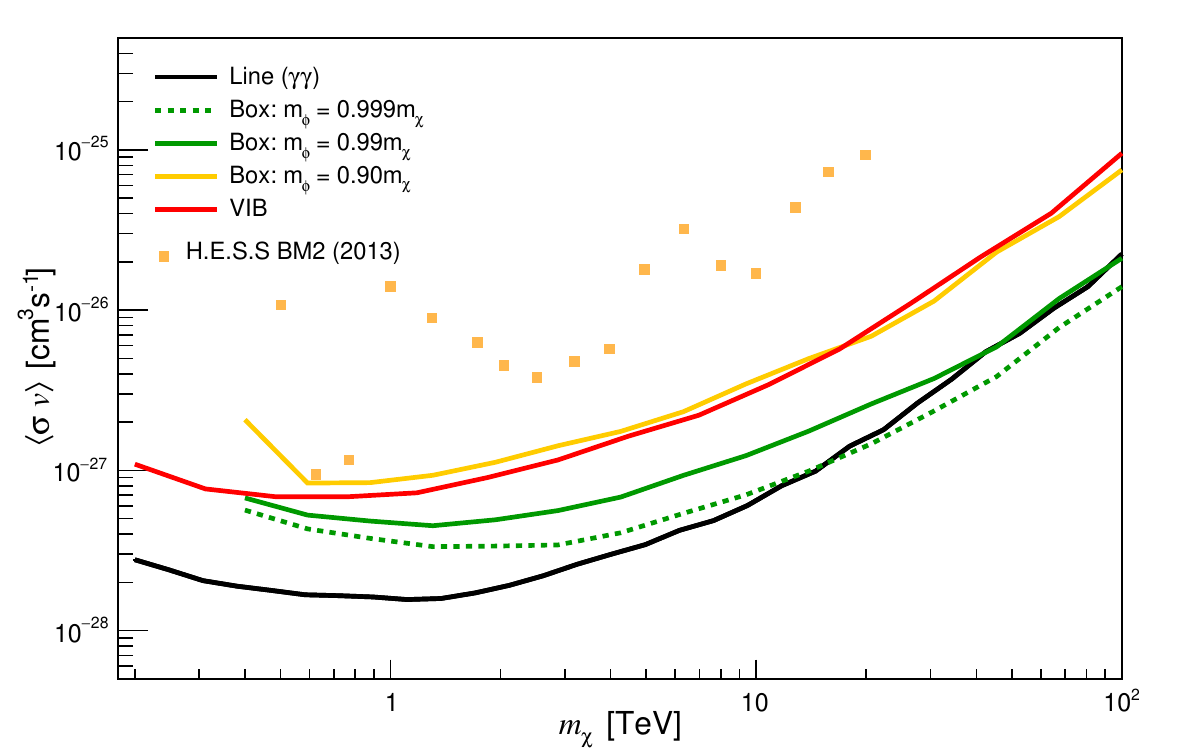}
\caption{
Expected median $95\%~C.L.$ exclusion limits, from GC observations, 
for the spectral shapes shown in \FIG\ref{fig:signalshapes}:
VIB (red), a relatively wide box with $m_\phi=0.9\,m_\chi$ (orange) and a narrow box with $m_\phi=0.99\,m_\chi$ (green).
For comparison, we also show the case of an extremely narrow box with $m_\phi=0.999\,m_\chi$ (dotted green)
and the result for a monochromatic line signal (black, same as in \FIG\ref{fig:limits_main}).  
Note that the analysis windows for the narrow box spectra are centred around $E=m_\chi/2$, 
the wide box spectrum is centred at the upper edge $E=(m_\chi +\Delta E )/2$,
while those for VIB and line spectra are centred around $E=m_\chi$; as a consequence, the lowest mass 
points that we include in our analysis are given by $m_\chi=0.4$\,TeV and $m_\chi=0.2$\,TeV, respectively.
We also indicate previous $95\%~C.L.$ limits obtained by H.E.S.S.~\cite{HESS:2013rld} 
for a signal shape model (BM2 from Ref.~\cite{Bringmann:2007nk})
that closely resembles the VIB signal studied here, rescaled to the DM profile adopted in this analysis for
the sake of comparison (see footnote \ref{foot_VIB_HESS} for further details).
}
\label{fig:limitDMshapes}
\end{figure}

We next assess the impact of deviations from an exactly monochromatic signal shape. As discussed in
\SEC\ref{section:spectral_signatures}, such deviations can appear quite commonly,
and are in fact intricately linked to the specific particle nature of the annihilating DM particles.
For definiteness, we consider here the same examples of such signal shapes as the ones introduced in 
\FIG\ref{fig:signalshapes}, and show in \FIG\ref{fig:limitDMshapes} the corresponding sensitivity of CTA
for our benchmark set of  assumptions for GC observations. 

The sensitivity to a VIB-like spectrum (red line) is very roughly a factor of $\sim5$ worse than that to a monochromatic 
signal (black line), consistent with previous findings~\cite{Bringmann:2011ye, HESS:2013rld}. 
The reason for this is a combination 
of three effects: {\it i)} the VIB signal is intrinsically weaker by a factor of 2 because there is only one photon
produced per DM annihilation, as opposed to two photons in the case of annihilation to $\gamma\gamma$, 
{\it ii)} the peak of the VIB signal occurs at slightly smaller energies than for a monochromatic signal, 
cf.~the left panel of \FIG\ref{fig:signalshapes},  where the (soft) background contribution is larger, and
{\it iii)} the VIB signal is less sharp than a line signal and hence not
quite as easily distinguishable from the (power-law) background. On the other hand, DM annihilation to a photon
pair is necessarily loop-suppressed, at order $\mathcal{O}(\alpha_{\rm em}^2)$, while the emission of a single photon
happens at $\mathcal{O}(\alpha_{\rm em})$. Depending on the DM model, the sensitivity of CTA to the VIB 
signature may thus still result in significantly more constraining limits than the sensitivity to a line signal.

Turning to the case of box-like signal shapes, there is an additional complication in that the intrinsic signal is 
not centred at $E=m_\chi$, as for VIB and $\gamma\gamma$, but at smaller energies (down to
$E=m_\chi/2$ for narrow boxes). The sensitivity to a box signal at $m_\chi=1$\,TeV, for example, should thus 
be compared to the sensitivity for a line signal at $m_\chi=500$\,GeV -- 
but only after multiplying the former by a factor of 4 because the signal strength is explicitly proportional to 
$m_\chi^{-2}$, cf.~Eq.~(\ref{DMflux}). On the other hand, there are four photons that are produced per 
annihilation, compared to two for the case of the $\gamma\gamma$ line. In summary, the sensitivity curve to 
an extremely narrow box -- which closely resembles a monochromatic line -- should in principle coincide 
exactly with the sensitivity curve for $\gamma\gamma$ after it has been shifted by a factor of 2 both 
downwards (towards smaller $\langle\sigma v\rangle$) 
and to the left (towards smaller $m_\chi$). For illustration we show in \FIG\ref{fig:limitDMshapes} the case of a 
very narrow box with $m_\phi=0.999\,m_\chi$ (green dotted line) which, indeed, follows this expectation to a 
very good accuracy.
Compared to the `monochromatic box limit' represented by the dotted green line, the sensitivity generally 
worsens as the box widens. This can be clearly seen for the explicit examples of a narrow box ($m_\phi=0.99\,m_\chi$, 
green line) and a wide box ($m_\phi=0.9\,m_\chi$, orange line) 
shown in the figure. For a narrow box, the origin of this sensitivity loss is 
simply that the signal becomes more and more smeared out, cf.~point {\it iii)} above. For a wide box
-- where the analysis window is centred on the upper end of the signal rather than on $m_\chi/2$, cf.~the
right panel of \FIG\ref{fig:signalshapes} -- an additional loss of sensitivity results from the fact that the low-energy
part of the signal is completely dominated by the background (and hence not even included in the analysis window 
anymore). 

In analogy to the concluding comment that we made about the sensitivity to a VIB-like signal, it is worth stressing that 
box-like signals are produced at leading order in perturbation theory, i.e.~without {\it any} generic suppression in 
$\alpha_{\rm em}$. This implies that CTA will be able to provide highly competitive limits on the class of
DM models that produce such a signal shape.  One way of illustrating this claim is to compare the sensitivity 
shown in \FIG\ref{fig:limitDMshapes} to the benchmark `thermal' 
annihilation cross section of $\langle\sigma v\rangle\sim2\cdot10^{-26}\,{\rm cm}^3/{\rm s}$ that is needed to
produce DM in the early universe, in the simplest models of thermal freeze-out (see, e.g., 
Ref.~\cite{Bringmann:2020mgx} for a recent 
discussion and precision determination of this quantity). 
We can thus conclude that CTA can actually have a significantly {\it better} sensitivity to TeV DM that is 
thermally produced by annihilations of the type $\chi\chi\to\phi\phi$ than for models where DM
directly annihilates to standard model particles (a case studied in detail in Ref.~\cite{CTA:2020qlo}).
For $\gamma\gamma$ and VIB signals, on the other hand, such a direct comparison is not as easily
possible since these signals are intrinsically suppressed by powers of $\alpha_{\rm em}$.

For comparison, we further include in the figure previous VIB limits obtained by H.E.S.S. \cite{HESS:2013rld}.\footnote{%
\label{foot_VIB_HESS}
Technically, the limit quoted here refers to a specific signal shape model introduced as `BM2' in 
Ref.~\cite{Bringmann:2007nk}, but that spectrum is VIB-dominated and closely resembles 
the signal spectrum we compare to here, cf.~Fig.~\ref{fig:signalshapes}, after convoluting with the instrument's energy 
resolution.
We obtain the limits shown in the figure by first converting the flux limits reported in Ref.~\cite{HESS:2013rld} 
to limits on $\langle\sigma v\rangle$, cf.~Eq.~(\ref{DMflux}). We then correct for the different assumptions about the
DM distribution by rescaling the result with the ratio of $J$-factors (computed for their RoI, and for the density
profile adopted in their and  in our analysis, respectively).
}
We are not aware of corresponding published limits for box-like spectra (but see Ref.~\cite{Ibarra:2015tya} 
for an earlier CTA sensitivity estimate).
Let us finally briefly comment on a significant theoretical activity in modelling the exact shape of the spectral endpoint feature 
for $\chi\chi\to\gamma\gamma$ annihilations, after taking into account radiative 
corrections~\cite{Ciafaloni:2010ti,Ovanesyan:2014fwa,Baumgart:2017nsr,Baumgart:2018yed,Beneke:2018ssm,
Beneke:2019vhz,Beneke:2019gtg, Bauer:2020jay,Beneke:2022eci}.
Since these corrections are necessarily model-dependent, at least to some extent, a detailed discussion is clearly beyond
the scope of this work. However, let us remark that the deviations from a monochomatic line are typically significantly
less pronounced than the case of the narrow box shown with a green solid line in Fig.~\ref{fig:signalshapes}. 
To a very good accuracy,
one can therefore obtain limits on such `generalized line signals' by simply convolving a given spectrum with the CTAO 
energy resolution, i.e.~a Gaussian of width $\sigma_{\rm res}$, and then rescaling our limits for $\gamma\gamma$ by the 
ratio of the resulting peak height to that for a monochromatic line, $2\times(2\pi \sigma_{\rm res}^2)^{-1/2}$. 
We expect the uncertainty associated  with this method to be less than the difference between  the solid and 
dotted green lines in Fig.~\ref{fig:limitDMshapes} -- and thus significantly less than the statistical uncertainty in the limit
prediction itself.

\section{Discussion}
\label{section:discussion}

In this section we explore the robustness of the results presented in \SEC\ref{section:results},
by studying how the individual benchmark assumptions that we made, cf.~\TAB\ref{table:benchmark},  
impact our final DM limits.
We focus here on our main target, the Galactic Centre, and the most decisive aspects with respect
to sensitivity projections for this target, namely the assumed DM density distribution (\ref{sec:profile_discussion}), 
the RoI masking (\ref{sec_mask:discussion}), the interstellar  emission modelling (\ref{sec_IEM_discussion}),  
and systematic uncertainty choices (\ref{sec:syst_discussion}).
In the Appendix, we further complement this by exploring the impact of the analysis window size
(\ref{app:energy-window}) as well as the RoI size and shape (\ref{app:roi}).

\subsection{Dark matter profiles} 
\label{sec:profile_discussion}

\begin{figure}[t]
\centering
  \includegraphics[width=0.8\textwidth]{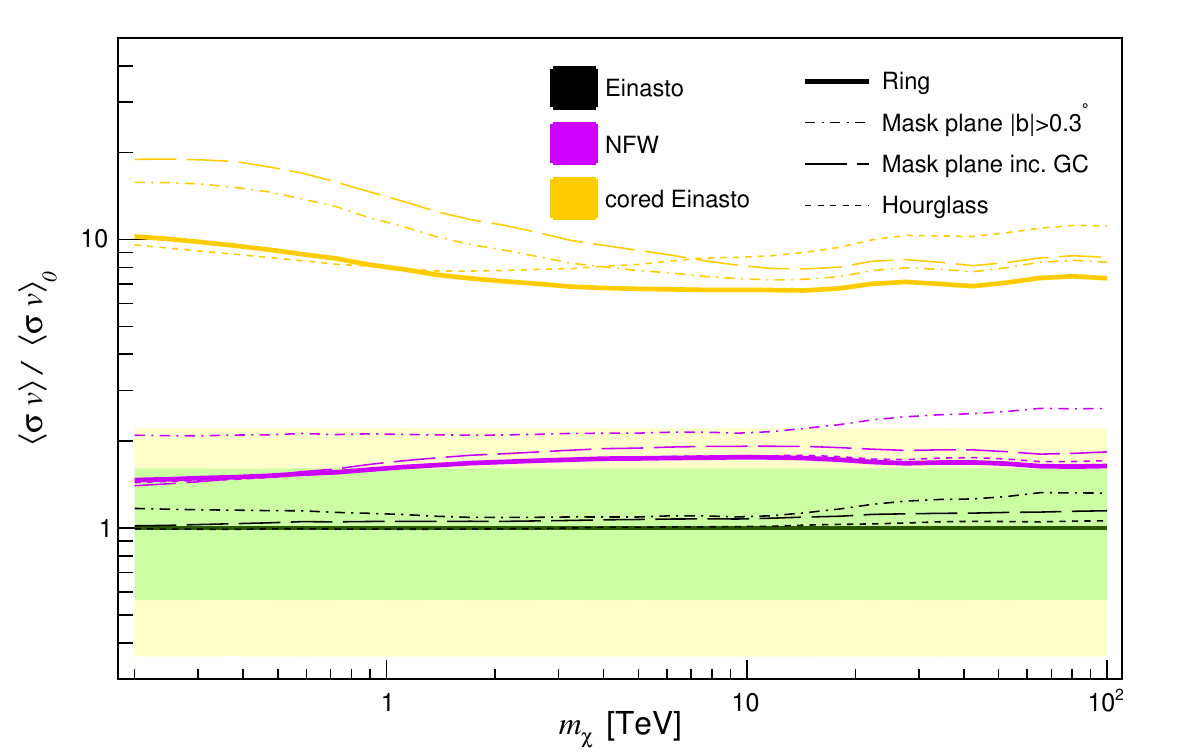}
\caption{
Impact of varying the DM profile (solid lines) and masking (other line styles) of the GC RoI on the
CTA DM sensitivity,
expressed as $95\%~C.L.$ exclusion limits normalized to the benchmark result (displayed as black solid line
both in \FIG\ref{fig:limits_main} and here). The light green and yellow bands show the $1\sigma$ and
$2\sigma$ variance of the expected limit for these benchmark settings (summarized in \TAB\ref{table:benchmark}).
All black lines refer to an Einasto profile, while magenta and orange lines
show the situations for an NFW and a cored Einasto profile, respectively, with profile parameters as defined in
\SEC\ref{section:galactic_centre}. 
Non-solid line styles correspond as indicated to different ways of masking the RoI, illustrated in \FIG\ref{fig:roimasks}.
}
\label{fig:limits-dmprofile-ratio}
\end{figure}

As described in \SEC\ref{section:galactic_centre}, 
the DM density profile is poorly constrained 
observationally in the inner region of  our galaxy, in particular within the inner $\sim0.7$\,kpc relevant for the 
RoI of our analysis. Motivated by high-performance N-body simulations, we chose the commonly
used Einasto profile as a benchmark assumption for the density profile. In \FIG\ref{fig:limits-dmprofile-ratio} we 
quantify how the sensitivity of CTA to a monochromatic DM signal worsens  in case the DM distribution follows 
instead the NFW profile (solid magenta line) or an Einasto profile with a core size of 1\,kpc (solid orange line). 
We find that the sensitivity is affected by less than a factor of 2 in the case of the NFW profile, well within the 
statistical spread of the expected $95\%\,C.L.$ limit that CTA will achieve. For a cored profile, on the other
hand, our sensitivity prediction would worsen by up to one order of magnitude. This loss of sensitivity is by far 
dominated by a corresponding decrease in the total $J$-factor, cf.~\TAB\ref{table:jfrac}, as is expected for an 
analysis comparing components with very different spectral shapes. Unlike in the case of a continuum 
signal~\cite{CTA:2020qlo}, in other words, the fact that the largely isotropic DM signal becomes 
morphologically degenerate  with the bright CR  background is much less important.

Incidentally, this observation also implies that it is straight-forward to translate the projected limits shown in 
\FIG\ref{fig:limits_main}, to a very reasonable accuracy, to the case of DM {\it decaying} via 
$\chi\to\gamma\gamma$. In this case one just has to replace 
$\frac12 \left\langle\sigma v\right\rangle (\rho_\chi/m_\chi)^2 \to  \Gamma \rho_\chi$ in Eq.~(\ref{DMflux}),
where $\Gamma$ is the decay rate for this channel. A limit of 
$\left\langle\sigma v\right\rangle<\left\langle\sigma v\right\rangle_{\rm max}$, therefore, 
is equivalent to a minimal lifetime of $\tau_{\chi\to\gamma\gamma}>
2m_\chi \left\langle\sigma v\right\rangle_{\rm max}^{-1} D/J$, where the `$D$-factor' 
$D\equiv \int_{\Delta\Omega} d\Omega\int\!\!d\ell\,\rho_\chi$ for decaying DM is defined
in analogy to the $J$-factor for annihilating DM. Note that this lifetime constraint applies to a DM particle with 
mass $2\,m_\chi$, i.e.~{\it twice}
the original mass.

\subsection{Region of interest}
\label{sec_mask:discussion}

\begin{figure}
\centering
	\includegraphics[trim=40 25 80 60,clip,width=0.85\textwidth]{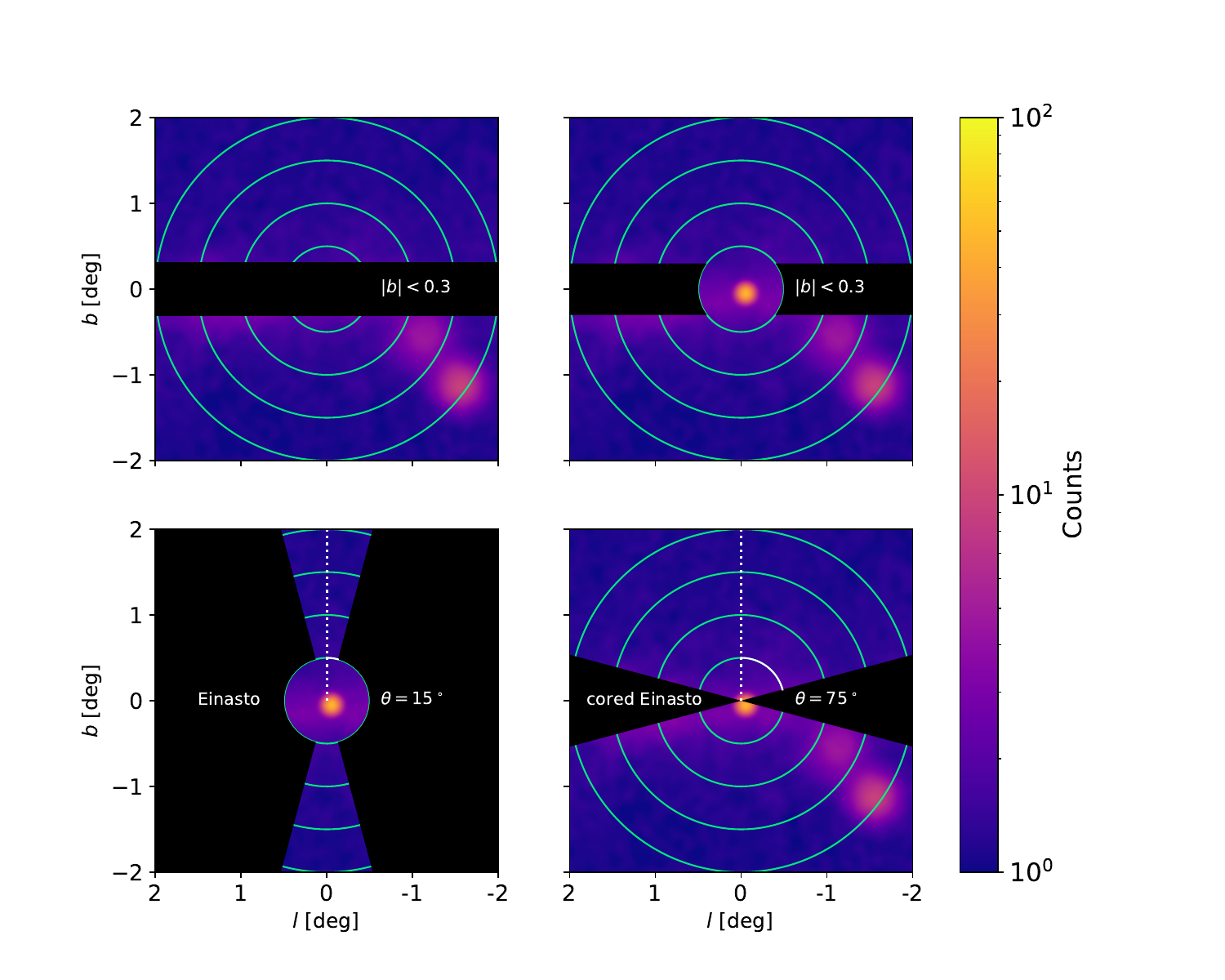}
\caption{Illustration of various masks of the RoI that we tested in our analysis. 
In Galactic coordinates, each of the four panels shows the 
region $(b,l)=(-2^\circ..2^\circ,-2^\circ..2^\circ)$.
The color scheme reflects the counts 
in the energy range $[1.55,2.51]$ TeV per pixel of size $0.01^\circ\times0.01^\circ$ with a Gaussian smoothing of $0.05^\circ$.
\textit{Top left:} Masking the Galactic plane.
\textit{Top right:} Masking the Galactic plane while including the inner central region.
\textit{Bottom left:} Hourglas shape, with opening angle S/N-optimized for an Einasto (or NFW) profile.
\textit{Bottom right:} Hourglas shape, with opening angle S/N-optimized for a cored profile. 
}
\label{fig:roimasks}
\end{figure}

While our benchmark analysis strategy includes the full RoI, a disc of radius $2^\circ$ centred on the GC, 
it is reasonable to ask wether increasing the RoI or masking regions with low signal-to-noise ratio (S/R), 
i.e.~bright backgrounds, 
could improve the sensitivity. As we discuss in more detail in Appendix \ref{app:roi}, increasing the RoI beyond 
$2^\circ$ would in fact hardly improve the sensitivity, but potentially lead to larger systematic uncertainties related
to the background modelling.
The more general question of optimizing the shape of the analysis region was studied in detail 
before, e.g.~Refs.~\cite{Serpico:2008ga,Bringmann:2012vr},  typically  resulting in the conclusion that 
analysis regions with hourglass-like shapes tend to provide maximal S/N. 
In the bottom panel of \FIG\ref{fig:roimasks} we show two such hourglass shapes for illustration, characterised by a 
parameter $\theta$ that describes the opening angle of the analysis region. In the bottom left panel,
the value of  $\theta=15^\circ$ is motivated by typical results from optimizing S/N for a cuspy profile (NFW or Einasto),
though we note in this case S/N does in fact not very strongly depend on  $\theta$; 
in the bottom right panel, $\theta=75^\circ$ is a more typical value that optimizes S/N for a cored profile. 
We indicate the impact of such a masking on our benchmark sensitivities with dotted lines in 
\FIG\ref{fig:limits-dmprofile-ratio}. 

An alternative to simply maximizing S/N is to chose a mask that aims at making one of our main analysis 
assumptions as realistic as possible, namely  that the background emission  can  be approximated by a 
power law in a narrow energy range. As  the Galactic plane is expected to contain a significant number of 
(subthreshold) sources that could affect the 
validity of this assumption, we thus consider a mask that fully covers the plane, $|b|<0.3^\circ$, 
as depicted in the top left panel of \FIG\ref{fig:roimasks}. The (very limited) impact of such a mask on the sensitivities is 
indicated with dash-dotted lines in \FIG\ref{fig:limits-dmprofile-ratio}. Finally, we also consider the option of 
masking the Galactic plane but including the GC in the analysis, cf.~the top right panel of 
\FIG\ref{fig:roimasks}, and show the impact on the DM sensitivity with dashed lines in 
\FIG\ref{fig:limits-dmprofile-ratio}. 

We observe that our sensitivities are largely robust to masking schemes, worsening by factors of at most two in extreme cases
due to the loss in photon statistics (which, in turn, is directly proportional to a corresponding reduction of the effective 
$J$-factor). This implies that line limits eventually derived from real data will also be very robust, 
only mildly affected by even very aggressive cuts in the analysis region in order to minimize the impact of underlying
modelling uncertainties.

\subsection{Background model dependence}
\label{sec_IEM_discussion}

\begin{figure}[t]
\centering
\includegraphics[width=0.8\linewidth]{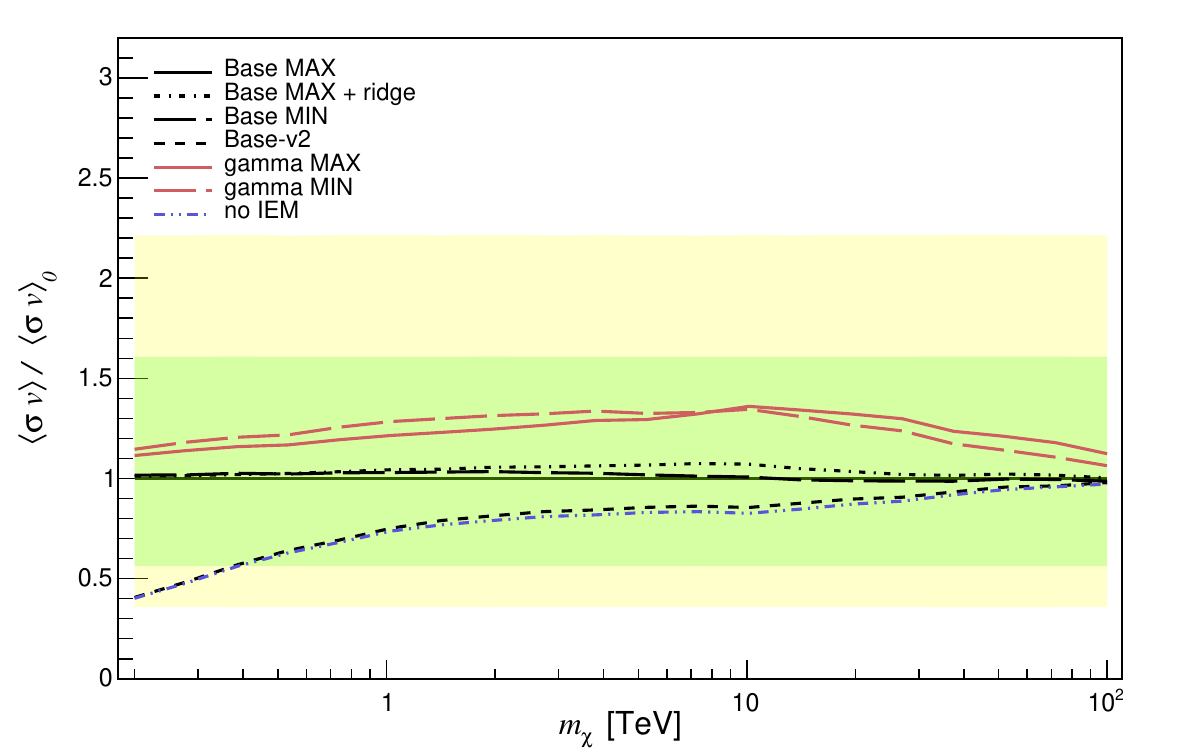}
\caption{Impact of varying the IEM model on the CTA DM sensitivity,
expressed as $95\%~C.L.$ exclusion limits normalized to the benchmark result (displayed as black solid line
both in \FIG\ref{fig:limits_main} and here). The different model setups are described in detail in \SEC\ref{section:galactic_centre}, 
with line styles as indicated in the legend. The light green and yellow bands show the $1\sigma$ and
$2\sigma$ variance of the expected limit for the benchmark settings (summarized in 
\TAB\ref{table:benchmark}).
 \label{fig:IEMs}}
\end{figure}

Modelling of  the interstellar emission is highly uncertain in the  Galactic plane, given presently available data,
and even more so in the inner region of  the 
Galactic Center. Thus, the question arises of how this affects the sensitivity predictions derived here.
As discussed in \SEC\ref{section:galactic_centre} 
we choose the Base MAX model as our 
benchmark analysis setting. In \FIG\ref{fig:IEMs} we explore how the predicted limits would change should a 
different model turn out to better describe the real data.

We observe that the difference with the Base MIN model is negligible, while in  the case of gamma models the 
sensitivity could worsen by up to $50\%$. In the case of the conservative IEM used in Ref.~\cite{Remy:2021cff},
dubbed Base-v2, the sensitivities would instead improve by up to $50\%$, especially at low energies. 
Note that this exercise optimistically assumes a perfect model for the emission which, however,
should not qualitatively affect our conclusions. In particular, the expected 
impact on the limits is of a similar order as the expected variation of the central limit prediction at the 
$1\sigma$ level, and hence not very significant.

The rather limited dependence of our results on the exact implementation of background modelling
is, in fact, one of the expected features of our analysis method. As long as the background does not itself 
contain sharp spectral features, the identification of these types of DM signals will remain relatively robust. In 
particular, the limit (or signal) significance will to a large degree only
be affected at the level of the noise contribution, i.e.~the overall background normalization. This is in contrast 
to other template-based analyses; see, e.g., Ref.~\cite{CTA:2020qlo} for a discussion of how the 
background modelling impacts the search for a DM signal with a smooth spectrum.

\subsection{Impact of instrumental systematics}
\label{sec:syst_discussion}

A realistic analysis will always be affected by some level of systematic uncertainty. 
This could have an instrumental origin, 
e.g.~related to event reconstruction or misclassification, or stem from modelling uncertainties. 
In this subsection we approach this issue in a general way and explore the impact of systematic uncertainty
following the parametric approach introduced in \SEC\ref{section:systematics}.

\begin{figure}[t]
\centering
\includegraphics[width=0.8\linewidth]{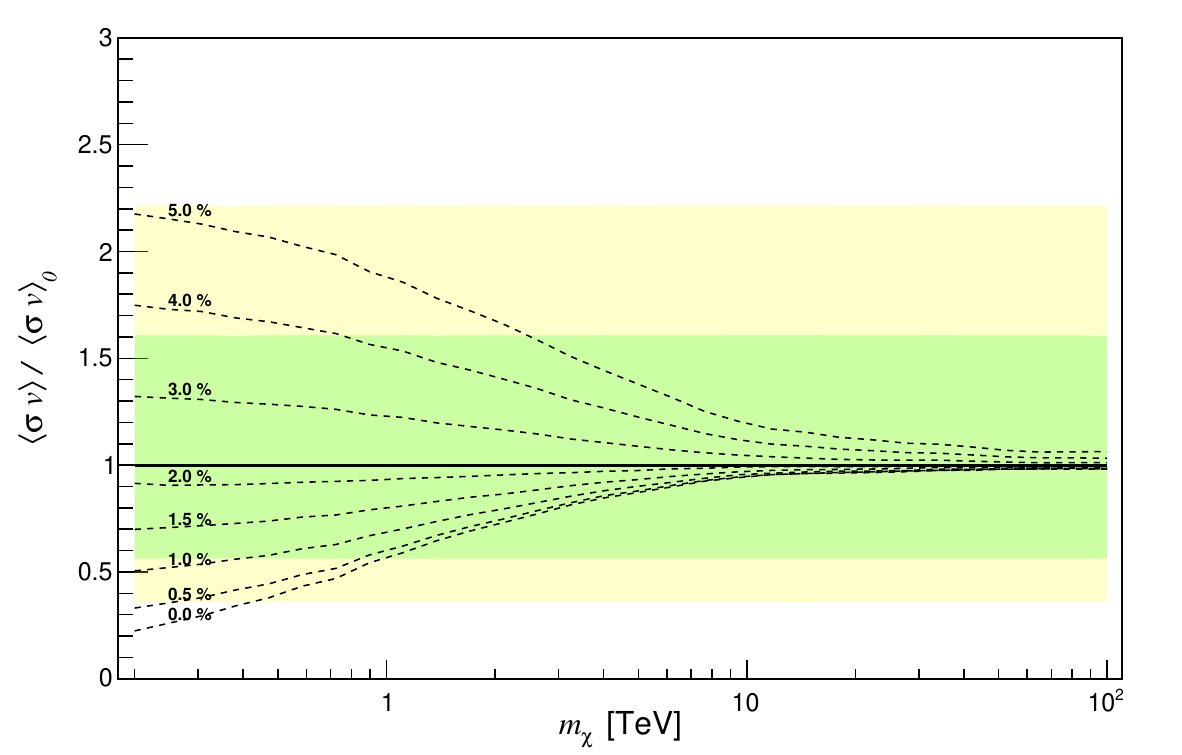}
\caption{%
Impact of varying the level of systematic uncertainty, cf.~Eq.~(\ref{eq:defsigma}), 
on the CTA sensitivity to a monochromatic DM signal,
expressed as $95\%~C.L.$ exclusion limits normalized to the benchmark result (displayed as black solid line
both in \FIG\ref{fig:limits_main} and here). The light green and yellow bands show the $1\sigma$ and
$2\sigma$ variance of the expected limit for the benchmark settings (summarized in \TAB\ref{table:benchmark}).
\label{figure:limit-ratio-systematic} }
\end{figure}

In \FIG\ref{figure:limit-ratio-systematic}, we show the effect of varying the overall normalization of the
covariance matrix, Eq.~(\ref{eq:defsigma}), which we refer to as `the systematic uncertainty' $\sigma$.
As expected, the figure demonstrates that systematic uncertainties only impact the limits at low
energies, where the total photon count is large. Increasing the systematic uncertainty from our 
benchmark value of 2.5\% to 5\%, for example, worsens the limits by up to a factor of 2.2 
(for a DM mass of $m_\chi=200$\,GeV). Not taking into account the effect of systematic uncertainties
at all, on the other hand, would result in limits that are too optimistic by up to a factor of 4.
Let us briefly mention that \FIG\ref{figure:limit-ratio-systematic} illustrates the effect of varying $\sigma$ for our 
benchmark analysis method of modelling the entire non-signal photon count as a power law; implementing 
instead the more aggressive modelling based on a power law of the {\it fluxes}, cf.~\SEC\ref{sec:sliding},
results in quantitatively almost identical results.

Our results are thus considerably less sensitive to instrumental uncertainties than what is 
familiar from generic DM spectra with a broader shape~\cite{CTA:2020qlo}. Again, the reason is that we chose
an analysis method that is very efficient in singling out sharp spectral features from an otherwise
feature-less `background'. Incidentally, this is also the explanation for 
why the sidebands in \FIG\ref{fig:limits_main} have a constant width, almost independent of the DM 
mass (as even more clearly visible in Figs.~\ref{fig:IEMs} and \ref{figure:limit-ratio-systematic});
we checked that this starts to change when increasing the level of systematic uncertainty to unrealistically
large values $\sigma\gg5$\%, leading to a broadening of the sidebands at small masses. 

In summary, systematic uncertainties dominate the overall uncertainties of the projected limits 
up to DM masses of a few TeV, from where statistical uncertainties begin to be more important. 
Even for sub-TeV DM masses, however, a mis-modelling of instrumental effects is not expected
to affect limits by more than a factor of $\sim$2 w.r.t.~to our results presented in \FIG\ref{fig:limits_main}.

\section{Conclusions}
\label{section:conclusion}

The Cherenkov Telescope Array Observatory has a great potential to probe thermally produced 
DM at TeV energies~\cite{CTA:2020qlo}. 
Due to  its superior energy resolution compared to current 
gamma-ray facilities, furthermore, it is expected to be the most sensitive instrument to identify possible sharp 
features in gamma-ray spectra at these energies, like monochromatic `line' signals. 
The detection of such features would provide smoking-gun evidence
for the decay or annihilation of DM particles, and may in fact reveal decisive information about the underlying microphysical 
model that describes these particles.

In this article
we have presented a detailed study to estimate the expected sensitivity of CTA to such distinct spectral features,
using up-to-date observational strategies and the latest IRFs taking into account updated telescope 
configurations. In particular, our analysis 
is based on a 500\,hr extensive Galactic Centre survey and 600\,hrs 
of dSph galaxy observations. For the latter, we follow an accompanying CTA consortium paper~\cite{CTAdSphsInPrep}
 focusing on DM in dSphs and work under the assumption that these 600\,hr of total observation time 
will be split evenly among six dSphs, three per hemisphere.  
We explore the commonly considered case of a monochromatic signal, originating for example from the loop-suppressed 
annihilations of a pair of DM particles into two photons, but also the possibility of more general spectral shapes that would constitute 
clear evidence for a DM signal. For the latter, we use two generic spectral templates that may originate from three-particle
final states including a single photon (`virtual internal bremsstrahlung') and the annihilation of DM into subsequently decaying
mediator particles (`box'-like spectra), respectively.

In addition to using the latest information related to the CTAO instrument and observational plans, as well as a rather broad focus 
on spectral features beyond the  commonly performed `line' searches, 
this work improves upon  previous CTA sensitivity projections in the following:

\begin{itemize}

\item We use state-of-the-art models of the astrophysical gamma-ray background in  the GC that include updated interstellar 
emission models and three known point sources. 

\item We perform a range of optimization studies and carefully explore various types of systematic uncertainties. In particular, we 
investigate the impact of various DM density profiles, regions of interest and masking, 
and of background and 
instrumental systematics. For the latter, we explicitly add an overall $2.5\%$ systematic uncertainty, 
on top of taking into account correlations in the expected instrumental uncertainties. 

\item We assess two main variants of the commonly adopted sliding energy window analysis technique to identify sharp spectral
features: {\it i)} locally modelling the total simulated count rate for the (instrumental and astrophysical) background as a simple power law, 
and {\it ii)}  an alternative method in which the 
astrophysical and instrumental background components are separated, noting that information about the latter is already 
contained in the IRFs; in  this case the power law is fit to the gamma-ray {\it flux} that  is then convoluted with the IRFs and 
added to the simulated CR counts. 
The former method is more conservative and constitutes our default analysis procedure. 
The  second approach improves DM sensitivity but depends on the IRF model -- and serves to
illustrate the potential gain in sensitivity that one may eventually hope for with real data and an exquisite understanding of the instrument in full operation mode.

\end{itemize}

Our main results are shown in Fig.~\ref{fig:limits_main}. In particular, the CTA sensitivity to spectral line features is expected to
improve upon current limits from ground-based experiments (notably H.E.S.S.) by a factor of $\sim 2$ at 1\,TeV, and by up to 
one order of magnitude in the multi-TeV range, which to a large degree is an effect of increased exposure.  
At high energies, even a 5\,$\sigma$ discovery of a signal is conceivable,
for DM annihilation cross sections just below current limits.
It should be stressed that a fiducial `Omega' configuration of CTAO 
would allow a further improvement of the limits by a factor of $\sim$\,2.
As discussed in Sec.~\ref{section:results_galactic_centre}, both constraining and discovery potential of CTA have profound 
implications for particle models of DM at the TeV scale, and we therefore also provide the full binned TS, 
cf.~Fig.~\ref{fig:likelihood-scan}, to consistently include the CTA sensitivity to monochromatic DM signals in, e.g., global scans
of the underlying parameter space of such models. 

We generally find that prospects to detect line-like DM signals in dSphs are significantly suppressed w.r.t.~what can
be achieved with GC observations (Fig.~\ref{fig:limits_dsph}). On the other hand, one should keep in mind that these targets
are very robust as far as modelling of the astrophysical background is concerned. As discussed in
Sec.~\ref{sec:profile_discussion},
they may therefore still constitute relevant complementary targets to detect monochromatic or similarly sharp spectral features in 
case the concentration of the DM density close to the GC turns out to be very unfavourable.
Projected limits on spectral features beyond the simplest possibility of a monochromatic line, cf.~Fig.~\ref{fig:limitDMshapes},
also appear very promising. In particular, the sensitivity to VIB-signals will improve in accordance with
what is expected for line signals; 
for scenarios where the dominant DM annihilation channel is into a pair of mediator particles, 
furthermore, CTA may even be able to test the thermal production of DM for masses up to around 50\,TeV.

In summary, this study complements Ref.~\cite{CTA:2020qlo}  on the CTA sensitivity to generic 
DM signals from the GC region in two important ways: by focussing on possible DM annihilation channels that
stress the discovery rather than the constraining power of the instrument and, related, by adopting a very different
analysis strategy that is specifically tailored to identify spectral (as opposed to  spatial) features.
The exciting combined message from these two works is that CTA is {\it guaranteed} to close significant parameter 
space of thermally produced DM and that, at the same time, a truly groundbreaking discovery remains in fact a 
fully viable {\it possibility}.

\vspace*{1cm}
\section*{Acknowledgments}

We gratefully acknowledge financial support from the following agencies and organizations:

\smallskip

State Committee of Science of Armenia, Armenia;
The Australian Research Council, Astronomy Australia Ltd, The University of Adelaide, Australian National University, Monash University, The University of New South Wales, The University of Sydney, Western Sydney University, Australia; Federal Ministry of Education, Science and Research, and Innsbruck University, Austria;
Conselho Nacional de Desenvolvimento Cient\'{\i}fico e Tecnol\'{o}gico (CNPq), Funda\c{c}\~{a}o de Amparo \`{a} Pesquisa do Estado do Rio de Janeiro (FAPERJ), Funda\c{c}\~{a}o de Amparo \`{a} Pesquisa do Estado de S\~{a}o Paulo (FAPESP), Funda\c{c}\~{a}o de Apoio \`{a} Ci\^encia, Tecnologia e Inova\c{c}\~{a}o do Paran\'a - Funda\c{c}\~{a}o Arauc\'aria, Ministry of Science, Technology, Innovations and Communications (MCTIC), Brasil;
Ministry of Education and Science, National RI Roadmap Project DO1-153/28.08.2018, Bulgaria; 
The Natural Sciences and Engineering Research Council of Canada and the Canadian Space Agency, Canada; 
ANID PIA/APOYO AFB230003, ANID-Chile Basal grant FB 210003, N\'ucleo Milenio TITANs (NCN19-058), FONDECYT-Chile grants 1201582, 1210131, 1230345, and 1240904; 
Croatian Science Foundation, Rudjer Boskovic Institute, University of Osijek, University of Rijeka, University of Split, Faculty of Electrical Engineering, Mechanical Engineering and Naval Architecture, University of Zagreb, Faculty of Electrical Engineering and Computing, Croatia;
Ministry of Education, Youth and Sports, MEYS  LM2018105, LM2023047, EU/MEYS CZ.02.1.01/0.0/0.0/16\_013/0001403, CZ.02.1.01/0.0/0.0/18\_046/0016007,  CZ.02.1.01/0.0/0.0/\allowbreak16\_019/0000754 and CZ.02.01.01/00/22\_008/0004632, Czech Republic; 
Academy of Finland (grant nr.317636 and 320045), Finland;
Ministry of Higher Education and Research, CNRS-INSU and CNRS-IN2P3, CEA-Irfu, ANR, Regional Council Ile de France, Labex ENIGMASS, OCEVU, OSUG2020 and P2IO, France; 
The German Ministry for Education and Research (BMBF), the Max Planck Society, the German Research Foundation (DFG, with Collaborative Research Centres 876 \& 1491), and the Helmholtz Association, Germany; 
Department of Atomic Energy, Department of Science and Technology, India; 
Istituto Nazionale di Astrofisica (INAF), Istituto Nazionale di Fisica Nucleare (INFN), MIUR, Istituto Nazionale di Astrofisica (INAF-OABRERA) Grant Fondazione Cariplo/Regione Lombardia ID 2014-1980/RST\_ERC, Italy; 
ICRR, University of Tokyo, JSPS, MEXT, Japan; 
Netherlands Research School for Astronomy (NOVA), Netherlands Organization for Scientific Research (NWO), Netherlands; 
University of Oslo, Norway; 
Ministry of Science and Higher Education, DIR/WK/2017/12, the National Centre for Research and Development and the National Science Centre, UMO-2016/22/M/ST9/00583, Poland; 
Slovenian Research Agency, grants P1-0031, P1-0385, I0-0033, J1-9146, J1-1700, N1-0111, and the Young Researcher program, Slovenia; 
South African Department of Science and Technology and National Research Foundation through the South African Gamma-Ray Astronomy Programme, South Africa; 
The Spanish groups acknowledge funds from "ERDF A way of making Europe" and the Spanish Ministry of Science and Innovation and the Spanish Research State Agency (AEI) via MCIN/AEI/10.13039/501100011033 through government budget lines PGE2021/28.06.000X.\allowbreak411.01, PGE2022/28.06.000X.411.01, PGE2022/28.06.000X.711.04, and grants PID2022-13\allowbreak7810NB-C22, PID2022-136828NB-C42, PID2022-139117NB-C42, PID2022-139117NB-C41, PID2022-136828NB-C41, PID2022-138172NB-C43, PID2022-138172NB-C42, PID2022-13911\allowbreak7NB-C44, PID2021-124581OB-I00, PID2021-125331NB-I00, PID2019-104114RB-C31,  PID\allowbreak2019-107847RB-C44, PID2019-104114RB-C32, PID2019-105510GB-C31, PID2019-104114RB-C33, PID2019-107847RB-C41, PID2019-107847RB-C43, PID2019-107847RB-C42; the "Centro de Excelencia Severo Ochoa" program through grants no. CEX2019-000920-S, CEX2020-001007-S, CEX2021-001131-S; the "Unidad de Excelencia Mar\'ia de Maeztu" program through grants no. CEX2019-000918-M, CEX2020-001058-M; the "Ram\'on y Cajal" program through grants RYC2021-032552-I, RYC2021-032991-I, RYC2020-028639-I and RYC-2017-22665; and the "Juan de la Cierva" program through grants no. IJC2019-040315-I and JDC2022-049705-I. La Caixa Banking Foundation is also acknowledged, grant no. LCF/BQ/PI21/11830030. They also acknowledge the project "Tecnolog\'ias avanzadas para la exploraci\'on del universo y sus componentes" (PR47/21 TAU), funded by Comunidad de Madrid regional government. Funds were also granted by the Junta de Andaluc\'ia regional government under the "Plan Complementario de I+D+I" (Ref. AST22\_00001) and "Plan Andaluz de Investigaci\'on, Desarrollo e Innovaci\'on" (Ref. FQM-322); by the "Programa Operativo de Crecimiento Inteligente" FEDER 2014-2020 (Ref.~ESFRI-2017-IAC-12) and Spanish Ministry of Science and Innovation, 15\% co-financed by "Consejer\'ia de Econom\'ia, Industria, Comercio y Conocimiento" of the Gobierno de Canarias regional government. The Generalitat de Catalunya regional government is also gratefully acknowledged via its "CERCA'' program and grants 2021SGR00426 and 2021SGR00679. Spanish groups were also kindly supported by European Union funds via the "Horizon 2020" program, grant no. GA:824064, and NextGenerationEU, grants no. PRTR-C17.I1, CT19/23-INVM-109, and "Mar\'ia Zambrano" program, BDNS: 572725. This research used computing and storage resources provided by the Port d'Informaci\'o Cient\'ifica (PIC) data center; 
Swedish Research Council, Royal Physiographic Society of Lund, Royal Swedish Academy of Sciences, The Swedish National Infrastructure for Computing (SNIC) at Lunarc (Lund), Sweden; 
State Secretariat for Education, Research and Innovation (SERI) and Swiss National Science Foundation (SNSF), Switzerland; Durham University, Leverhulme Trust, Liverpool University, University of Leicester, University of Oxford, Royal Society, Science and Technology Facilities Council, UK; 
U.S. National Science Foundation, U.S. Department of Energy, Argonne National Laboratory, Barnard College, University of California, University of Chicago, Columbia University, Georgia Institute of Technology, Institute for Nuclear and Particle Astrophysics (INPAC-MRPI program), Iowa State University, the Smithsonian Institution, V.V.D. is funded by NSF grant AST-1911061, Washington University McDonnell Center for the Space Sciences, The University of Wisconsin and the Wisconsin Alumni Research Foundation, USA.

\smallskip

The research leading to these results has received funding from the European Union's Seventh Framework Programme (FP7/2007-2013) under grant agreements No~262053 and No~317446.
This project is receiving funding from the European Union's Horizon 2020 research and innovation programs under agreement No~676134.

\appendix
 \newpage
 \section{Analysis details}
\label{app}

\subsection{The width of the sliding energy window}
\label{app:energy-window}

\begin{figure}
\centering
\includegraphics[width=0.8\linewidth]{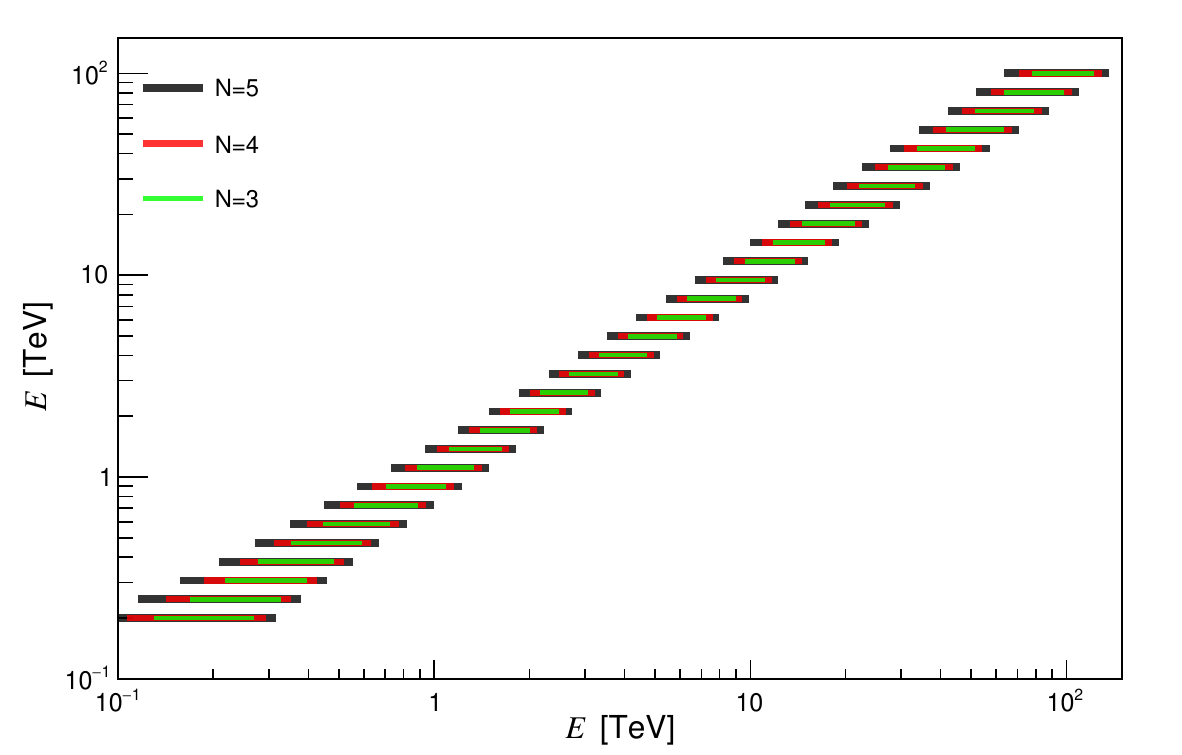}
\caption{Energy window sizes $\Delta = 2N\times \sigma_\mathrm{res}(E_0)$ for $N=3,4,5$, where 
$\sigma_\mathrm{res}(E_0)$ is the energy resolution shown in Fig.~\ref{fig:cta}. The benchmark setting
adopted in our analysis is given by $N=4$.}
\label{fig:energywindow}
\end{figure}

For the analysis presented in the main part of this article we adopted a sliding energy window with a width
of $\Delta = 8\sigma_\mathrm{res}(E_0)$ around a signal centered at $E_0$, where $\sigma_\mathrm{res}(E_0)$
is the energy resolution of CTAO at that energy (as depicted in Fig.~\ref{fig:cta}). Fig.~\ref{fig:energywindow} 
illustrates this choice, along with the effect of increasing or decreasing $\Delta$ with respect to the energy
resolution. 

\begin{figure}
\centering
\includegraphics[width=0.495\linewidth]{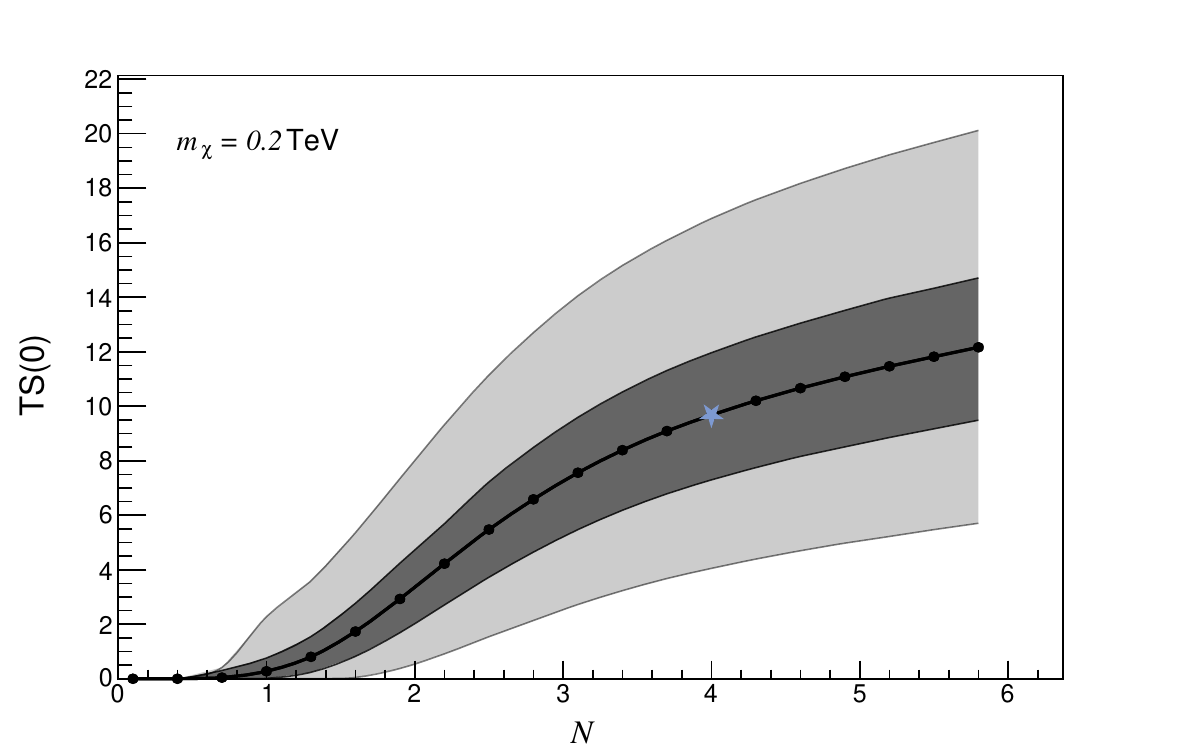}
\includegraphics[width=0.495\linewidth]{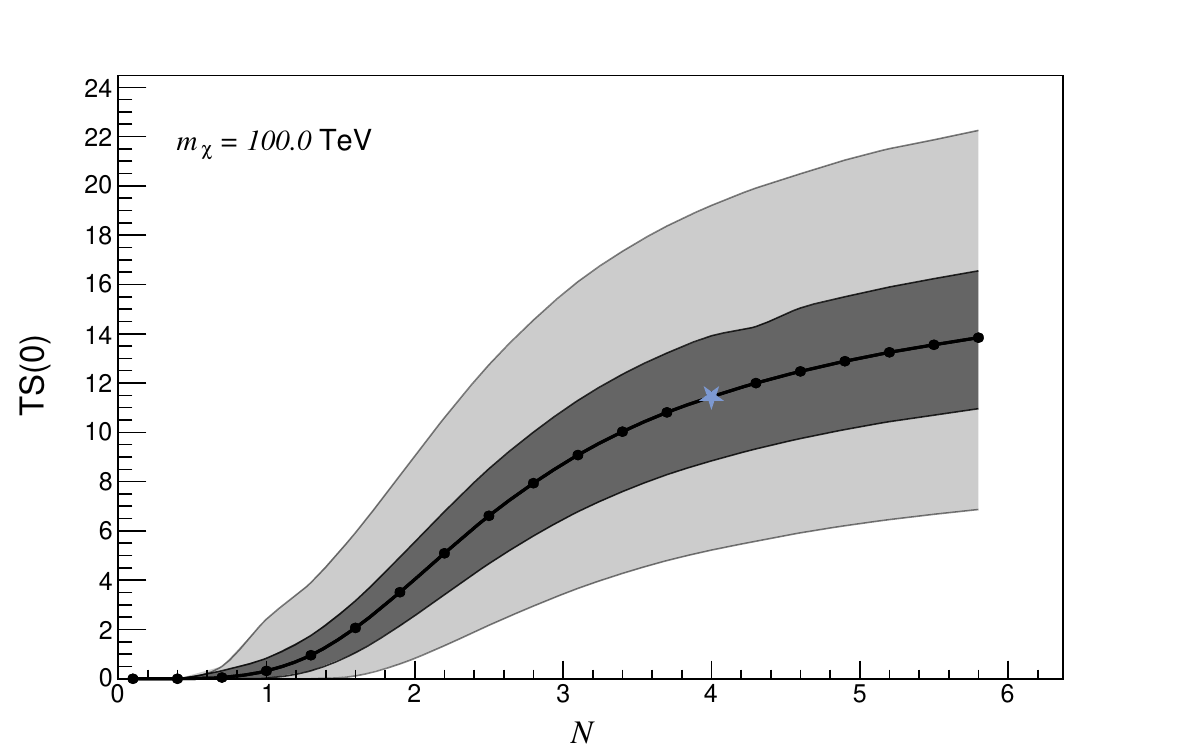}
\caption{%
Effect of varying the sliding energy window size, cf.~Fig.~\ref{fig:energywindow}, on the
test statistic under the background-only hypothesis, but with an (arbitrarily normalized) 
monochromatic DM signal present in the data. The left and right panel show the case of a monochromatic signal
at $E_0=0.2$\,TeV and $E_0=100$\,TeV, respectively. The star at $N=4$ indicates the value adopted in our analysis. 
For $N\gtrsim4$ the gradient becomes less steep, such that the increased statistical power in identifying the
signal no longer outweighs the likewise increasing uncertainty connected to the underlying background modelling
(see text for further discussion).
}
\label{fig:tswindows}
\end{figure}

In this appendix we address the question of how to optimize the sliding energy window size 
for the purpose of our  analysis, i.e.~how to chose $N$ in
\be
\Delta = 2N\times \sigma_\mathrm{res}(E_0)\,.
\label{eq:energy-window}
\ee
It is clear that the identification of a sharp spectral feature {\it and} a power-law background at lower
and higher energies will fail if the analysis window is too small compared to the energy resolution,
i.e.~for $N\lesssim1$. In fact, one should expect that the determination of the background 
power law will monotonically improve as $N$ is increased, and as a result the determination of the exact 
signal normalization should improve as well. In Fig.~\ref{fig:tswindows} we confirm this expectation by plotting the test statistics
under the background-only hypothesis, but with a monochromatic DM signal present in the data; 
for illustration, we choose here a signal at $E_0=0.2$\,TeV ($E_0=100$\,TeV) in the left (right) panel.
Naively, the fact that TS$(0)$ continuously rises with $N$ would then 
suggest that the optimal analysis approach is to formally take the $N\to\infty$ limit, i.e.~to include the
entire energy range observable by CTAO in the analysis.

However, this would not only be computationally unreasonably expensive -- due to a proliferation of
nuisance parameters capturing systematic uncertainties -- but is in fact at odds with the very idea of the 
sliding energy window technique, which is based on the 
assumption of a very simple (power-law) description of the background inside the analysis window. 
The point is that by {\it increasing} $N$ one will always formally increase the statistical power to determine
the model parameters, while by {\it decreasing} $N$ the assumption of a power law will necessarily improve,
thus removing systematic uncertainty and background-model dependence (mathematically speaking, in the limit $N\to0$,
the assumption of a power-law background
becomes exact). Conversely, for an energy range that is too wide, the simplistic assumption
of a power law will result in unsatisfactory background modelling.

Fortunately, inspection of Fig.~\ref{fig:tswindows} reveals a clear transition between two regimes, which we
use as guiding principle for choosing the `optimal' window size: for $N\lesssim4$, the information gain from increasing $N$
is still substantial, while for $N\gtrsim4$ the TS only increases rather modestly. Recalling that the statistical
significance of the derived limit scales roughly as $\sqrt{TS(0)}$, increasing the sliding energy window further
thus hardly affects the limits anymore. 
For  $N\lesssim4$, furthermore, a power law can still be expected to describe 
the actual background very well (see also Appendix \ref{App:Asimov}). 
We find a qualitatively very similar behaviour across the entire range of
observable energies, and that the exact choice of $N\approx4$ has only a minor impact. 
Let us note that similar criteria to determine
the optimal window size have been adopted before, e.g.~in Ref.~\cite{Bringmann:2011ye} in terms of the relative 
change directly in the expected DM limit, rather than a change in TS$(0)$, 
when allowing for generic deviations of some fiducial background model from the
power-law assumption used in the analysis.

\subsection{Choosing the Galactic centre region of interest}
\label{app:roi}

\begin{figure}[t]
\centering
\includegraphics[width=0.8\textwidth]{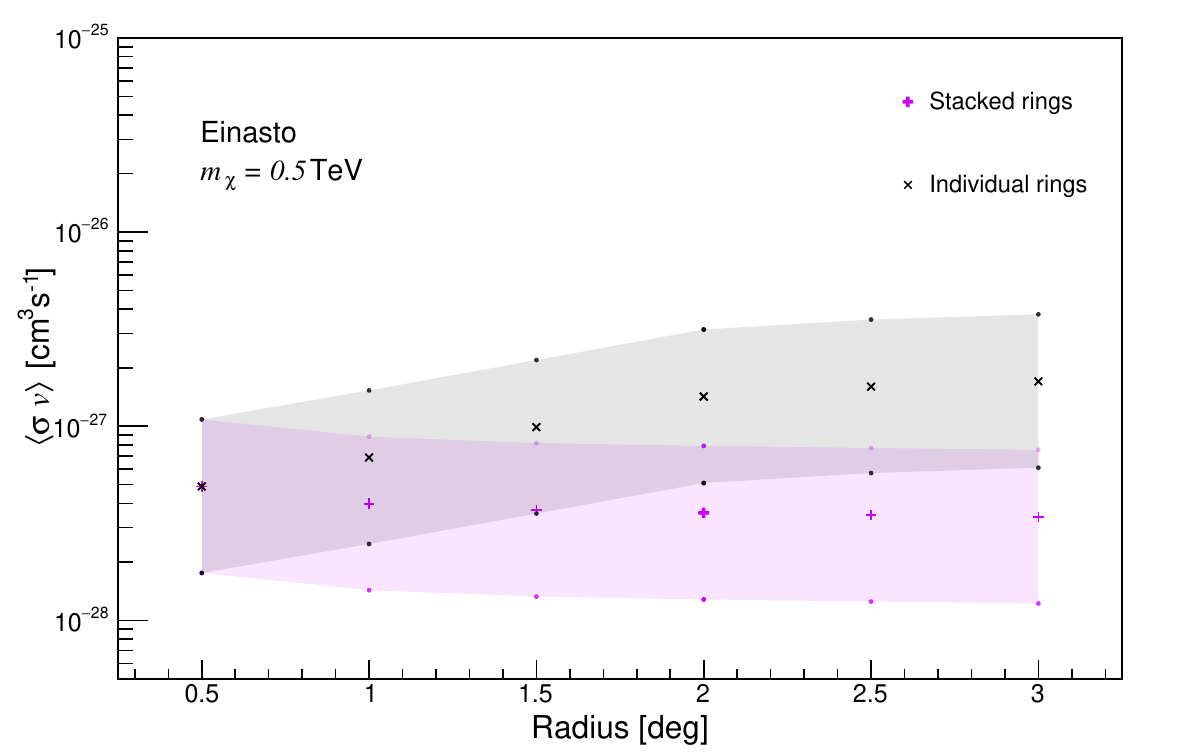}
\caption{Median $95\%\,C.L.$ exclusion limits (black `$\times$') and their $1\,\sigma$ variance (grey shading) 
on the DM annihilation cross section that result from RoIs consisting of individual rings centered on the GC,  with width 
$0.5^\circ$ and outer edge at the stated radius. Purple `$+$' symbols and shadings indicate the corresponding limits
when instead combining all rings up to the stated radius. All limits are based on a DM mass of 
$m_\chi=0.5\mathrm{\ TeV}$ DM and an Einasto profile for the DM distribution.
}
\label{fig:flux_prbin}
\end{figure}

For our analysis, as illustrated in \FIG\ref{fig:background_spatial_bin}, we consider a spherical RoI with radius 
$2^\circ$ centered on the GC. In this appendix we motivate this choice. For this purpose, we show in \FIG\ref{fig:flux_prbin}
how individual rings contribute to the constraining power of the analysis.  As expected, the sensitivity monotonically
increases (purple band) with the total size of the RoI -- but it is also clear that it saturates relatively quickly.
This is because the constraining power from the {\it individual} rings deteriorates when going to larger radii. 
Three effects are responsible for this behaviour: {\it i)} an increased background (or noise) photon count due to the
larger area of outer rings, {\it ii)} a $J$-factor that slightly decreases beyond about $2^\circ$, see also  \TAB\ref{table:jfrac},
and {\it iii)} a decreased CTAO exposure further away from the GC, beyond about $2^\circ$, 
given the adopted observational strategy. 
On the other hand, similar to the discussion in Appendix \ref{app:energy-window}, it is desirable to keep the RoI
small in order to minimize systematic uncertainties in our background modelling. 

\FIG\ref{fig:flux_prbin} is based on an Einasto profile and a DM mass of $m_\chi=0.5$\,TeV, but we note that 
other profiles or DM masses result in a qualitatively very similar behaviour. In particular, increasing the size of the 
RoI beyond 2 degrees hardly impacts the sensitivity in any of the cases we have considered. 
In our analysis, we hence fix for simplicity the GC RoI to $2^\circ$ for all masses and over all energies
(for the case of dSphs we choose $0.5^\circ$, following an analogous reasoning).
We note that already a simple S/N analysis arrives at a similar conclusion, namely that
for GC line searches it is favourable to focus on RoIs with a scale of the order of a few degrees~\cite{Serpico:2008ga}.

Let us finally mention that a corresponding S/N optimization can also be performed for the masking, with S (N) denoted as 
$\mu^\chi$ ($\mu^\chi+\mu^{\rm bg}$) in \SEC\ref{sec:statistics}, resulting 
in the hourglass shapes shown in \FIG\ref{fig:roimasks}. In this case, however, we find that the S/N ratio only has
a rather weak dependence on the opening angle $\theta$, so the exact value of $\theta$ does not affect our analysis
in any appreciable way.

\subsection{Asimov Dataset vs.~Monte Carlo realizations}
\label{App:Asimov}

In this Appendix we derive our analytic estimates of the median and the variance of 
the sensitivity limits, based on evaluating the Asimov data set, and verify these estimates by direct comparison to MC 
simulations.
The defining property of the Asimov data set $A$ is that its best-fit parameter values coincide with 
the true model parameter values realized in nature (or taken as input values for MC simulations).
This implies
\begin{equation}
\mathcal{L}_A(\hat{\nu},\hat{\bm{\theta}}) = \mathcal{L}_A(\nu_{\rm true},\bm{\theta}_{\rm true})\,,
\end{equation}
where $\nu_{\rm true}$ is the {\it true} signal strength and $\bm{\theta}_{\rm true}$ are the values of all nuisance parameters.

In Eq.~(\ref{eq:MLR}) we introduced the standard log-likelihood test statistic $\text{TS}(\nu)$ as a function of the 
{\it hypothesized} signal strength $\nu$, for any given data set. Under the Wald
approximation, TS takes the form of a parabola around the best-fit value.
For a signal strength that is physically constrained to be non-negative (i.e.~$\nu \geq 0$), the likelihood then
takes the form of a truncated Gaussian, with~\cite{cowan:asimov}
\begin{equation}
\text{TS}(\nu)\simeq\Aqnu \equiv
\begin{cases}
 \frac{\nu^2}{\Asigma^2}-\frac{2\hat{\nu}\nu}{\Asigma^2}
 & \nu\geq \hat{\nu} \quad\&\quad \hat{\nu} < 0 \\
 \frac{(\nu-\hat{\nu})^2}{\Asigma^2} 
 & \nu\geq \hat{\nu} \quad\&\quad \hat{\nu} \geq 0 \\
 0
 & \nu < \hat{\nu}
 \end{cases}\,,
 \label{eq:q_wald}
\end{equation}
 where the standard deviation $\sigma_A$ is to a very good accuracy independent of the best-fit value 
 $\hat\nu$. In practice, we can most
 easily extract $\sigma_A$ by evaluating the above equation on an Asimov data set without signal, for which $\hat{\nu}=0$,
 resulting in  
$\Asigma = {\nu}\big({{\Aqnu}^{A,0} }\big)^{-1/2}$
for any given value of $\nu$. We further note that an alternative way of stating \EQ(\ref{eq:q_wald}) is by formally
solving for the assumed signal strength,
\begin{equation}
\Anq = 
\begin{cases}
\hat{\nu} + \sqrt{\hat{\nu}^2 + \bar{\nu}^2}
 & \hat{\nu} < 0\\
\hat{\nu} + \bar{\nu} &\hat{\nu}\geq 0 
 \end{cases},
 \label{eq:Q}
\end{equation}
where we have introduced $ \bar{\nu} \equiv \sqrt{\Aqnu}\Asigma$.

Let us now consider the case with no signal, i.e.~$\nu_{\rm true}=0$. The best-fit value $\hat \nu$
in any given dataset is then still a random variable, distributed according to a normal distribution
{$f^{\hat{\nu}}=\mathcal{N}(0,\Asigma)$} with variance $\Asigma$. 
The value of $\nu$ that produces a given value of 
$\Aqnu$ (e.g.~$\Aqnu=2.71$ for a 95\%\,upper limit) thus also becomes a random variable,
with distribution\footnote{%
One can derive this relation by using the fact $\nu$ in \EQ(\ref{eq:Q}) is a monotonically increasing function of 
$\hat\nu$. This implies that their {\it cumulative} distributions must agree, $F_{\Aqnu}^\nu(\nu)\equiv\int^\nu_0 d\nu'f_{\Aqnu}^\nu (\nu')\stackrel{!}{=}\int^{\hat\nu(\nu)}_{\hat\nu(0)} d\hat\nu'f^{\hat\nu} (\hat\nu')\equiv F^{\hat\nu}(\hat\nu(\nu))$,
where $\hat\nu(\nu)$ is the inverse of \EQ(\ref{eq:Q}). Therefore,
$f_{\Aqnu}^\nu(\Anq)
=
d F_{\Aqnu}^\nu(\Anq)
/{d\Anq}
= 
dF^{\hat{\nu}}(\hat{\nu}(\Anq))
/{d\Anq} = 
f(\hat{\nu}(\Anq))\cdot
d(\hat{\nu}(\Anq))
/{d\Anq}
\label{eq:f_nu}
$,
from which \EQ(\ref{eq:f_nu_full}) directly follows.
} 
\begin{equation}
f_{\Aqnu}^\nu = 
\frac{1}{\sqrt{2\pi}\sigma_A}
\begin{cases}
\frac{1}{2}[1 + \frac{\bar{\nu}^2}{\Anq^2}]
\exp\left[{-\frac{(\frac{1}{2}\Anq - \frac{\bar{\nu}^2}{2\Anq} -\bar\nu)^2}{2\sigma_A^2}}\right]
&\hat{\nu} < 0 \\
\exp\left[{-\frac{(\nu-\bar\nu)^2}{2\sigma_A^2}}\right]
 &\hat{\nu}\geq 0
 \end{cases}\,.
\label{eq:f_nu_full}
\end{equation}
Note that the required signal strength to set an upper limit based
on the best-fit value $\hat{\nu}$ thus has an {\it asymmetric} distribution, as a direct 
consequence of the constraint $\Anq\geq0$. For comparison, the distribution of upper limits
for an unconstrained signal strength would simply be
\be
\label{eq:sidebands_symm}
f_{q_\nu}^\nu =\mathcal{N}(\bar{\nu},\Asigma)\,,
\ee
i.e.~as in the second line of Eq.~(\ref{eq:f_nu_full}) but without the restriction to $\hat\nu\geq0$.

An important implication of an asymmetric distribution is the appearance of asymmetric sidebands that describe the variance
of the expected limits.  Recalling that $\hat\nu$ follows a normal distribution with $\langle\hat\nu\rangle=0$, 
we can directly  
read off the `$N\sigma$-bands' of $f_{\Aqnu}^\Anq$ from \EQ(\ref{eq:Q}). Namely, we expect the limit on the signal
strength as derived
from a given data realization to lie within
\be
\nu\in\left[\langle\nu\rangle-\Delta\nu^-,\langle\nu\rangle+\Delta\nu^+\right]\,,
\ee
where
\bea
\langle\nu\rangle &=& \bar\nu \label{eq:asav}\,,\\
\Delta\nu^+ &=&    
 {N}\Asigma \label{eq:asmax}\,,\\
\Delta\nu^- &=& - N\Asigma +
\sqrt{(N\Asigma)^2 + \bar{\nu}^2} 
-  \bar{\nu}\,.
\label{eq:asmin}
\eea

\begin{figure}[t]
\centering
\includegraphics[width=0.49\textwidth]{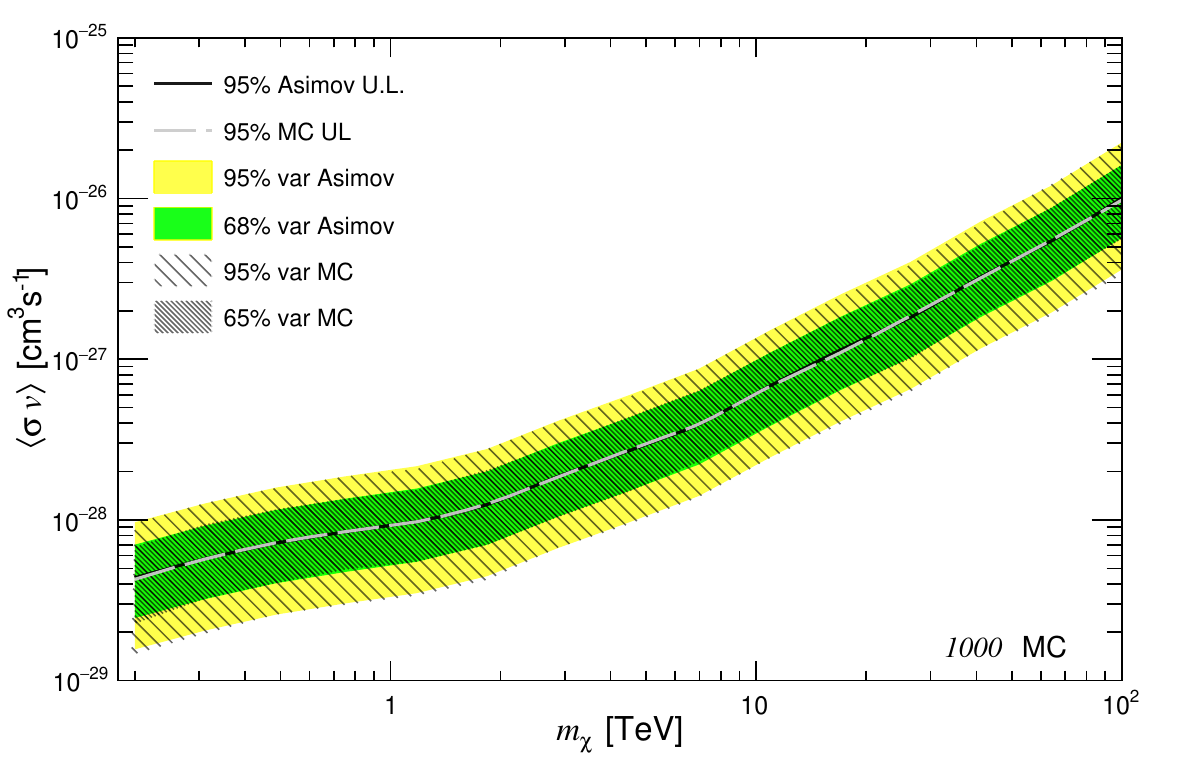}
\includegraphics[width=0.49\textwidth]{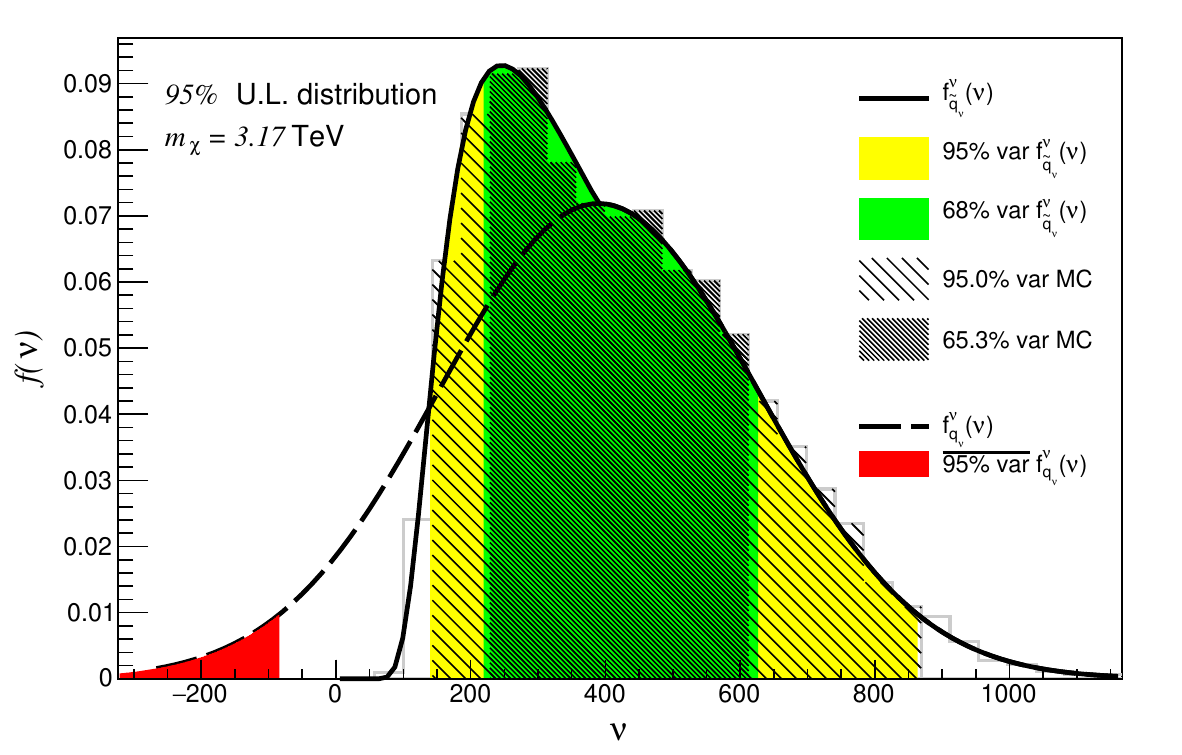}
\caption{\footnotesize{
\textit{Left.} Median and variance of upper limits (UL) from 1000 MC simulations (light dashed line and hatched areas) 
and analytic expressions based on 
the Asimov data set (solid black line and colored areas).
\textit{Right.} The solid line shows the distribution $f_{\tilde q_\nu}^\nu$
of $95\%$ upper limits as given in Eq.~(\ref{eq:f_nu_full}), for $m_\chi=1.6$\,TeV, and the histogram
shows the same quantity as derived from \EQ (\ref{eq:upperlimit}) used on MC data. 
As explained in the text, the skewness of the distribution is due to non-negative signals, $\nu\!\geq\!0$. Hatched bands 
show $68\%$ and $95\%$ quantiles of the MC simulations, respectably, while green and yellow areas show the corresponding
quantities based on Eqs.~(\ref{eq:asmax}, \ref{eq:asmin}).
The dashed line, finally, corresponds to the distribution expected for an unconstrained signal strength parameter; 
here, the region colored in red indicates limits that are more than $2\sigma$ smaller than the mean expectation.}
}
\label{fig:asimov}
\end{figure}

To confirm the validity of our analytic expressions based on the Asimov data set, we computed the upper limit on the signal 
strength at $95\%\,C.L.$ for 1000 MC data sets. These sets 
were generated as Poisson realizations of a power law with a spectral index of $-2.4$ and normalized to the total photon count 
of a GC simulation. 
In the left panel of \FIG\ref{fig:asimov}, we show the median as well as 1\,$\sigma$ and 2\,$\sigma$ sidebands of these 
limits.\footnote{%
The precision to which one can determine quantiles for the distribution of constraints is limited by the bin size. 
In order to stress this aspect, 
we refrain from interpolating between bins, and instead quote the percentage of the distribution that is 
covered by entire bins (closest to $1\,\sigma$ and $2\,\sigma$ bands, respectively).
}
For comparison, we also show these quantities as computed from Eqs.~(\ref{eq:asav}-\ref{eq:asmin}),
with $\Asigma$ estimated from the Asimov data set as $\Asigma = {\nu}\big({{\Aqnu}^{A,0} }\big)^{-1/2}$ for
various pairs of $(\nu, \Aqnu)$  and $\Aqnu=\text{TS}(\nu)$ as given in Eq.~(\ref{eq:MLR});
here, the Asimov data set is treated as an `MC toy' \textit{without} any statistical fluctuations.
Clearly, the agreement is excellent.

In the right panel of \FIG\ref{fig:asimov}, we compare the distribution of $95\%$ upper limits found in the MC simulations 
(histograms)
to the analytical expression in \EQ(\ref{eq:f_nu_full}) (solid line), for a DM signal located at $m_\chi=1.6$\,TeV. 
The skewness of the distributions is clearly visible and, again, the agreement is very good. For comparison, we also 
show with dashed lines the very different distribution that would result if the signal normalization could also be negative, 
i.e.~Eq.~(\ref{eq:sidebands_symm}). The area outside the $95\%\,C.L.$ of this distribution (marked in red)
would in fact only cover {\it negative} signal normalizations.

\subsection{Systematic uncertainty}
\label{app:corr}

\begin{figure}
\centering
\includegraphics[width=0.48\linewidth]{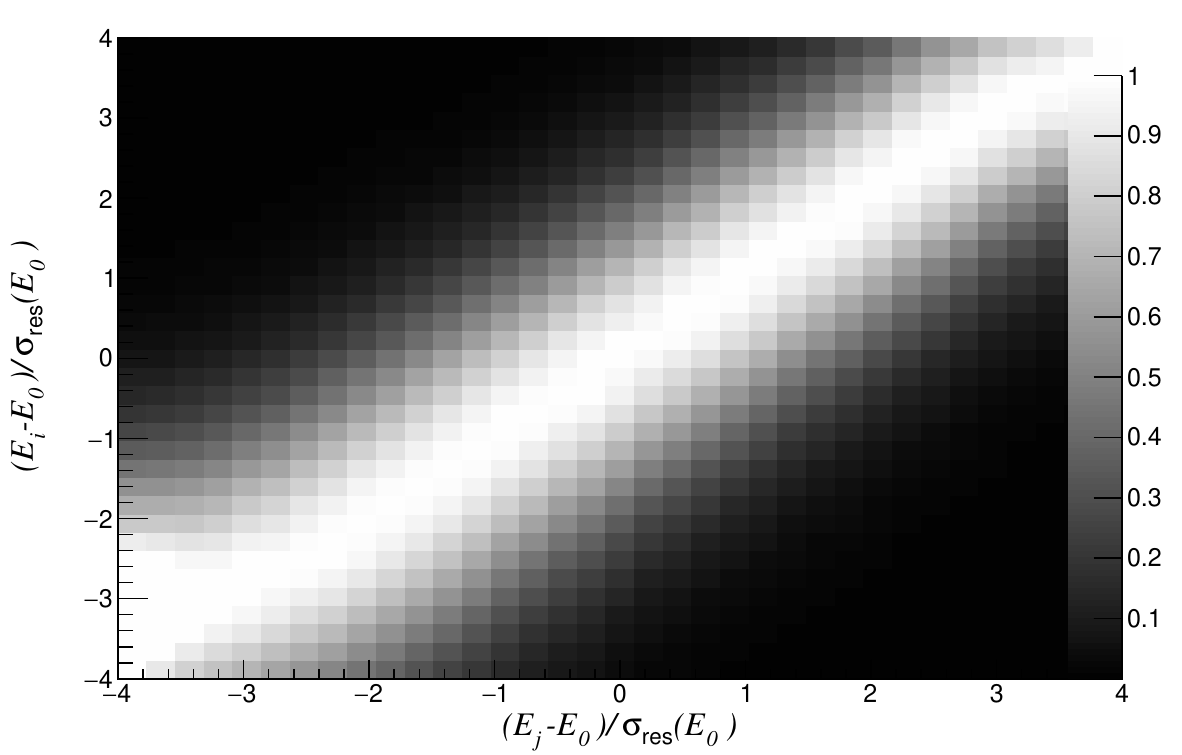}
~~~
\includegraphics[width=0.48\linewidth]{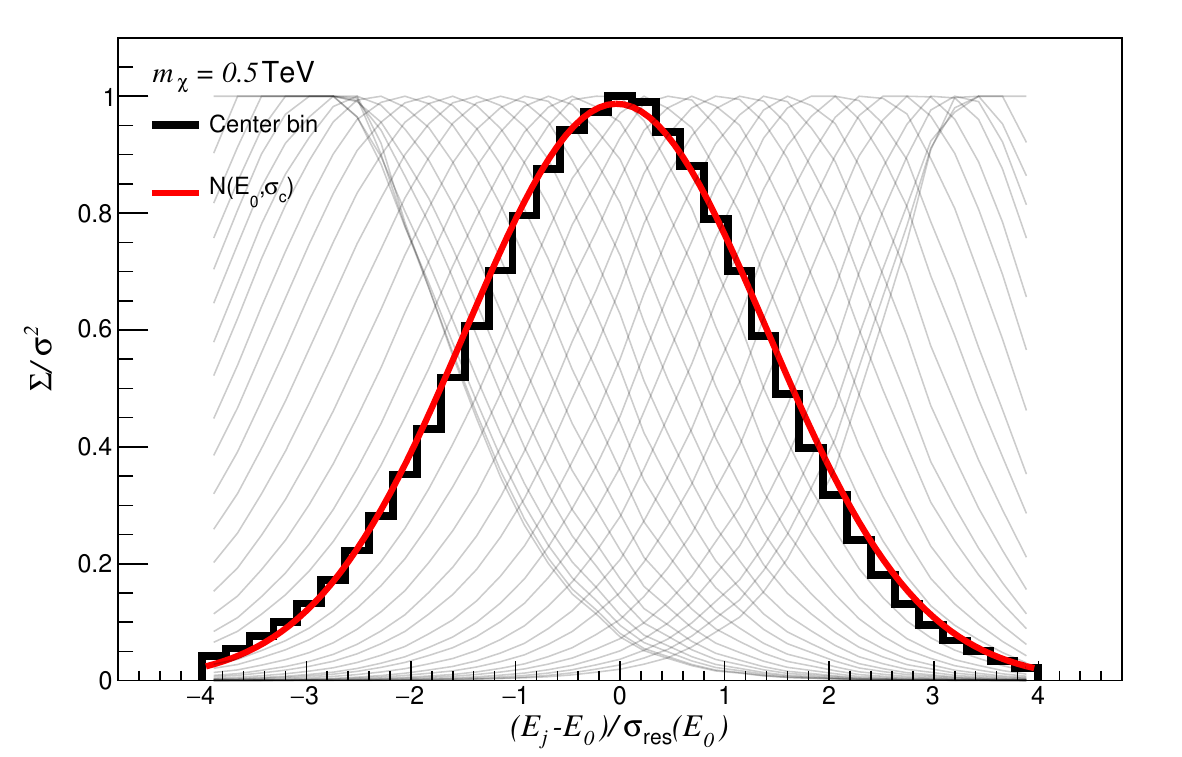}
\caption{{\it Left.} The covariance matrix of Gaussian noise convoluted with the energy response matrix of CTAO, rescaled
by the variance $\sigma$ of the noise.
Photon energy bins $i,j$ are centered around the true mean energy $E_0$, which for the purpose of this plot we set to $E_0=0.5$\,TeV, and stated in units of the energy resolution  
$\sigma_\mathrm{res}(E_0)$ shown in Fig.~\ref{fig:cta}.
{\it Right.} 
The solid black line shows the center bin of the rescaled covariance matrix, $\Sigma/\sigma^2=\widetilde{M} \widetilde{M}^T$, 
and the red line is the best-fit normal distribution to approximate this function, with variance $\sigma_c=1.44$. For comparison,
thin black lines show 
$(\widetilde{M} \widetilde{M}^T)_{ij}$ for fixed bins $i$ that do {\it not} correspond to the central energy bin $E_0$.}
\label{fig:noise_correlation}
\end{figure}

In this Appendix we discuss {\it i)} the overall level of uncertainty and {\it ii)} 
the form of the covariance (or correlation)
matrix $\Sigma_{ii'}$ introduced in Eq.~(\ref{eq:defsigma}).
Starting with the latter, we recall that this quantity is needed to account for systematic uncertainties due to noise correlations, 
cf.~the likelihoods in 
Eqs.~(\ref{eq:likelihoodSyst}, \ref{eq:likelihoodDSPH}) that we use in our analysis.
We start from the energy response matrix $M_{ij}$, which is contained in the IRF in the form of a $2$-dimensional histogram 
describing the likelihood to reconstruct an energy (bin) $i$ as a function of true energy $j$,
and normalize it as $\widetilde{{M}}_{ij} \equiv {M}_{ij} / \sum_{k} {M}_{kj}$. 
We then generate random noise vectors ${\bm{\epsilon}}$, with each of the components $\epsilon_i$ drawn from a normal
distribution $N(0,\sigma)$ with variance $\sigma$, and convolute them with the normalized energy response matrix to
give $\tilde{\bm{\epsilon}}\equiv\widetilde{{M}}{\bm{\epsilon}}$. We work under the assumption that 
observed photon counts ${\bm \eta}$ are subject to intrinsic fluctuations due to such Gaussian fluctuations, i.e.~that
they can be modelled as ${\bm \eta}=\langle{\bm \eta}\rangle + \tilde{\bm{\epsilon}}$.
The covariance matrix thus becomes
\begin{equation}
\Sigma(\bm{{\eta}},\bm{{\eta}}) 
\equiv\Big\langle
(\bm{{\eta}} -\langle\bm{{\eta}}\rangle)
(\bm{{\eta}} - \langle\bm{{\eta}}\rangle)^T
\Big\rangle
=
\widetilde{M}
\langle\bm{\epsilon},\bm{\epsilon}^T\rangle
\widetilde{M}^T
=
\sigma^2
\widetilde{M}
\widetilde{M}^T\,,
\end{equation}
which we show as a contour plot in the left panel of \FIG\ref{fig:noise_correlation}. For the purpose of this
figure, we choose a central `pivot' energy $E_0=0.5$\,TeV to define bins $i=0$ and $j=0$.
 
In the right panel of the figure  we directly plot $(\widetilde{M}\widetilde{M}^T)_{ij}$ as a function 
of Energy $E_j$, for various discrete choices of $E_i$. We highlight (solid black line) the central 
bin distribution, i.e.~$E_i=E_0$. As illustrated by the red line, the correlation matrix is very well fitted 
by a Gaussian, in this case with variance of $\sigma_c=1.46$ expressed in units of the energy resolution
$\sigma_\mathrm{res}(E_0)$ at energy $E_0$. 
We find that the best-fit value of this dimensionless variance only varies by an amount of 
the order of $10\%$ when considering 
different energies. Such variations do not have any significant impact on our sensitivity results, motivating
us to consistently fix the correlation length at $1.5\times \sigma_\mathrm{res}$ as stated in \EQ(\ref{eq:defsigma}).

We note in passing that the treatment of systematic uncertainties described here is complementary to
 the focus on mostly {\it spatial} correlations in the template fitting adopted in Ref.~\cite{CTA:2020qlo}.
That analysis, in particular, is tailored to spatial pixels of the order of the PSF and energy bins somewhat
larger than the energy resolution; in our analysis, on the other hand, the spatial bins
are much larger than the PSF,
and the energy bins are significantly smaller than the energy resolution.

\begin{figure}
\centering
\includegraphics[width=0.8\linewidth]{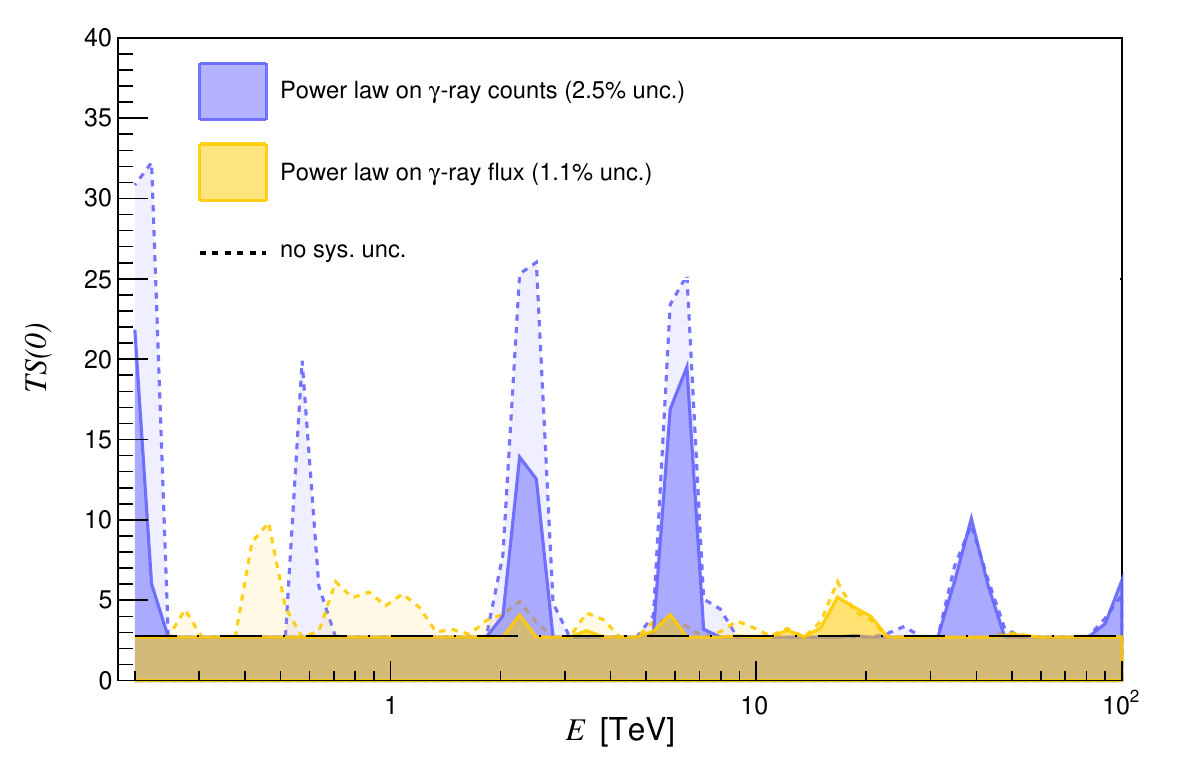}
\caption{The value of the test statistic that is 
necessary to establish a $95\%$ upper exclusion limit for the two different background 
models described in the main text, for an assumed signal at energy $E_0$.
These models describe, respectively, a power law on the total counts (blue) and only 
on the intrinsic gamma-ray spectrum (orange),  which then is convoluted with the IRFs. 
Dotted lines show these values without assuming any systematic uncertainty in the
 analysis, while solid lines show the effect of adding an overall systematic uncertainty $\sigma$ at the indicated level. 
For comparison, the black dashed line shows the value expected for a $\frac{1}{2}\chi^2_1$ distribution.
Results are based on $300$ MC toy simulations of the GC survey.
}
\label{fig:tsdistribution}
\end{figure}

\medskip
\noindent
Let us now turn to the overall systematic uncertainty. As described in \SEC\ref{section:systematics}, we choose a 
value of 2.5\,\% for this quantity in order to correct for the fact that the test statistic, TS$(\nu)$, will realistically speaking
not exactly follow the $\frac{1}{2}\chi^2_1$ distribution one expects if all signal (and background) components are modelled 
perfectly. Here we motivate this choice by computing 
the $95\%$ quantile of the TS$(0)$ distribution from a set of $300$ dedicated MC simulations of the GC analysis.
In \FIG\ref{fig:tsdistribution} we show the result for the two different background models described in \SEC\ref{sec:sliding},
as a function of the monochromatic signal energy $E_0$. Concretely, these two models are given by a power law directly on the 
counts (blue) and a power law on the `intrinsic' gamma ray spectrum (orange) that
only afterwards is convoluted with the IRFs, respectively. Dotted lines show the results without including any systematic 
uncertainty, parameterized by the $\eta_i$ in the construction of the likelihoods outlined in \SEC\ref{sec:statistics}, 
while solid lines show the effect of adding systematic uncertainty at the indicated level.

Compared to the expectation for a $\frac{1}{2}\chi^2_1$ distribution, namely a flat value of $2.71$, one can clearly identify 
several deviations. These can be traced back to the fact that the power-law assumption of the background is not perfect.
Adding some level of overall systematic uncertainty to each of the counts smears out local deviations from a simple power law
and therefore, as also clearly visible in the figure, leads to a reduction of these deviations. In other words, the larger the
assumed statistical uncertainty, the more does the test statistic follow a $\frac{1}{2}\chi^2_1$ distribution (which we assume in 
the analysis, see also Appendix \ref{App:Asimov}).
As expected, modelling only the intrinsic gamma-ray background as a power law (orange) provides a significantly better
description of the counts. This, however, rests on the assumption of essentially perfect IRF modelling -- which clearly 
is challenging to 
achieve in practice. Our benchmark analysis strategy therefore consists in being much more agnostic and instead
modelling the total counts as a power law (blue). This leads to several deviations, most notably at $\sim5$\,TeV, where different 
spectral cuts in the transition region between MSTs and SSTs are expected to give rise to sharp variations in the spectrum
(see also \FIG\ref{fig:background_spatial_bin}, right panel). We consider it sufficient to mitigate these deviations by
introducing an overall uncertainty of $\sigma=0.025$, resulting in a TS distribution that only shows significant 
($\gtrsim4\,\sigma$) deviations
around the mentioned $\sim5$\,TeV feature (which could be addressed by a dedicated treatment in the presence of real data).
We also note that even an extremely conservative choice of $\sigma=0.05$ would at most decrease the estimated sensitivity
by a factor of 2 at the lowest DM masses, cf.~\FIG\ref{figure:limit-ratio-systematic} in the main text.

\newpage
\bibliographystyle{JHEP}
\bibliography{bibl}

\end{document}